\documentclass[preprint,showkeys,secnumarabic,amsfonts,showpacs,amsmath,amssymb]{revtex4}
\usepackage[dvips]{color}
\usepackage{array}
\usepackage{rotating}
\usepackage{epsfig}
\usepackage{amsmath}
\usepackage{graphicx}
\begin{document}
\title{ Searching for a Cosmological Preferred Axis in complicated class of cosmological models: Case study $f(R, T)$ model}

\author{A. Salehi}
\email{salehi.a@lu.ac.ir} \affiliation{Department of Physics, Lorestan University, Lorestan, Iran}
\author{S. Aftabi}
\email{sajjad.aftabi@gmail.com} \affiliation{Young Researchers and Elite Club, Rasht Branch, Islamic Azad University, Rasht, Iran}

\date{\today}

\begin{abstract}
 \noindent \hspace{0.35cm}
Recent astronomical observations show that the universe may be anisotropic on large scales. The Union2 SnIa data hint that the universe has a preferred direction. If such a cosmological privileged axis indeed exists, one has to consider an anisotropic expanding Universe instead of the isotropic cosmological model. In this paper, we present a detailed analysis of the dark energy dipole in $f(R,T)=f_{1}(R)+f_{2}(T)$ Cosmological Model using three types of dipole fit (DF) method which are (I)dipole + monopole fitting for distance modulus(DMFDM), (II)dipole + monopole fitting for luminosity distance(DMFLD) and (III) general dipole fitting for luminosity distance(GDFLD). We have found the maximum anisotropic deviation direction for (DMFDM) method as  $(l, b)=(315^{+25}_{-25},-23^{+14}_{-15})$, for (DMFLD) as $(l, b)=(l, b)=(315^{+35}_{-37},-23^{+18}_{-18})$, and  for (GDFLD) method as $(l, b)=(317^{+32}_{-32},-23^{+18}_{-18})$ which are located very close to each other. We compare our model with the $CPL$, $\Lambda CDM$ and $\omega CDM$ models. Constraints on $(l, b)$ in $f(R, T)$ model are not much different from the cases of the $CPL$, $\Lambda CDM$ and $\omega CDM$ models. Moreover, the results are consistent with other studies.

\end{abstract}



\maketitle
\newpage
\section{introduction}
Cosmological principle is one of the basic assumptions of modern cosmology. According to the cosmological principle, the Universe is homogenous and isotropic on scales larger than a few hundred Mpc, which is consistent with currently observational data sets such as the Cosmic Microwave Background (CMB) radiation data from the Wilkinson Microwave Anisotropy Probe (WMAP)(\cite{Komats}-\cite{Hinshaw}). However, recent observational evidence  included  Large Scale Velocity Flows(\emph{DarkFlow}) (\cite{Watkins}) anisotropy in the Values of the Fine Structure Constant $\alpha$ ($\alpha$ \emph{Dipole})(\cite{Webb1}, \cite{Mariano}), anisotropy in Accelerating Expansion Rate (\emph{Dark Energy Dipole})(\cite{a122}, \cite{a128}), and other effects (\cite{s18}-\cite{Antoniou}-\cite{Perivolaropoulos}) indicate that the Universe may be anisotropic on large scales. A number of authors have investigated the anisotropies of the cosmic acceleration (\cite{Gordon1}--\cite{a122}), which was motivated in several aspects. In particular, several groups such as \cite{Schwarz}--\cite{Kalus}  have applied the hemisphere comparison method to study the anisotropy of $\Lambda CDM$, $\omega CDM$ and the dark energy model with $CPL$ parametrization. Some previous works payed attention to  study the anisotropic expansion of the universe (Dark Energy Dipole) using the SNIa data and found statistically significant evidence for anisotropies.\\
More recently, \cite{Antoniou} have applied the hemisphere comparison method to the standard $\Lambda CDM$ model and found that the hemisphere of maximum accelerating expansion is in the direction $(l,b)=(309^{-23}_{+23},18^{-10}_{+11})$ with Union2 data. \cite{Schwarz} took use of the hemisphere comparison method to fit the $\Lambda CDM$ model to the supernovae data on several pairs of opposite hemispheres, and a statistically significant preferred axis was found.\\
\cite{Chang1} have investigated the anisotropic Cosmological model in the Randers space-time. They found the preferred direction as $(l, b) =(306, -18)$. \cite{Cai0} have  taken the deceleration parameter $q_{0}$ as the diagnostic to quantify the anisotropy level in the $\omega CDM$ model.\\
\cite{Cai} constructed a direction-dependent dark energy model based on the isotropic background described by the $\Lambda CDM$, $\omega CDM $ and $CPL$ models and employed  the Union2 dataset to constrain the anisotropy direction and strength of modulation. They found the best-fitting value of the maximum deviation direction from the isotropic background is not sensitive to the details of isotropic dark energy models.\\
\cite{Wang} have studied dipolar anisotropic expansion with cosmographic parameters. They found $(l,b)=(309^{\circ}, -8.6^{\circ})$.\\
\cite{Yang} chose two simple cosmological models, $\Lambda CDM$ and $\omega CDM$ for the hemisphere comparison approach, and $\Lambda CDM $ for the dipole fit. In the first approach, they used the matter density and the equation of state of dark energy as the diagnostic qualities in the $\Lambda CDM$ and $\omega CDM$, respectively. In the second method, they employed distance modulus as the diagnostic quality in $\Lambda CDM$. They found a preferred direction of $(l,b)=(307^{\circ}, -14^{\circ})$.\\
In testing for anisotropy or consistency with isotropy, we can ask which cosmological probes are most sensitive in what redshift ranges to such a hypothetical anisotropy, i.e. what constraints could be put on angular variations in the local dark energy equation of state.\\
We cannot make a convincing conclusion from only one dataset, model or method about the origin of the anisotropy. Anisotropy may come from systematic uncertainty, as well as the intrinsic property of the universe. If the privileged axes derived from different datasets, different methods and different cosmological models are close to each other, we can safely conclude that anisotropy is an intrinsic property of the Universe. As we mentioned above several studies payed attention to find a preferred axis of the Universe in isotropic background described by the $\Lambda CDM$, $\omega CDM $ and $CPL$ models; however, possibility of existence a privileged axis for the Universe may enhance if modified cosmological models such as f(R,T) could predict it and produce closely result. Searching for preferred cosmological axis of the Universe, we focus on generalized gravity model $f(R,T)$.\\

Recently, the observations of high redshift type Ia supernovae, the surveys of clusters of galaxies (\cite{Reiss}- \cite{Riess2}), Sloan digital sky survey ({\bf SDSS})(~\cite{Abazajian}) and Chandra X--ray observatory (~\cite{Allen}) reveal the universe accelerating expansion and that the density of matter is very much less than the critical density. Also, the observations of Cosmic Microwave Background (CMB) anisotropies (\cite{Bennett}) indicate that the universe is flat and the total energy density is very close to the critical one (\cite{Spergel}). The observations though determines basic cosmological parameters with high precisions and strongly indicates that the universe presently is dominated by a smoothly distributed and slowly varying dark energy (DE) component, but at the same time they poses a serious problem about the origin of DE (\cite{Tsujikawa}). The most The 'cosmological constant' as  the best candidate for explaining cosmic acceleration in literature faces serious problems such as fine-tuning and a huge discrepancy between theory and observations (\cite{Copeland}-\cite{Tsujikawa2}). On the other hand, modification of the geometrical part of the Einstein-Hilbert action by replacing an arbitrary function of the Ricci scalar $R$ (\cite{Nojir}) has constructed well-developed dark energy models. This phenomenological approach is called as the Modified Gravity. Using the Modified Gravity we can strongly explain the rotation curves of galaxies, the motion of galaxy clusters, the Bullet Cluster, and cosmological observations without the use of dark matter or Einstein's cosmological constant (\cite{Nojiri}-\cite{Felice}). Cosmic inflation, mimic behavior of dark matter and current cosmic acceleration being compatible with the observational data are other successful predications of the $f(R)$ theories (\cite{Nojiri}-\cite{Starobinsky}).\\

A generalization of $f(R)$ modified theories of gravity was proposed in \cite{Bertolami} studies, by including in the theory an explicit coupling of an arbitrary function of the Ricci scalar R with the matter Lagrangian density $L_{m}$. A specific application of the latter $f(R,L_{m})$ gravity was proposed in \cite{Poplawski} studies, which may be considered a relativistically covariant model of interacting dark energy, based on the principle of least action. The cosmological constant in the gravitational Lagrangian is a function of the trace of the stress-energy tensor, and consequently the model was denoted "$\Lambda(T )$ gravity". It was argued that recent cosmological data favor a variable cosmological constant, which are consistent with $\Lambda(T )$ gravity, without the need to specify an exact form of the function $\Lambda(T )$ (\cite{Poplawski}). $\Lambda(T )$ gravity is more general than the Palatini $f(R)$ gravity, and reduces to the latter when we neglect the pressure of the matter.\\

In this paper, a class of the Modified Gravity theories in which the gravitational action contains a general function $f(R, T )$, where R denotes the Ricci scalar and T is the trace of the energy-momentum tensor, has been considered. \cite{Harko} introduced this type of the Modified Gravity,$f(R, T )$, which obtained significant outcomes: the reconstruction of cosmological solutions, where late-time acceleration was accomplished by \cite{Houndjo} and the energy conditions was analyzed by \cite{Alvarenga}. \cite{Sharif} studied the thermodynamics of Friedmann-Lemaˆýtre-Robertson-Walker (FLRW) spacetimes. Moreover, the occurrence possibility of future singularities was studied by (\cite{Houndjo2}). Besides these achievements, a serious shortcoming in this kind of theory has been the non-conservation of the energy-momentum tensor. To circumvent this problem, in this paper, we show that $f(R, T )$ functions can always be constructed in a way to be consistent with the energy-momentum tensor standard conservation. In this regard, we can  assume separable algebraic functions of the form $ f(R, T ) = f_{1}(R) + f_{2}(T )$ in which the function $f_{2}(T )$ is obtained by imposing the conservation of the energy-momentum tensor. Now, in order to search for dipolar asymmetry, we construct an anisotropic dark energy model and aim to detect the maximum anisotropy direction. Furthermore, we consider the impact of redshift on the direction by using the redshift tomography method, with the  Union2 data. Finally, we compare our results for the f(R,T) model with $\Lambda CDM ,\omega CDM $ and $CPL$ models and also some previous studies.\\

The paper is structured as follows: In Section 2, we obtain the field equations of $f(R, T)$ gravity and analyze the stability of the dynamical system of the f(R,T) model. In section 3, we discuss free parameters of the model in some detail and constrain these free parameters using observational data. In section 4, we investigate the scalar perturbations in $f(R, T ) =f_1(R)+f_2(T)$ type theories. In order to search for Dark Energy Dipole in the model using observational data, in section 5 we describe some important anisotropy models and method. Then we introduce and extend types of Dipole-Fitting method in order to investigate possible anisotropy from the data. We compare the results of these three types of DF method used for the f(R,T) model with each other in Section 6. Moreover, in order to explore the possible redshift dependence of the anisotropy, we have implemented a redshift tomography analysis in Section 7. In section 8, we have performed the anisotropy analysis for $CPL$ , $\Lambda CDM$ and $\omega CDM$ models. We have applied $DMFLD$ method to find the anisotropy of $\Lambda CDM$ and $\omega CDM$  and the dark energy model with $CPL$ parametrization in order to make a comparison between these models and the $f(R,T)$ model. Finally, in section 9, we conclude, summarize and compare the results of this work with some of the recent studies of \cite{Mariano}, \cite{Webb1}, \cite{Cai}, searching for evidence for a preferred cosmological axis.\\

\section{FIELD EQUATIONS OF $f(R,T)$ MODEL}
The action of $f(R, T)$ gravity is of the form
\begin{align}\label{action}
S=\int \sqrt{-g} d^{4} x \left[\frac{1}{16 \pi G} f(R,T^{(\textrm{m})})+L^{\textrm{(m)}}+L^{\textrm{(rad)}} \right],
\end{align}
where $R$ is Ricci scalar, $f(R,T^{\textrm{(m)}})$ is an arbitrary function of the Ricci scalar and $T^{\textrm{(m)}}$, $L^{\textrm{(m)}}$ and $L^{\textrm{(rad)}}$ are the Lagrangian of the dust matter and radiation, $g$ is the determinant of the metric, $T^{\textrm{(m)}}\equiv g^{\mu \nu}
T^{\textrm{(m)}}_{\mu \nu}$ is the trace of the energy--momentum tensor and we set $c=1$.
By varying the action (\ref{action}), with respect to the metric tensor $g_{\mu\nu}$, the field equations are obtained as
\begin{align}\label{field tensor}
f_{R}(R,T) R_{\mu \nu}-\frac{1}{2} f(R,T) g_{\mu \nu}+\\ \nonumber \Big{(} g_{\mu \nu}
\square - \bigtriangledown_{\mu} \bigtriangledown_{\nu}\Big{)}f_{R}(R,T) =\\  \nonumber \Big{(}8
\pi G+ f_{T}(R,T)\Big{)} T^{\textrm{(m)}}_{\mu \nu}+8 \pi G  T^\textrm{(rad)}_{\mu \nu},\
\end{align}

where $\nabla $  denotes the covariant derivative and
\begin{align}
\square \equiv\bigtriangledown^{\mu}\bigtriangledown_{\mu},\ \ f_{T}(R,T) \equiv \frac{\partial f(R,T)}{\partial T},\ \ \\ f_{R}(R,T) \equiv \frac{\partial f(R,T)}{\partial R},\ \ g^{\alpha \beta} \frac{\delta T^{\textrm{(m)}}_{\alpha \beta}}{\delta g^{\mu \nu}}=-2T^{\textrm{(m)}}_{\mu \nu}.\nonumber
\end{align}\\
contracting of equation (\ref{field tensor}) yields
\begin{equation}\label{trace tensor}
f_{R}(R,T)R+3 \square f_{R}(R,T)-2f(R,T)=\Big{(}8 \pi G+ f_{T}(R,T)\Big{)}T.
\end{equation}

Now, in this model, we assume the perfect fluid and the spatially
flat Friedmann--Lema\^{\i}tre--Robertson--Walker (FLRW) metric
\begin{equation}\label{metricFRW}
ds^{2}=-dt^{2}+a^{2}(t) \Big{(}dx^{2}+dy^{2}+dz^{2}\Big{)},
\end{equation}
where $a(t)$ is the scale factor. Let us
 rewrite~(\ref{field tensor}) as a standard form similar to GR, i.e.
\begin{equation}\label{FRW}
G_{\mu \nu}=\frac{8 \pi G}{f_{R}(R,T)} \left(T^{\textrm{(m)}} _{\mu \nu}+T^{\textrm{(rad)}} _{\mu \nu}+T^{\textrm{(eff)}} _{\mu \nu}\right),
\end{equation}

where
\begin{align}\label{Generalized einstein}
T^{\textrm{(eff)}} _{\mu \nu} \equiv \frac{1}{8 \pi G}[\frac{1}{2}
\Big{(}f(R,T)  -f_{R}(R,T)R\Big{)}g_{\mu \nu} +\\ \nonumber \Big{(} \bigtriangledown_{\mu}
\bigtriangledown_{\nu}- g_{\mu \nu} \square\Big{)}f_{R}(R,T)+f_{T}(R,T) T^{\textrm{(m)}}_{\mu \nu}]
\end{align}
Regarding the Bianchi identity, obviously in $f(R,T)$ gravity, the above effective energy--momentum tensor is~not conserved. Thus, by applying the conservation of the energy--momentum tensor of the whole matter, i.e. $\nabla^{\mu}T^{\textrm{(m)}}_{\mu \nu}=0=\nabla^{\mu}T^{\textrm{(rad)}}_{\mu \nu}$, the following constraint must be held. That is
\begin{equation}\label{source}
\frac{3}{2} H(t) f_{T}(R,T)=\dot f_{T}(R,T),
\end{equation}
where $H(t)$ is the Hubble parameter and dot denotes the derivative with respect to the time $t$. Equations (\ref{field tensor}) and (\ref{trace tensor}), by assuming metric (\ref{metricFRW}), give
\begin{align}\label{first}
3H^{2}f_{R}(R,T)+\frac{1}{2} \Big{(}f(R,T)-f_{R}(R,T)R\Big{)}\\ \nonumber+3\dot{f_{R}} (R,T)H=\Big{(}8
\pi G +f_{T}(R,T)\Big{)}\rho^{\textrm{(m)}}+8 \pi G\rho^{\textrm{(rad)}} ,
\end{align}
as the Friedmann--like equation, and
\begin{eqnarray}\label{second}
2f_{R}(R,T) \dot{H}+\ddot{f_{R}} (R,T)-\dot{f_{R}} (R,T) H=\\ \nonumber -\Big{(}8 \pi G +f_{T}(R,T)\Big{)}\rho^{\textrm{(m)}}-\frac{32}{3} \pi G\rho^{\textrm{(rad)}},
\end{eqnarray}
as the Raychaudhuri--like equation.\\
\cite{Harko} gave three classes of these models\\
                              $f(R,T)= \left\{
\begin{array}{ll}
 R + 2f(T )\\ f_{1}(R) + f_{2}(T )\\ f_{1}(R) + f_{2}(R )f_{3}(T)
\end{array}
\right.
$\\

In this paper, we have focused on the second class $f(R,T)=f_{1}(R)+f_{2}(T)$. Now, by rewriting equations (\ref{first}) and (\ref{second}) for this model, one can obtain
\begin{align}\label{eom1}
\frac{R}{6 H^{2}}- \frac{\dot{f_{1}'}(R)}
{H f_{1}'(R)}-\frac{f_{1}(R)}{6H^{2} f_{1}'(R)}-\frac{f_{2}(T)}{6 H^{2} f_{1}'(R)}\\ \nonumber +\frac{8 \pi G \rho^{\textrm{(m)}}}{3H^{2} f_{1}'(R)} +\frac{f_{2}'(T) \rho^{\textrm{(m)}}}
{3H^{2} f_{1}'(R)}+\frac{8 \pi G \rho^{\textrm{(rad)}}}{3H^{2} f_{1}'(R)}=1
\end{align}
and
\begin{align}\label{eom2}
\frac{\dot{H}}{H^{2}}=-\frac{\ddot{f_{1}'}(R)}{2H^{2} f_{1}'} +\frac{\dot{f_{1}'}(R)}{2H f_{1}'(R)} \\ \nonumber -\frac{4
\pi G \rho^{\textrm{(m)}}}{H^{2} f_{1}'(R)}-\frac{f'_{2}(T) \rho^{\textrm{(m)}}}{2H^{2} f_{1}'(R)}
-\frac{16\pi G \rho^{\textrm{(rad)}}}{3H^{2} f_{1}'(R)}.
\end{align}
For more simplicity, we introduce a few independent new variables as
\begin{align}
&\zeta\equiv-\frac{\dot{f_{1}'}(R)}{H f_{1}'(R)},\ \
\eta\equiv-\frac{f_{1}(R)}{6 H^{2} f_{1}'(R)},\ \
\vartheta\equiv \frac{R}{6 H^{2}},\ \
\\& \nonumber
\xi\equiv-\frac{f_{2}(T)}{3 H^{2} f_{1}'(R)},\ \
\chi\equiv\frac{8\pi G \rho^{\textrm{(rad)}}}{3 H^{2} f_{1}'(R)},\ \
\nu\equiv-\frac{Tf'_{2}(T)}{3 H^{2} f_{1}'(R)}\ \
\end{align}
where the prime denotes differentiating with respect to the argument and we have used $R=6(\dot{H}+2H^2)$ for metric (\ref{metricFRW}). By applying the conservation equation (\ref{source}) for the minimal combination, $f(R, T ) = f_{1}(R) + f_{2}(T )$ gives
\begin{align}\label{e1}
Tf_{2}^{''}(T)=-\frac{1}{2}f_{2}^{'}(T)
\end{align}
This constraint restricts its form to a particular one, namely,
\begin{align}\label{ft}
f(R, T ) = f_{1}(R) + c_{1}\sqrt{-T}+c_{2}
\end{align}
where $c_{1} $ and $c_{2}$ are constants with respect to T. The conservation of the energy–momentum tensor also leads to the case in which the variable $\nu$ is a function of $\xi$, namely,$\nu=\frac{\xi}{2}$. Therefore, these six variables will reduce to five independent variables once the constraint equation (\ref{source}) is applied.
\begin{align}
&\frac{d \zeta}{d N}= -1+\zeta (\zeta-\vartheta) -3\eta -\vartheta -\frac{3}{2} {\xi}+\chi \label{minima 1},\\
&\frac{d \eta}{d N}= \frac{\zeta \vartheta}{\alpha} +\eta\left(4+\zeta -2\vartheta \right)\label{minima 2},\\
&\frac{d \vartheta}{d N}=- \frac{\zeta \vartheta}{\alpha} +2\vartheta \left(2 -\vartheta\right)\label{minima 3},\\
&\frac{d \xi}{d N}=\xi\left(\frac{5}{2}+\zeta-2\vartheta\right)\label{minima 4},\\
&\frac{d \chi}{d N}=\chi\left(\zeta-2\vartheta\right)\label{minima 5},
\end{align}
Where $N$ represents derivatives with respect to $\ln a$ and $\alpha \equiv \frac{R f_{1}''(R)}{f_{1}'(R)}$ which for constant value of $\alpha$ gives
\begin{align}\label{fr}
f_{1}(R)=C_{1R}R^{\alpha+1}+C_{2R}
\end{align}
where $C_{1R}$ and $C_{2R}$ are constants.

\begin{table*}
\centering \caption{The fixed points solutions of the dynamical system problem of $f(R,T)=f_{1}(R)+f_{2}(T)$.}
\begin{minipage}{200mm}
\begin{tabular}{l l l l}\hline\hline
Fixed points     &Coordinates ($\zeta$, $\eta$, $\vartheta$, $\xi$, $\chi$)
&eigenvalue &$stability$\\[0.5 ex]
\hline
$P_{1}$ & $\left(0,0,0,0,1\right)$&$\left(\frac{5}{2},1,-1,4,4\right)$&$ saddle$\\[0.5 ex]
$P_{2}$ & $\left(-4,5,0,0,0\right)$&$\left(\frac{-3}{2},-5,-4,-3,\frac{4(\alpha+1)}{\alpha}\right)$&$ stable for $ $-1<\alpha<0$\\[0.5 ex]
$P_{3}$ & $\left(1,0,0,0,0\right)$&$\left(1,2,5,\frac{7}{2},\frac{4(\alpha-1)}{\alpha}\right)$& $unstable$ \\[0.5 ex]
$P_{4}$ & $\left(-1,0,0,0,0\right)$&$\left(3,\frac{3}{2},-2,-1,\frac{4\alpha+1)}{\alpha}\right)$& $saddle$\\[0.5 ex]
$P_{5}$ & $\left(-\frac{5}{2},0,0,~\frac{7}{2},0\right)$&$\left(\frac{-7}{2},\frac{-5}{2},\frac{-3}{2},\frac{3}{2},\frac{8\alpha+5}{2\alpha}\right)$&$saddle$\\[0.5 ex]
$P_{6}$ & $\left(\frac{4\alpha}{1+\alpha},-\frac{2\alpha}{(1+\alpha)^2},\frac{2\alpha}{1+\alpha},0,-\frac{5\alpha^{2}+2\alpha+1}{(1+\alpha)^2}
\right)$&$\left(\frac{5}{2},1,4,\frac{\alpha-1+\delta_{4}}{2\alpha+1},\frac{\alpha-1-\delta_{4}}{2\alpha+1}\right)$&$saddle$\\
$P_{7}$ & $\left(0,-1,2,0,0\right)$&$\left(\frac{-3}{2},-4,-3,\frac{-3\alpha+\delta_{3}}{2\alpha},-\frac{3\alpha+\delta_{3}}{2\alpha}\right)$&$ stable focus 0<\alpha<\frac{16}{25} $\\[0.5 ex]
$ $ & & &$  stable \frac{16}{25}<\alpha<1$\\[0.5 ex]
$P_{8}$&$\left(\frac{2(1- \alpha)}{1 + 2\alpha},\frac{1-4 \alpha}{\alpha (1+2\alpha)},-\frac{(1-4\alpha)(1+\alpha)}{\alpha(1+2\alpha)},0,0\right)$
&$\left(-\frac{2(5\alpha^2+2\alpha-1)}{\alpha(2\alpha+1)},-\frac{8\alpha^2+3\alpha-2}{\alpha(2\alpha+1)},\frac{1-4\alpha}{\alpha},-\frac{2(\alpha^2-1)}{\alpha(2\alpha+1)},-\frac{10\alpha^2+3\alpha-4}{2\alpha(2\alpha+1)}\right)$&$stable for \alpha<-1$\\[0.5 ex]
$ $ & & &$  0 <\alpha<\frac{-1}{2}$\\[0.5 ex]
$ $ & & &$   \alpha>1$\\[0.5 ex]
$P_{9}$&$\left(\frac{3 \alpha}{1 + \alpha},~-\frac{1 + 4 \alpha}{2 (1 + \alpha)^2},\frac{1+4\alpha}{2(1+\alpha)},0,0\right)$&$\left(3,\frac{3}{2},-1,\frac{-3\alpha+\delta_{2}}{4\alpha(\alpha+1)},-\frac{3\alpha+\delta_{2}}{4\alpha(\alpha+1)}\right)$&$saddle$\\[0.5 ex]
$P_{10}$&$\left(\frac{3\alpha}{2(1+\alpha)},-\frac{5+8\alpha}{4(1+\alpha)^2},\frac{5+8\alpha}{4(1+\alpha)},\frac{4-\alpha(3 + 10 \alpha)}
{4(1+ \alpha)^2},0\right)$& $\left(\frac{-5}{2},\frac{-3}{2},\frac{3}{2},\frac{1}{8\alpha(\alpha+1)}(-6\alpha^2-9\alpha+\delta_{1}),
-\frac{1}{8\alpha(\alpha+1)}(6\alpha^2+9\alpha+\delta_{1})\right)$&$saddle$\\[0.5 ex]
 \hline\hline
\end{tabular}
\label{table:1}
\end{minipage}
\end{table*}

Note  that  the second order nonlinear differential equations of systems was simplified  to first order differential equations by introducing a few new variables. It is interesting to consider the behavior of systems around the equilibrium points($
\frac{d \zeta}{d N}=0
\frac{d \eta}{d N}= 0
\frac{d \vartheta}{d N}=0
\frac{d \xi}{d N}=0
\frac{d \chi}{d N}=0$) using jacobian stability analysis.\\
   The Jacobin stability of a dynamical system can be regarded as the robustness of the system to small perturbations of the whole trajectory. This is a very
convenable way of regarding the resistance of limit cycles to small perturbation of trajectories. It gives us the possibility of study all the evolutional
paths admissible for all initial conditions (\cite{amin1}-\cite{amin4}). It is especially important in cosmology where there is the problem of initial conditions. Using the dynamical systems methods one hopes to answer the question of what is the range of initial conditions and parameters of the system for which the subsequent evolution is compatible with current
\begin{align}
J=
\begin{pmatrix}
2\zeta-\vartheta & -3 & -\zeta-1 & \frac{-3}{2} & 1\\
\frac{\vartheta}{\alpha}+\eta & 4+\zeta-2\vartheta & \frac{\zeta}{\alpha}-2\eta & 0 & 0\\
\frac{-\vartheta}{\alpha} & 0 & \frac{-\zeta}{\alpha}+4-4\vartheta & 0 & 0\\
\xi & 0 & -2\xi & \frac{5}{2}+\zeta-2\vartheta & 0\\
\chi & 0 & -2\chi & 0 & \zeta-2\vartheta\\
\end{pmatrix},
\end{align}
\begin{align}
\delta_{1}=\sqrt{676\alpha^4+700\alpha^3-55\alpha^2-16\alpha} \nonumber\\
\delta_{2}=\sqrt{256\alpha^4+160\alpha^3-31\alpha^2-16m} \nonumber\\
\delta_{3}=\sqrt{25\alpha^2-16\alpha} \nonumber\\
\delta_{4}=\sqrt{81\alpha^2+30\alpha-15} \nonumber\\
\end{align}\\
Evaluating  the Jacobian matrix at the steady state and computing the corresponding eigenvalues of them, we can investigate stability or instability based on the real parts of the eigenvalues. Table \ref{table:1} shows the property of critical points of dynamical system. Fig.1 and Fig.2 demonstrate the attractor property of the dynamical system in the 3-Dimensional phase plane from different perspectives. Imprecisely speaking, the trajectories of the phase space approach to a fixed point if all eigenvalues get negative values, and recede from a fixed point if all eigenvalues have positive values. However, the fixed points occurring in the former and the latter sets are called the stable and unstable points, respectively. The fixed points with both positive and negative eigenvalues are called saddle points, and those trajectories which approach to a saddle fixed point along some eigenvectors may recede from it along some other eigenvectors.\\

\begin{figure}
\centering
\includegraphics[scale=.3]{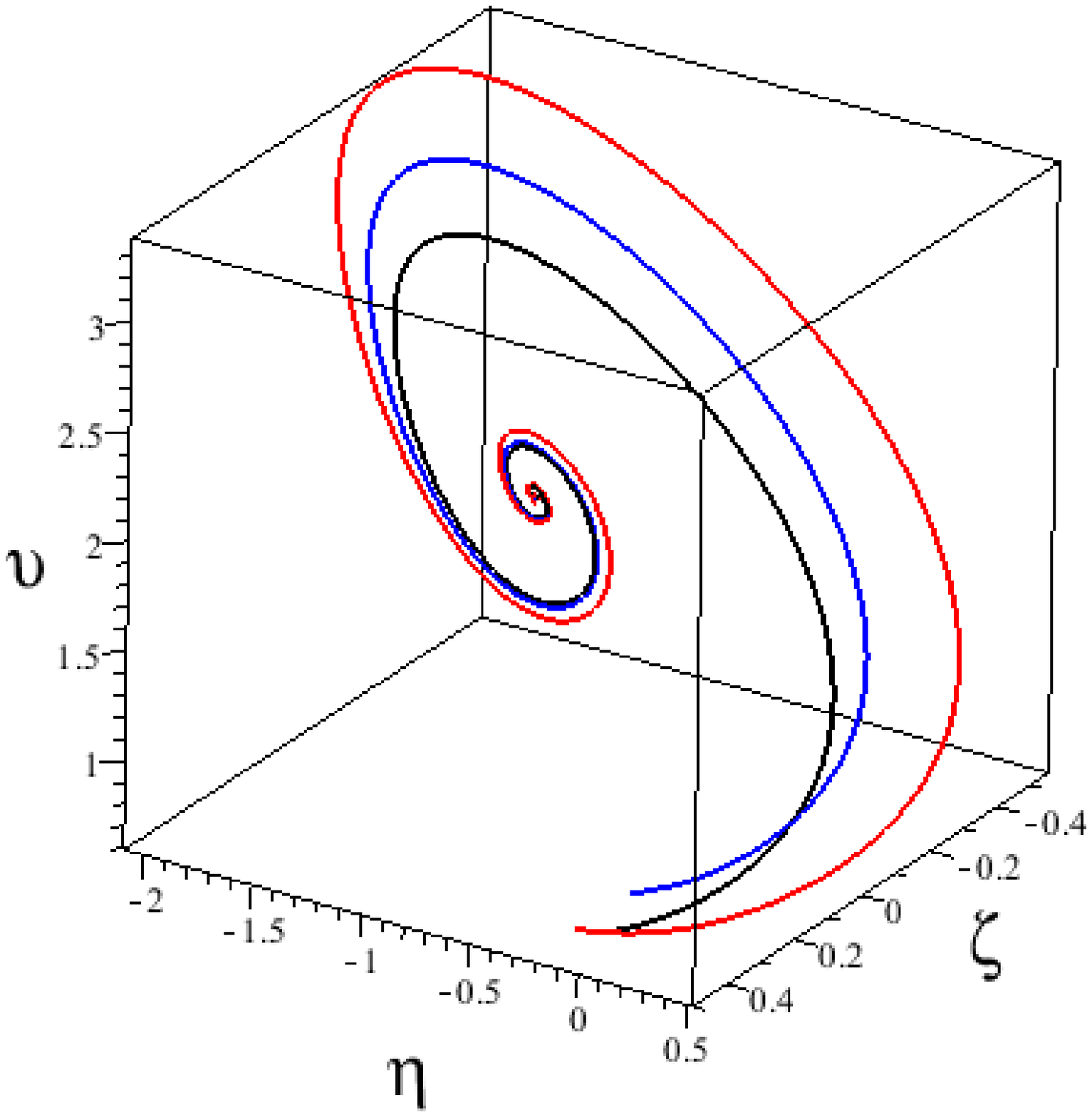}\hspace{0.1 cm}\\
Fig. 1: The attractor property of the dynamical system in the 3-Dimensional phase plane. \
\label{Figure:1}
\end{figure}
\begin{figure}
\centering
\includegraphics[scale=.3]{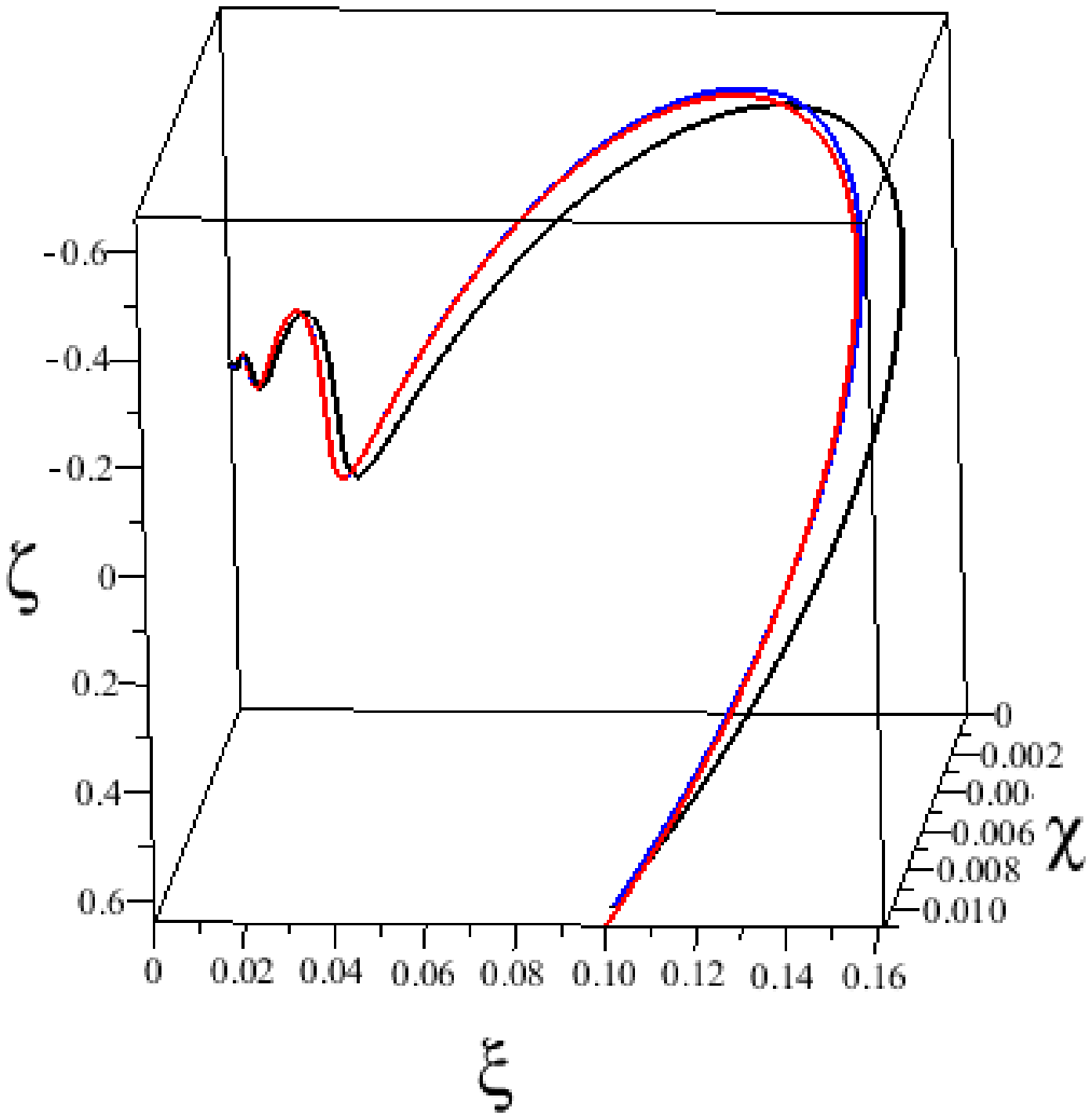}\hspace{0.1 cm}\\
Fig. 2: The attractor property of the dynamical system in the 3-Dimensional phase plane.
\label{Figure:2}
\end{figure}

\section{constrain  on  parameters of the model}
In pervious section, we  investigated stability of dynamical system by introducing the dimensionless parameters $\{\zeta,\eta,\vartheta,\xi,\chi,\nu\}$, it is obvious that the critical points and eigenvalues dependent only on the free parameter $\alpha$ (see Table \ref{table:1}). Also,  we can see from equation \ref{minima 1} to \ref{minima 5} that parameter $\alpha$ is the only parameter which has been explicitly revealed  in the set of equations and has directly affected the dynamical system; however,  there are  some  parameters such as $\{C_{1R},C_{1T},C_{2R},C_{2T}\}$ which have not appeared in the set of equations \ref{minima 1} to \ref{minima 5} and have been masked by the dimensionless new variables; but they can affect on dynamics of the system. In fact, the new variables dependent on these parameters. For example, we can rewrite the variable $\eta$ as $\eta\equiv-\frac{f_{1}(R)}{6 H^{2} f_{1}'(R)}=-\left(\frac{1}{1+\alpha}+\frac{C_{2R}}{C_{1R}}R^{-(\alpha+1)}\right)\vartheta$. We can see that this variable dependents on $C_{1R},C_{2R}$ and $\alpha$. It is important to note that in a dynamical system with a set of equations both free parameters and initial conditions determine the dynamics of the system. Free parameters affect critical points and initial conditions affect the trajectories of variables in phase space. Here, the parameter $\alpha$ is the only free parameter. Although the parameters $\{C_{1R},C_{1T},C_{2R},C_{2T}\}$ have not appeared in the equations explicitly, they can affect value of initial conditions. In order to study the effect of these parameters and constrain them with observation, we reveal these parameters in new set of equations. In this respect, we introduce some other new variables as\\

\begin{align}\label{e13}
x_{1}=H,x_{2}=R,x_{3}=\rho^{m},x_{4}=\rho^{rad},x_{5}=f_{R}(R,T),x_{6}=\dot{f_{R}} (R,T)
\end{align}
Equation (\ref{first}) ,(\ref{ft}) and (\ref{fr}) give the following constraints between  $\dot{f_{R}} (R,T)$ and the variables  $x_{1} $to $x_{5}$, namely
\begin{align}
&\dot{f_{R}} (R,T)=\frac{1}{3x_{1}}\{(1-\frac{\sqrt{6}C_{1T}}{12x_{1}})x_{3}+x_{4}-\frac{C_{1R}}{2}
(\frac{x_{5}}{C_{1R}(\alpha+1)})^{\frac{\alpha+1}{\alpha}}\nonumber\\
&-x_{5}\left(3x_{1}^{2}-\frac{1}{2}(\frac{x_{5}}{C_{1R}(\alpha+1)})^{\frac{1}{\alpha}}\label{mi4}\right)-\frac{1}{2}(C_{2R}+C_{2T})-\frac{C_{1T}\sqrt{6}x_{1}}{2}\}\\
&x_{2}=(\frac{x_{5}}{C_{1R}(\alpha+1)})^{\frac{1}{\alpha}}\label{mi5}
\end{align}
Now,using (\ref{mi4}) and (\ref{mi5}), for the autonomous equations of motions, we obtain
\begin{align}
&\frac{dx_{1}}{dN}=\frac{( \frac{x_{5}}{C_{1R}(\alpha+1)})^{\frac{1}{\alpha}}-12x_{1}^{2}}{6x_{1}} \label{mini1},\\
&\frac{dx_{3}}{dN}=-3x_{3}\label{mini2},\\
&\frac{dx_{4}}{dN}=-4x_{4}\label{mini3},\\
&\frac{dx_{5}}{d N}=\frac{1}{3x_{1}^{2}}\{(1-\frac{\sqrt{6}C_{1T}}{12x_{1}})x_{3}+x_{4}-\frac{C_{1R}}{2}
(\frac{x_{5}}{C_{1R}(\alpha+1)})^{\frac{\alpha+1}{\alpha}}\nonumber\\
&-x_{5}\left(3x_{1}^{2}-\frac{1}{2}(\frac{x_{5}}{C_{1R}(\alpha+1)})^{\frac{1}{\alpha}}\label{mini4}\right)-\frac{1}{2}(C_{2R}+C_{2T})-\frac{C_{1T}\sqrt{6}x_{1}}{2}\}
\end{align}

\begin{figure}
\centering
\includegraphics[scale=.4]{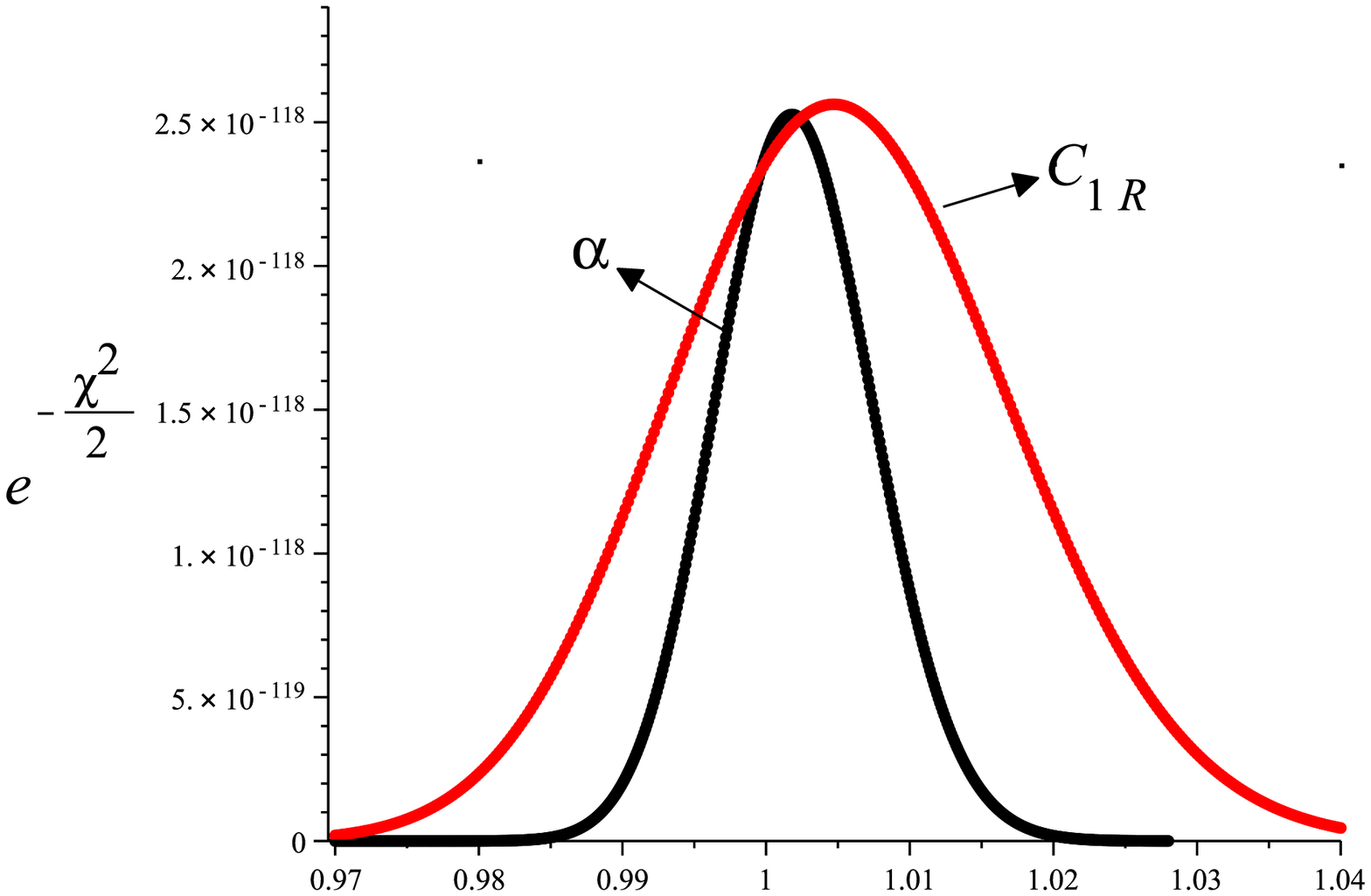}\hspace{0.1 cm}\includegraphics[scale=.4]{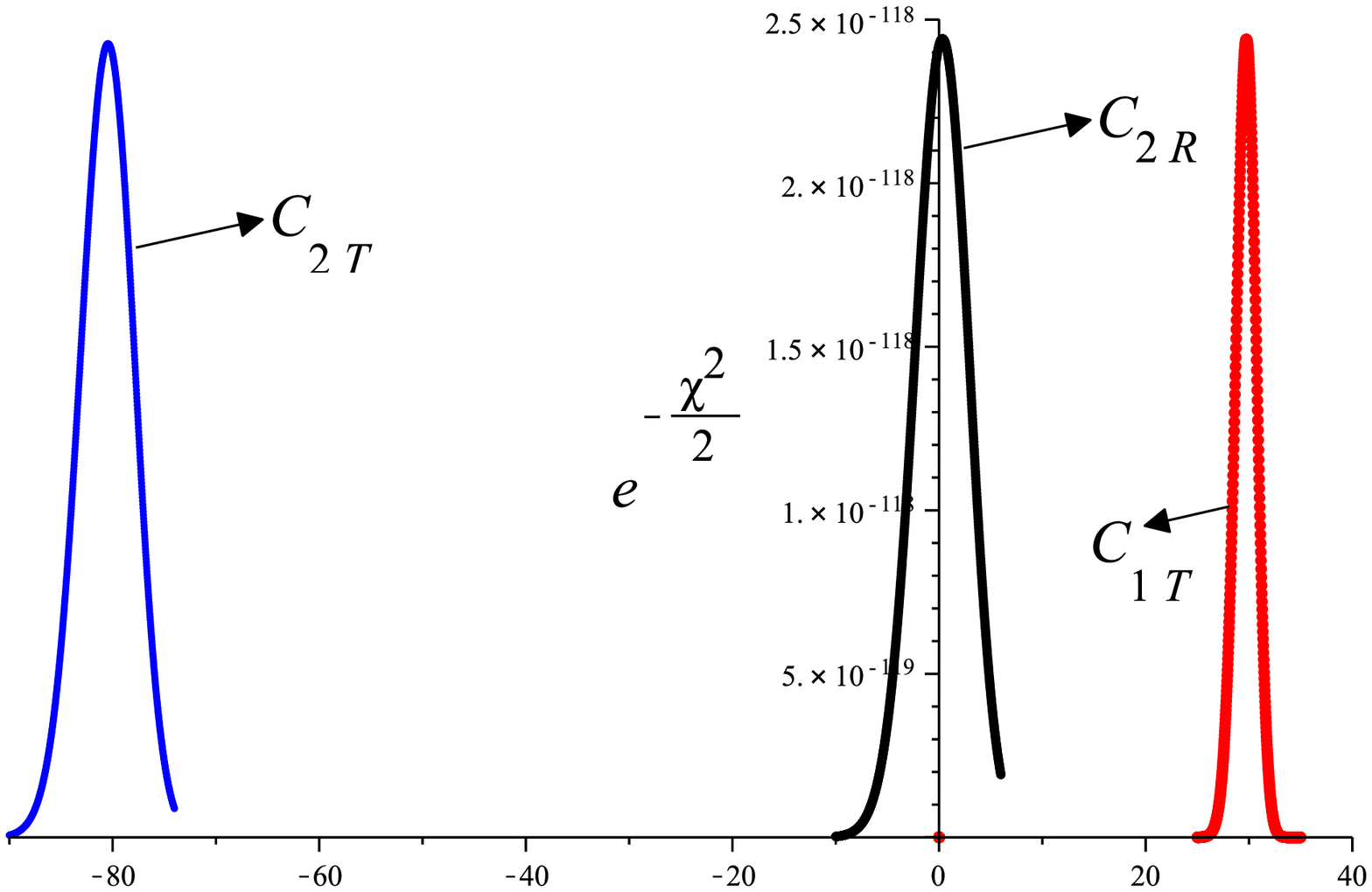}\hspace{0.1 cm}\\
Fig. 3: One dimensional likelihood for parameters $\alpha$,$C_{1R}$, $C_{2R}$, $C_{1T}$ and $C_{2T}$\\
\label{Figure:3}
\end{figure}

Therefore, we have a dynamical system with four independent variables and five free parameters$(\alpha,$$C_{1R}$ , $C_{2R}$,$C_{1T}$ and $C_{2T}$. Note that this system of equations is corresponding to equations (\ref{minima 1}) to (\ref{minima 5}). We have best-fitted these parameters using SNe Ia data by $\chi^2$ method. The likelihood for these parameters have been shown in Fig.\ref{Figure:3}.\\

\section{Scalar PERTURBATIONS IN $f(R, T ) =f_1(R)+f_2(T)$ type theories}\label{Section3}
Let us consider the scalar perturbations of a flat FRW metric in the longitudinal gauge and in conformal time:
\begin{eqnarray}
{\rm d}s^2=a^2(\eta)\left[(1+2\Phi){\rm d}\eta^2-(1-2\Psi){\rm d}{\bf x}^2\right]\,\,,
\label{manu17}
\end{eqnarray}
where $\Phi \equiv \Phi(\eta,{\bf x})$ and $\Psi \equiv \Psi(\eta,{\bf x})$ are the scalar perturbations. The components of perturbed energy-momentum tensor in this gauge are given by
\begin{eqnarray}
\hat{\delta} T^{0}_{\;0}= \hat{\delta}\rho = \rho_0 \delta \,,\quad \hat{\delta} T^{i}_{\;j}=-\hat{\delta} p\;\delta^{i}_{\;j}=-c^2_s\rho_0\delta^{i}_{\;j}\delta\,,\quad \hat{\delta} T^{0}_{\;i}=-\hat{\delta} T^{i}_{\;0}=-\left(1+c_s^2\right)\rho_0\partial_iv\,, \label{manu18}
\end{eqnarray}
where $v$ denotes the potential for the velocity perturbations. The first order perturbed equations in a dust matter dominated universe, $c^2_s=0$ will be obtained as \cite{Alvarenga2}

\begin{eqnarray}\label{sif}
\Phi-\Psi=-\frac{f_{1R_0R_0}}{f_{1R_0}}\hat{\delta}R\,\,\,,
\label{ij}
\end{eqnarray}
with
\begin{eqnarray}
\hat{\delta} R=-\frac{2}{a^2}\Big[3\Psi''+6\left(\mathcal{H}'+\mathcal{H}^2\right)\Phi
+3\mathcal{H}\left(\Phi'+3\Psi'\right)-k^2\left(\Phi-2\Psi\right)\Big]\,\,.
\end{eqnarray}
\begin{eqnarray}
\Big[3\mathcal{H}\left(\Phi'+\Psi'\right)+k^2\left(\Phi+\Psi\right)+3\mathcal{H}'\Psi
-\left(3\mathcal{H}'-6\mathcal{H}^2\right)\Phi\Big]f_{1R_0}+\left(9\mathcal{H}\Phi-3\mathcal{H}\Psi+3\Psi'\right)f'_{1R_0}\nonumber\\=-a^{2}\delta\rho_{0}[\kappa^2-f_{2T_{0}}]
\label{00}
\end{eqnarray}
\begin{eqnarray}
\Big[\Phi''+\Psi''+3\mathcal{H}\left(\Phi'+\Psi'\right)+3\mathcal{H}'\Phi+
\left(\mathcal{H}'+2\mathcal{H}^2\right)\Psi\Big]f_{1R_0}+\left(3\mathcal{H}\Phi
-\mathcal{H}\Psi+3\Phi'\right)f'_{1R_0}\nonumber\\
+\left(3\Phi-\Psi\right)f''_{1R_0}=\frac{1}{2}a^{2}\delta\rho_{0}f_{2T_{0}}\,,
\label{ii}
\end{eqnarray}
\begin{eqnarray}
\left(2\Phi-\Psi\right) f'_{1R_0}+\Big[\Phi'+\Psi'+\mathcal{H}\left(\Phi+\Psi\right)\Big]f_{1R_0}=-a^{2}v\rho_{0}(\kappa^2-f_{2T_{0}})\,\,,
\label{0i}
\end{eqnarray}
\begin{eqnarray}
\delta'-k^2v-3\Psi' \,=\,0
\label{conservation_1_Alvaro_bis}
\end{eqnarray}
and
\begin{eqnarray}\label{vi}
\Phi+\mathcal{H}v+v'\,=\,\frac{f_{2T_0}}{2(\epsilon^2-f_{2T_0})}\left(3\mathcal{H}v - \delta\right)
\label{conservation_2_Alvaro_bis}
\end{eqnarray}

where $\kappa^{2}=8 \pi G$, the prime holds for the derivative with respect to $\eta$, $\mathcal{H}\equiv a'/a$ and
the subscript $0$ holds for unperturbed background quantities:
$R_0$ denotes the scalar curvature corresponding to the unperturbed metric,
$\rho_0$ the unperturbed energy density, with $f_{10}\equiv f_1(R_0)$, $f_{1R_0}\equiv {\rm d}f_1(R_0)/{\rm d}R_0$, $f_{20}\equiv f_2(T_0)$, $f_{2T_0}\equiv {\rm d}f_2(T_0)/{\rm d}T_0$ and $c_s^2=p_0/\rho_0$
, $f_{1R_0R_0}={\rm d}^2f_1(R_0)/{\rm d}R_0^2$

\subsection{Solution of the equations using dynamical system}
The complete set of equations that describes the general linear perturbations for the  model have been presented in pervious section. These equations are a set of nonlinear second order differential equations with a large number of variable for which there is no analytical solution except for simplest cases and only numerical analysis can be performed. Our purpose is to convert second order differential equation to first order by introducing some new variables. There are various reasons for doing this, one being that a first order system is much easier to solve numerically. Also, it allows us to investigate the behavior of the system in phase space. Phase planes are useful in visualizing the behavior of the system particularly in oscillatory systems where the phase paths can "spiral in" towards zero, "spiral out" towards infinity, or reach neutrally stable situations called centres. This is a useful method to determine whether dynamics of a system are stable or not.\\
The structure of phase space of the field equations is simplified by defining a few variables and parameters. These variables are generally defined as
\begin{eqnarray}
\chi_{1}&=&\frac{\Phi'}{\Phi\mathcal{H}}\\
\chi_{2}&=&\frac{k}{\mathcal{H}}\\
\chi_{3}&=&\frac{f'_{1R_{0}}}{\mathcal{H}f_{1R_{0}}}\\
\chi_{4}&=&\frac{\delta}{\Phi}\\
\chi_{5}&=&\frac{\rho_{0}a^{2}}{f_{1R_{0}}}\\
\chi_{6}&=&\frac{f_{2T_{0}}}{\mathcal{H}^{2}}\\
\chi_{7}&=&\frac{\Psi}{\Phi \mathcal{H}}\\
\chi_{8}&=&\frac{\Psi}{\Phi}
\end{eqnarray}

Now, for the autonomous equations of motions, we obtain
\begin{eqnarray}\label{x1n}
\frac{d\chi_{1}}{dN}&=&\Gamma-\chi_{1}^{2}-\chi_{1}\varepsilon\\
\frac{d\chi_{2}}{dN}&=&-k\varepsilon\\
\frac{d\chi_{3}}{dN}&=&\beta-\chi_{3}^{2}-\varepsilon\chi_{3}\\
\frac{d\chi_{4}}{dN}&=&\Pi-\chi_{4}\chi_{1}\\
\frac{d\chi_{5}}{dN}&=&-\chi_{5}-\chi_{5}\chi_{3}\\
\frac{d\chi_{6}}{dN}&=&\frac{3}{2}\chi_{6}-2\varepsilon\chi_{6}\\
\frac{d\chi_{7}}{dN}&=&\Xi-\chi_{7}\chi_{1}-\varepsilon\chi_{7}\\
\frac{d\chi_{8}}{dN}&=&\chi_{7}-\chi_{8}\chi_{1}\label{x8n}
\end{eqnarray}
Where $N = ln a$ thus, $\frac{d}{dN}=\frac{1}{\mathcal{H}}\frac{d}{d\eta}$. Also, we have used the following parameters
\begin{eqnarray}
\frac{\mathcal{H}'}{\mathcal{H}^{2}}&=&\varepsilon\\
\frac{\Phi''}{\Phi\mathcal{H}^{2}}&=&\Gamma\\
\frac{\Psi''}{\Phi\mathcal{H}^{2}}&=&\Xi\\
\frac{\delta'}{\Phi\mathcal{H}}&=&\Pi\\
\frac{2a^{2}f_{1R_{0}}}{f_{1R_{0}R_{0}}}&=&\Omega
\end{eqnarray}

After some calculation from equations(\ref{sif})-(\ref{vi}), we can obtain the above parameters in terms of the new variables as \\
   \begin{eqnarray}\label{par1}
\varepsilon&=&\frac{1}{1-\chi_{8}}\left[\chi_{1}+\chi_{7}+ \frac{1}{3}\chi^{2}_{2}(1+\chi_{8})+(3-\chi_{8}+\chi_{7})\chi_{5}-\frac{1}{\kappa2}\chi_{5}\chi_{4}(\kappa^2\chi^{2}_{2}-k\chi_{6})\right]\\
\Xi&=&-\frac{2}{1-\chi_{8}}\left[\chi_{1}+\chi_{7}+\frac{1}{3}\chi^{2}_{2}(1+\chi_{8})+(3-\chi_{8}+\chi_{7})\chi_{5}-\frac{1}{k^{2}}\chi_{5}\chi_{4}(\kappa^2\chi^{2}_{2}-k^2\chi_{6})\right]\\ \nonumber
&-&\chi_{1}-3\chi_{7}+\frac{1}{3}\chi^{2}_{2}-\chi_{7}\chi_{1}+\frac{1}{3}\chi^{2}_{2}(1-2\chi_{8})+\frac{\Omega}{3}(1-\frac{1}{k^2}\chi^{2}_{2}\chi_{8})\\
\Gamma&=&-\Xi-3\varepsilon(1+\frac{1}{3}\chi_{8})-3\chi_{1}-3\chi_{7}-2\chi_{8}-(3-\chi_{8}+3\chi_{1})\chi_{5} +\beta(\chi_{8}-3)\\ \nonumber &+&\frac{1}{2}\chi_{4}\chi_{5}\chi_{6}\\
\Pi&=&\frac{-k^{2}\chi^{2}_{2}\left[(2-\chi_{8})\chi_{3}+\chi_{1}+\chi_{7}+1+\chi_{8}\right]}{\chi_{5}(\kappa^{2}\chi^{2}_{2}-k^{2}\chi^{6})}+3\chi_{7}\\
\Omega&=&3\alpha(1+\varepsilon)\frac{k^{2}}{\chi^{2}_{2}}\label{par5}
\end{eqnarray}

Where we have supposed that $\beta=\frac{f''_{1R_{0}}}{f_{1R_{0}}\mathcal{H}^{2}}$. By substituting equations (\ref{par1})-(\ref{par5}) into equations (\ref{x1n})-(\ref{x8n}), the complete set of equations that describes the behavior of the system in terms of new variables will be provided.\\
In general ,the critical points and eigenvalues of the system will be obtained in terms of $\beta, k$. Here, we have obtained critical points of the system for $\beta=1, k=0.3$(see table \ref{table:2n}).The corresponding eigenvalues are as:\\
\begin{table*}
\caption{Critical Points of the system}  
\centering 
\begin{tabular}{|c|c|c|c|c|c|c|c|c|} 
\hline\hline 
$Critical Points$ \ & $\chi_{1}$ \ &  $\chi_{2}$ \ & $\chi_{3}$ \ & $\chi_{4}$ \ & $\chi_{5}$& $\chi_{6}$ \ & $\chi_{7}$ \ &$\chi_{8}$ \\
\hline 
$P1$  & $8.6$ &$5.7$ & $-1.0$ & $-0.0$ & $-3.5$ & $0.0$ & $-0.1$ & $-0.0$ \\ 
\hline 
$P2$  & $8.6$ &$5.7$ & $-1.0$ & $-0.1$ & $0.0$ & $0.0$ & $-8.8$ & $-1.0$ \\
\hline 
$P3$  & $8.6$ &$-5.7$ & $-1.0$ & $-0.0$ & $-3.5$ & $0.0$ & $-0.1$ & $-0.0$ \\
\hline 
$P4$  & $8.6$ &$-5.7$ & $-1.0$ & $-0.1$ & $0.0$ & $0.0$ & $-8.8$ & $-1.0$ \\
\hline 
\end{tabular}\\
\label{table:2n} 
\end{table*}
\begin{align}
Ev1=
\begin{pmatrix}
-11.7+86.6i \\
-11.7-86.6i \\
-7.0+6.9i \\
-7.0-6.9i \\
-8.7 \\
2.3 \\
8.1\times10^{-8}\\
1.5\\
\end{pmatrix},
Ev2=
\begin{pmatrix}
-20.3 \\
2.8 \\
-9.9+5.9i \\
-9.9-5.9i \\
-2.5 \\
-7.3\\
-3.1\times10^{-7}\\
1.5\\
\end{pmatrix},
Ev3=
\begin{pmatrix}
-11.7+86.6i \\
-11.7-86.6i \\
-7.0+6.9i \\
-7.0-6.9i \\
-8.7 \\
2.3 \\
8.1\times10^{-8}\\
1.5\\
\end{pmatrix},
Ev4=
\begin{pmatrix}
-20.3 \\
2.8 \\
-9.9+5.9i \\
-9.9-5.9i \\
-2.5 \\
-7.3\\
-3.1\times10^{-7}\\
1.5\\
\end{pmatrix},
\end{align}

\begin{tabular*}{2.5 cm}{cc}
\includegraphics[scale=.45]{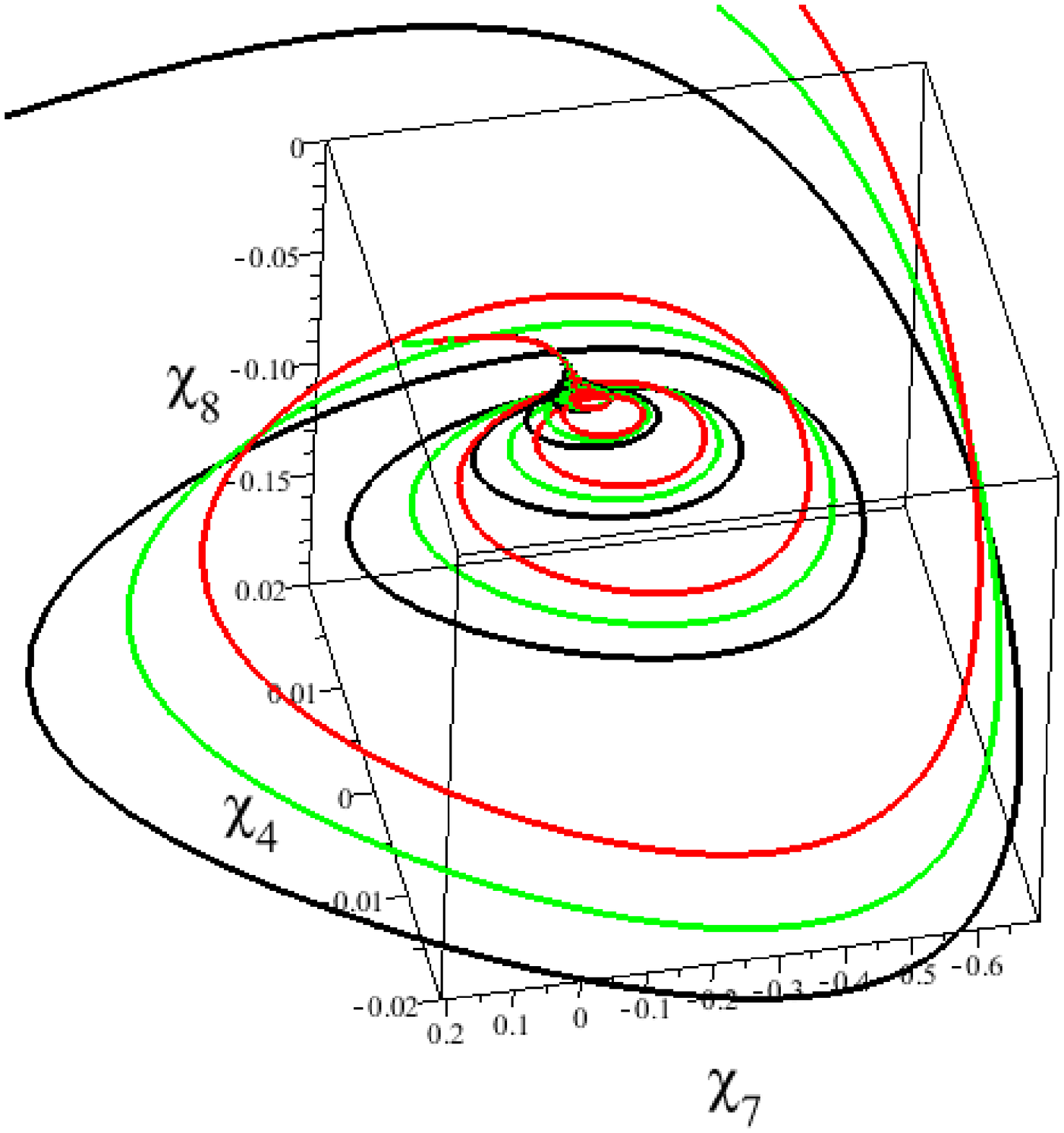}\hspace{0.1 cm}\includegraphics[scale=.45]{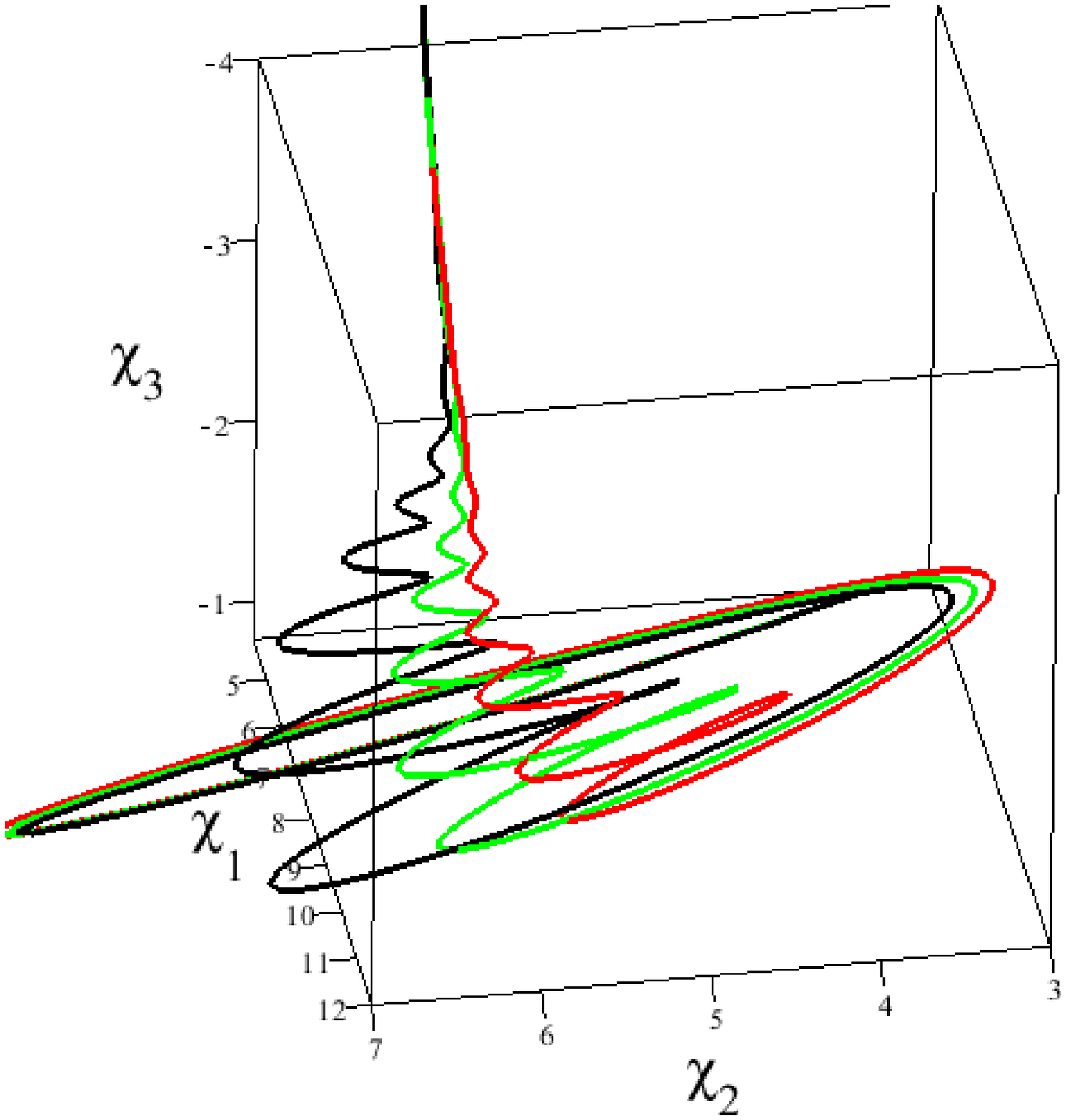}\hspace{0.1 cm} \\
Fig. 4: Attractor behavior of the system for $\beta=1, k=0.3$ \\
\label{Figure:4n}
\end{tabular*}\\

Due to the fact that there are complex values in some matrix elements of the eigenvalues, the dynamical system shows attractor behavior. The attractor behavior of the system has been shown in Fig. 4. Note that the attractor behavior in phase space implies that the system oscillates and moves toward steady state in a critical point.\\

\begin{tabular*}{2.5 cm}{cc}
\includegraphics[scale=.3]{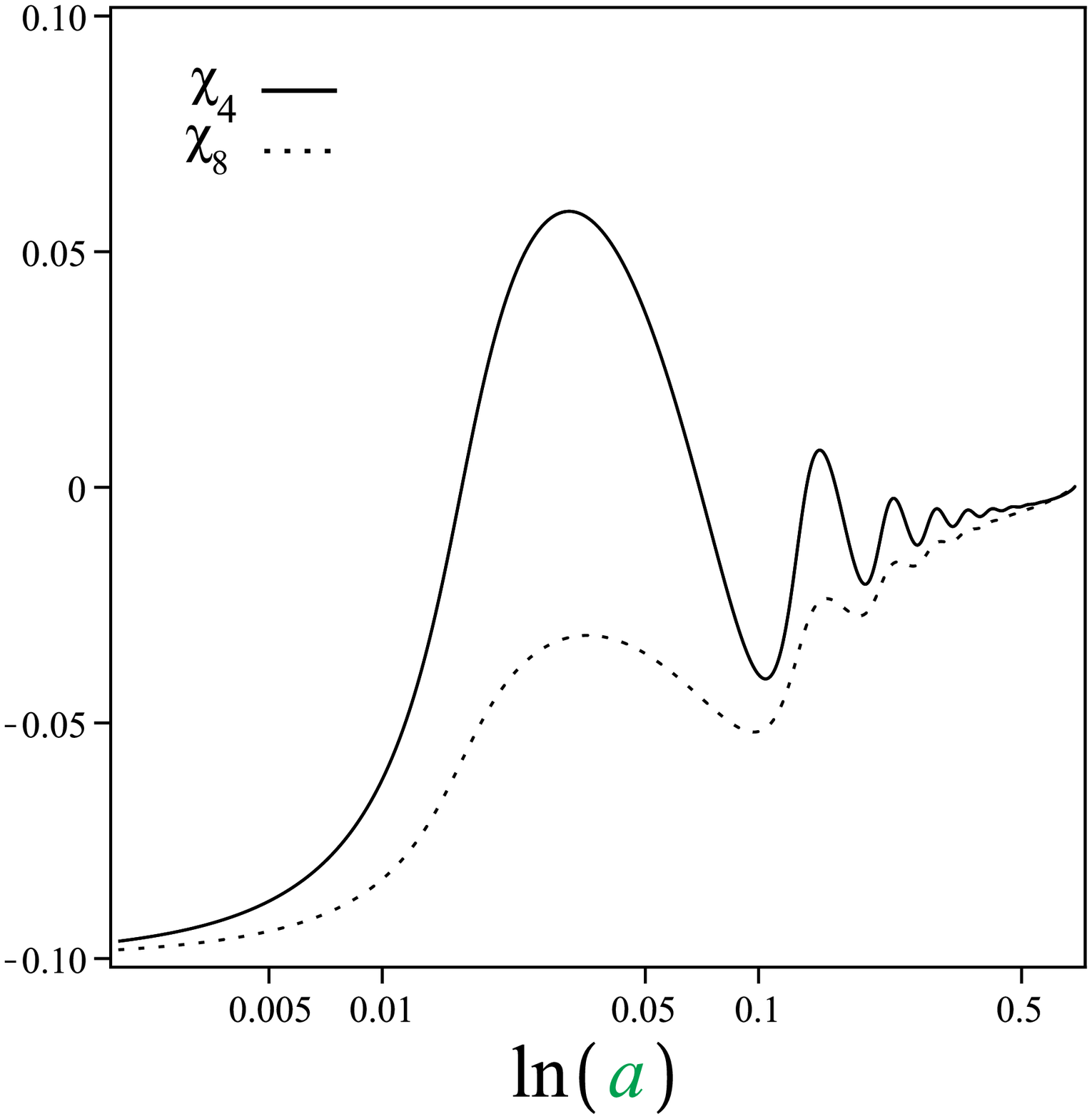}\hspace{0.1 cm}\includegraphics[scale=.3]{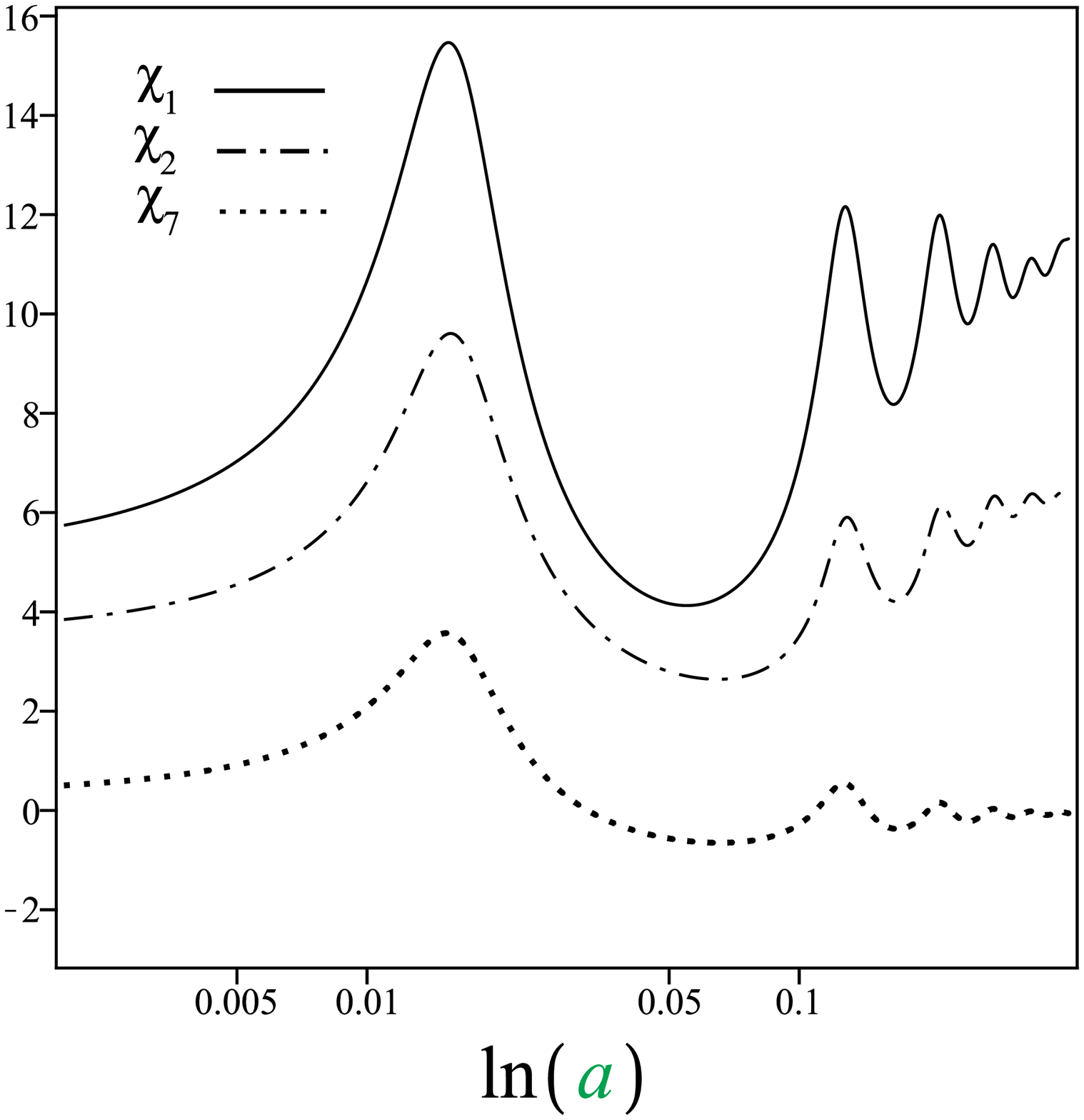}\hspace{0.1 cm}\includegraphics[scale=.3]{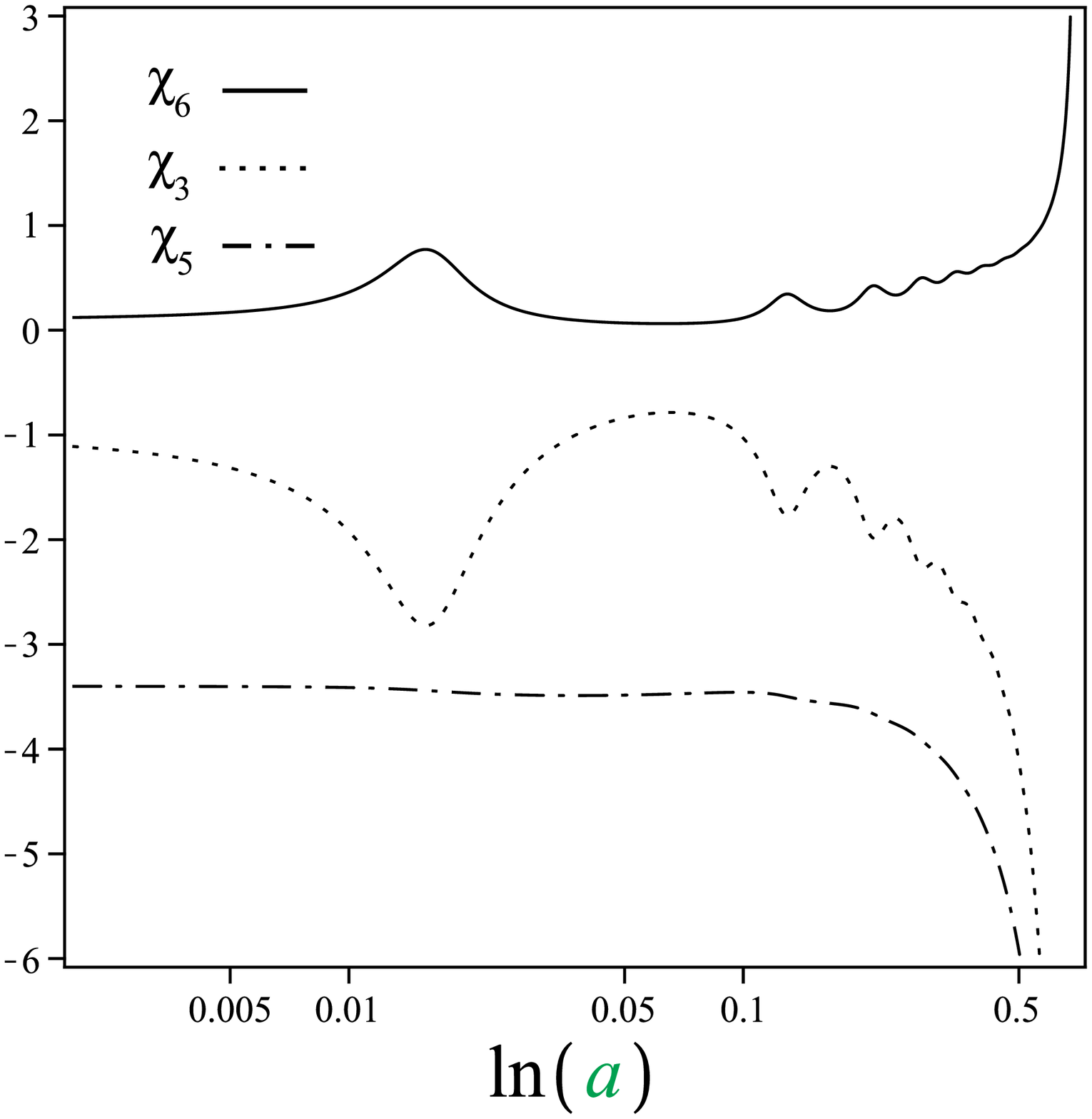}\hspace{0.1 cm} \\
Fig. 5: The oscillating behavior of the system $\beta=1, k=0.3$ \\
\label{Figure:5n}
\end{tabular*}\\

The oscillating behavior of the system has been shown in Fig. 5. Moreover,  we are interested in behavior of the parameters $\delta,\Psi,\Phi$. Therefore, we can reconstruct them from new variables as\\
\begin{eqnarray}\label{parr1}
\frac{1}{\Phi}\frac{d\Phi}{dN}&=&\chi_{1}\\
\frac{1}{\Phi}\frac{d\Psi}{dN}&=&\chi_{7}\\
\frac{1}{\Phi}\frac{d\delta}{dN}&=\Pi&
\end{eqnarray}

Fig. 6. shows the oscillating behavior of the parameters $\delta, \Phi, \Psi$.

\begin{figure}[t]
\includegraphics[scale=.5]{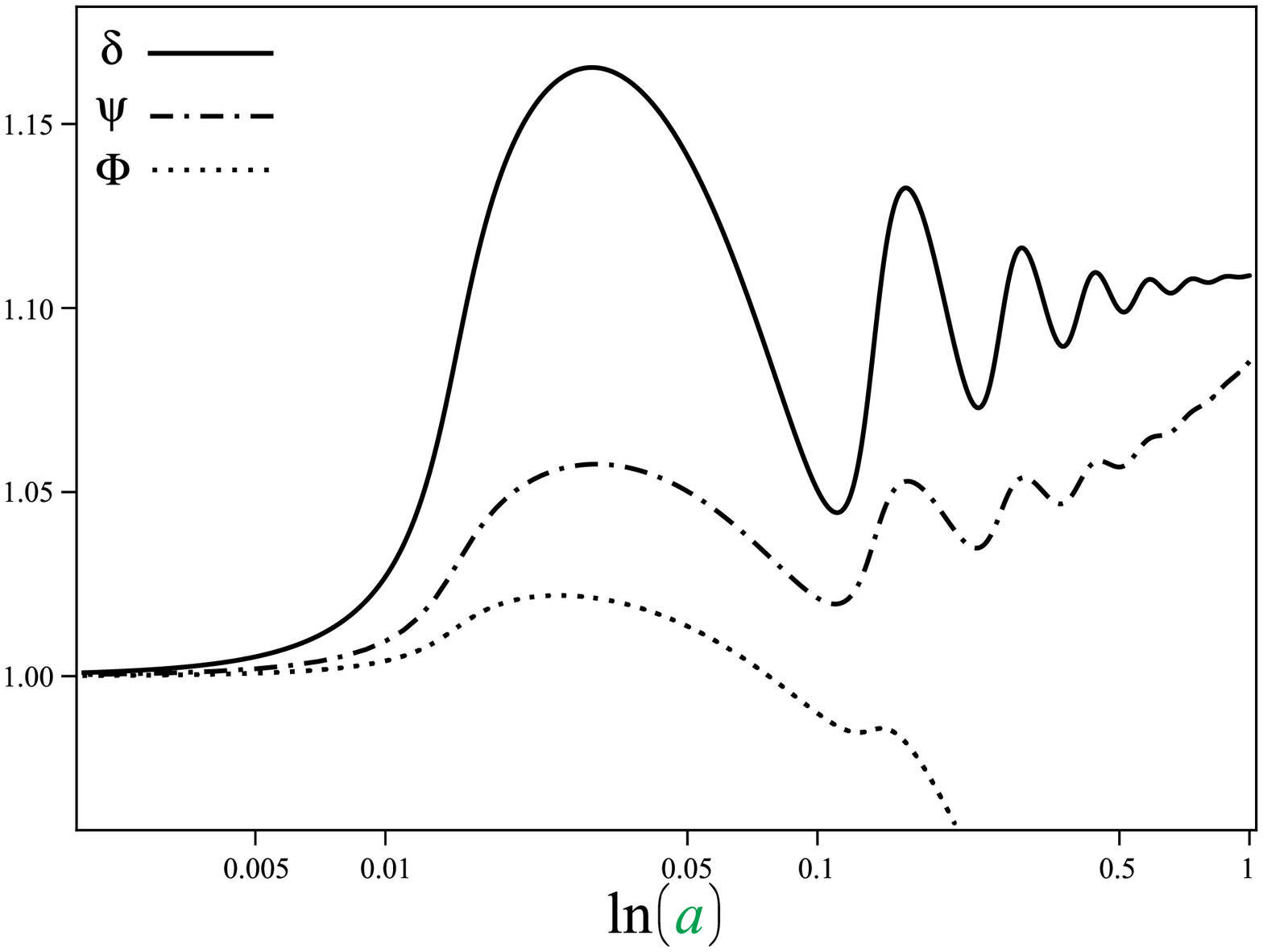}\hspace{0.1 cm} \\
Fig. 6: The oscillating behavior of the parameters $(\delta,\Phi,\Psi).$\\
\label{Figure:6n}
\end{figure}

\subsection{Solution for$\frac{f_{1R_{0}R_{0}}}{f_{1R_{0}}}\rightarrow0$}
In this section, we solve the equations for the $\frac{f_{1R_{0}R_{0}}}{f_{1R_{0}}}\rightarrow0$. Note that this limit in $f(R)$ theories  corresponds to the large
scalaron mass  which defined as \cite{Eingorn} \cite{Muller} \cite{Faraoni};
\begin{eqnarray}\label{mass}
m^{2}=\frac{a^{2} f_{1R_{0}}}{3 f_{1R_{0}R_{0}}}
\end{eqnarray}[ref].

Applying this condition to equation (\ref{sif}) yields $\Psi=\Phi$. Therefore, the equations (\ref{sif})-(\ref{conservation_2_Alvaro_bis}) are simplified as follows

\begin{eqnarray}\label{cons}
(6\Phi'\mathcal{H}+2k^2 \Phi+6\mathcal{H}^{2}\Phi)f_{1R_{0}}+(6\mathcal{H}\Phi+3\Phi')f'_{1R_{0}}
=-a^{2}\delta\rho_{0}[\kappa^2-f_{2T_{0}}]
\end{eqnarray}
\begin{eqnarray}\label{phiz0}
\left(2\Phi''+6\mathcal{H}\Phi'+4\mathcal{H}'\Phi+2\mathcal{H}^{2}\Phi\right)f_{1R_{0}}+
\left(2\mathcal{H}\Phi+3\Phi'\right)f'_{1R_{0}}+2\Phi f''_{1R_{0}}=\frac{1}{2}a^{2}\delta\rho_{0}f_{2T_{0}}
\end{eqnarray}
\begin{eqnarray}\label{nu}
(2\Phi'+2\Phi\mathcal{H})f_{1R_{0}}+\Phi f'_{1R_{0}}=-a^{2}v\rho_{0}(\kappa^2-f_{2T_{0}})
\end{eqnarray}

\begin{eqnarray}
\delta'-k^2v-3\Phi' \,=\,0
\label{conservation_1_Alvaro_bis0}
\end{eqnarray}
and
\begin{eqnarray}
\Phi+\mathcal{H}v+v'\,=\,\frac{f_{2T_0}}{2(\kappa^2-f_{2T_0})}\left(3\mathcal{H}v - \delta\right)
\label{conservation_2_Alvaro_bis0}
\end{eqnarray}
From equation (\ref{cons})we have
\begin{eqnarray}\label{cons2}
1+\frac{\Phi'}{\Phi\mathcal{H}}+\frac{k^{2}}{3\mathcal{H}^{2}}
+\frac{f'_{1R_{0}}}{\mathcal{H}f_{1R_{0}}}+\frac{\Phi'}{2\Phi\mathcal{H}}\frac{f'_{1R_{0}}}{\mathcal{H}f_{1R_{0}}}
=\frac{\delta}{6\Phi}\frac{\rho_{0}a^{2}}{f_{1R_{0}}}\left(\frac{-\kappa^{2}}{\mathcal{H}^{2}}
+\frac{f_{2T_0}}{\mathcal{H}^{2}}\right)
\end{eqnarray}

Hence, the autonomous Equation of Motion for the  independent variables can be obtained via
\begin{eqnarray}
\frac{d\chi_{1}}{dN}&=&\Gamma-\chi_{1}^{2}-\varepsilon\chi_{1}\\
\frac{d\chi_{2}}{dN}&=&-k\frac{\mathcal{H}'}{\mathcal{H}^{2}}\\
\frac{d\chi_{3}}{dN}&=&\beta-\chi_{3}^{2}
-\varepsilon\chi_{3}\\
\frac{d\chi_{4}}{dN}&=&\Pi-\chi_{4}\chi_{1}\\
\frac{d\chi_{5}}{dN}&=&\chi_{5}-\chi_{5}\chi_{1}\\
\frac{d\chi_{6}}{dN}&=&\frac{3}{2}\chi_{6}-2\varepsilon\chi_{6}
\end{eqnarray}
Also, from equation (\ref{cons2}) we obtain
\begin{eqnarray}\label{cons3}
1+\chi_{1}+\frac{\chi^{2}_{2}}{3}+\chi_{3}+\frac{1}{3}\chi_{1}\chi_{3}=
\frac{1}{6k^{2}}\chi_{4}\chi_{5}(k^{2}\chi_{6}-\kappa^{2}\chi_{2}^{2})
\end{eqnarray}

Applying constraint (\ref{cons3}), the system reduces to a system with five independent variables

\begin{tabular*}{2.5 cm}{cc}
\includegraphics[scale=.45]{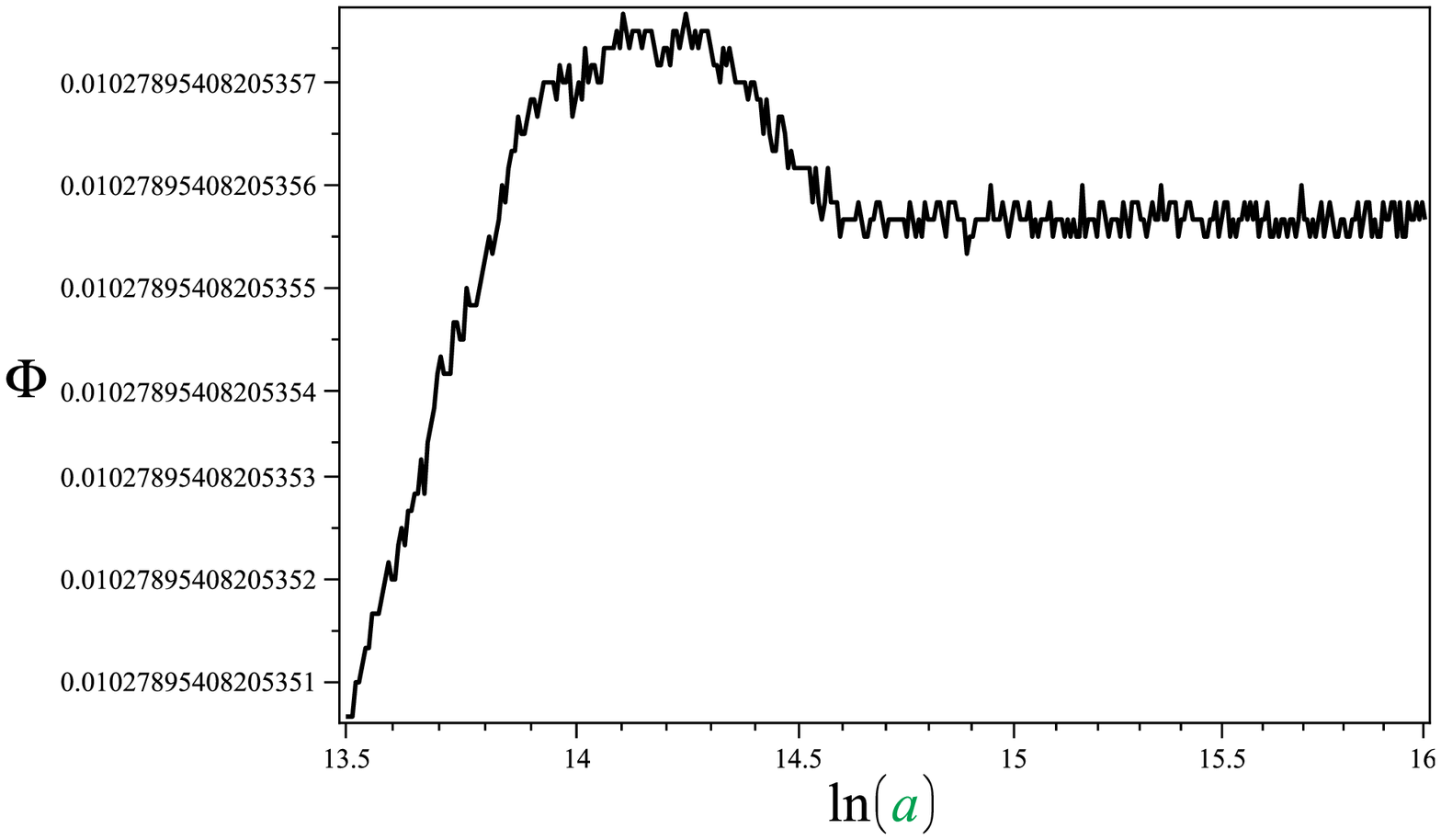}\hspace{0.1 cm}\includegraphics[scale=.45]{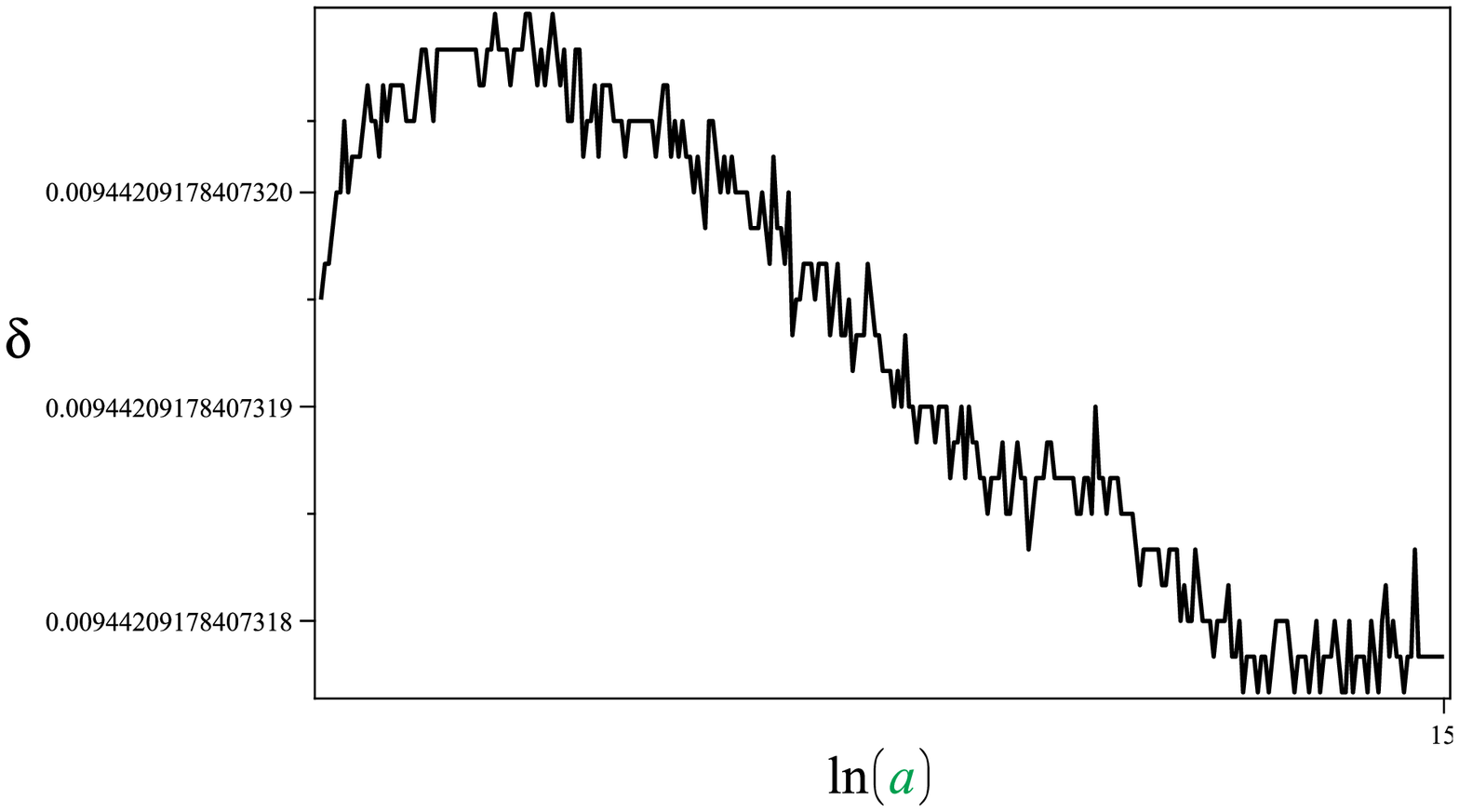}\hspace{0.1 cm} \\
Fig. 7: Fluctuation of the parameters $\Phi, \delta$ for $f(R,T)$ model when $\frac{f_{1R_{0}R_{0}}}{f_{1R_{0}}}\rightarrow0$\\
\label{Figure:7n}
\end{tabular*}\\

By setting $x_{6}=0$, behavior of the dynamical system in $f(R)$ theory will be provided. Here, we have plotted two dimensional, three dimensional phase space and evolution of variables for $f(R)$ theory in Fig. 8. \\
\begin{tabular*}{2.5 cm}{cc}
\includegraphics[scale=.3]{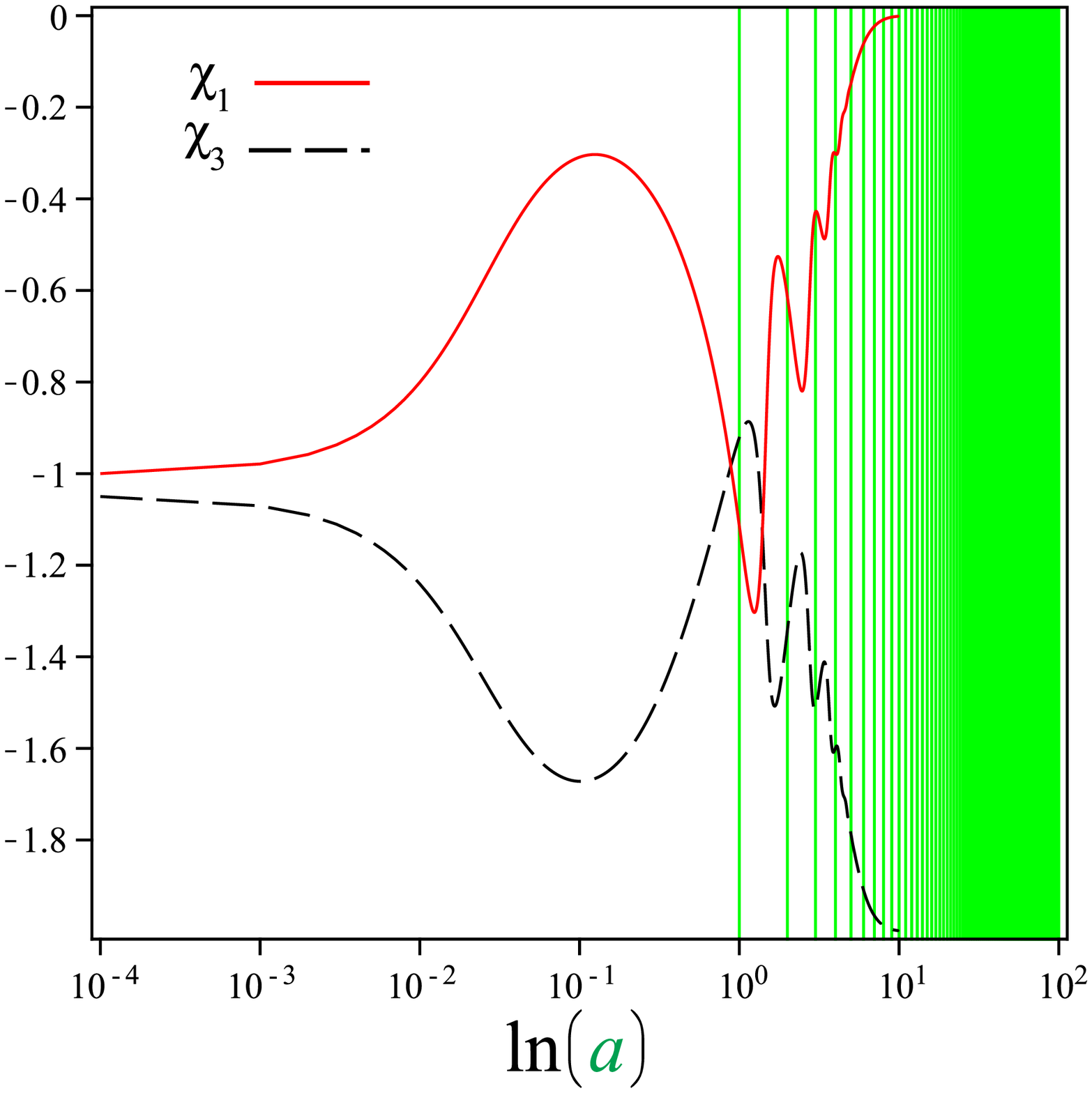}\hspace{0.1 cm}\includegraphics[scale=.3]{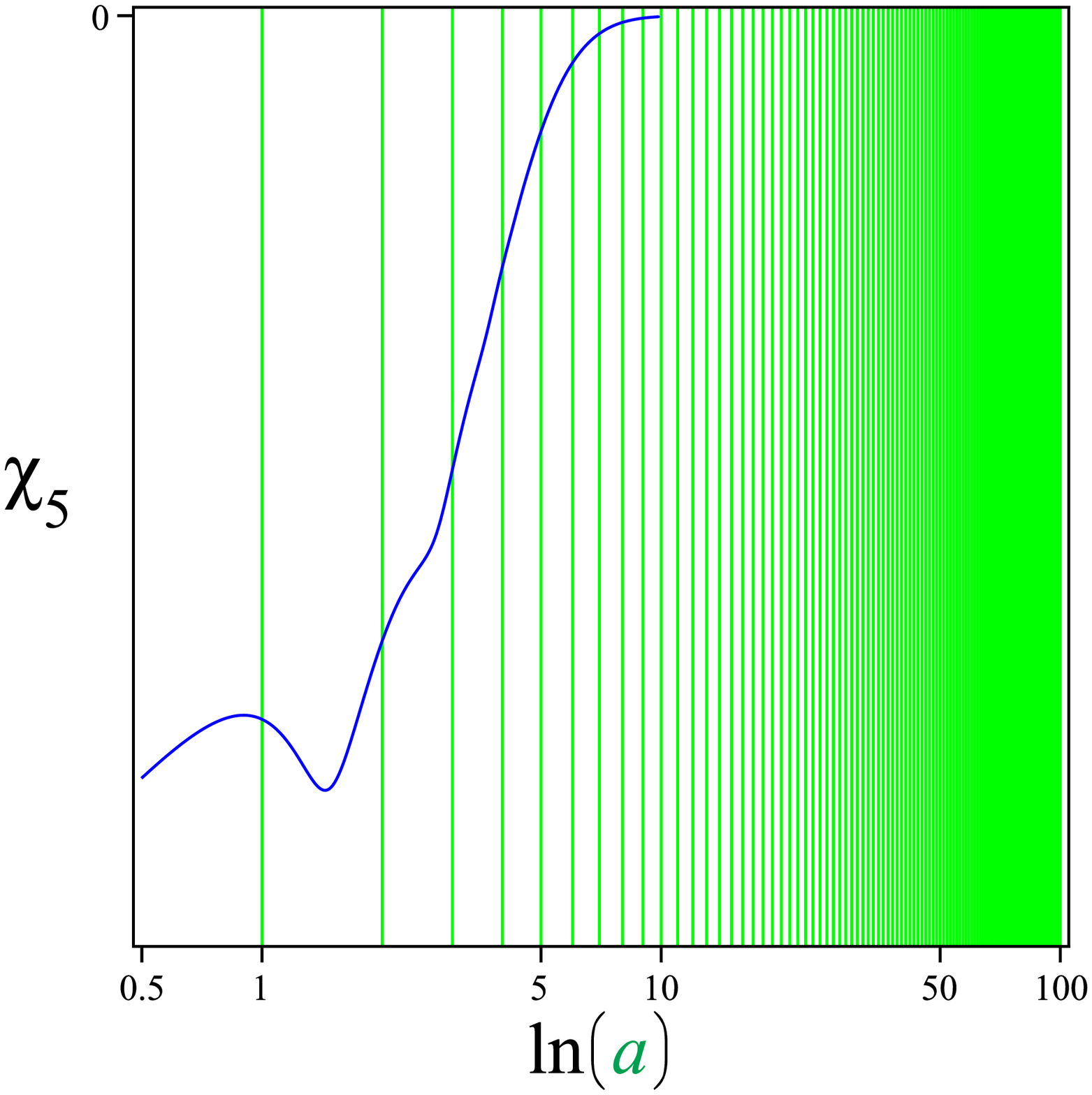}\hspace{0.1 cm}\includegraphics[scale=.3]{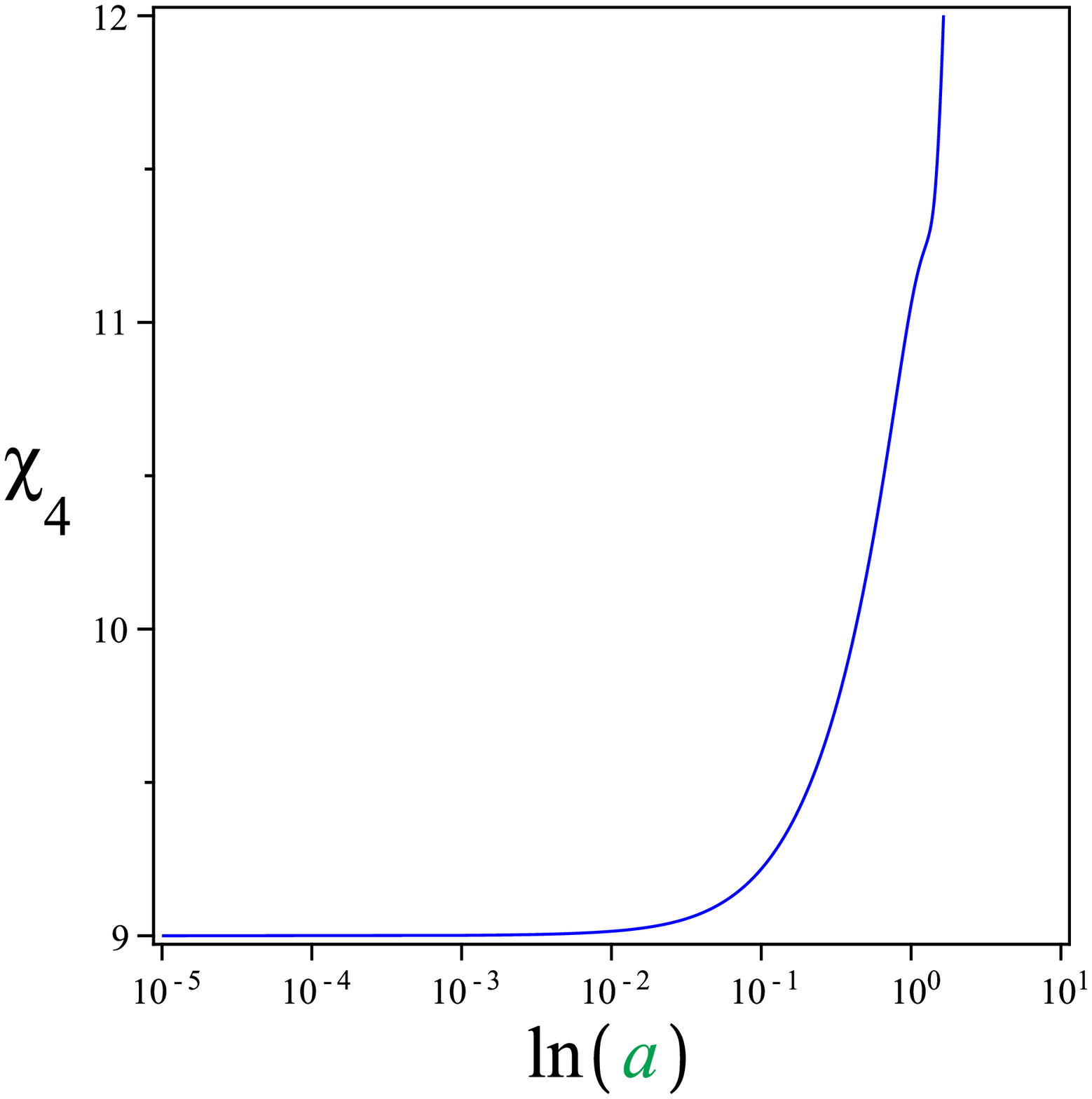}\hspace{0.1 cm} \\
Fig. 8: Evolution of variables for $f(R)$ theory when $\frac{f_{1R_{0}R_{0}}}{f_{1R_{0}}}\rightarrow0$ \\
\label{Figure:8n}
\end{tabular*}\\

\begin{tabular*}{2.5 cm}{cc}
\includegraphics[scale=.5]{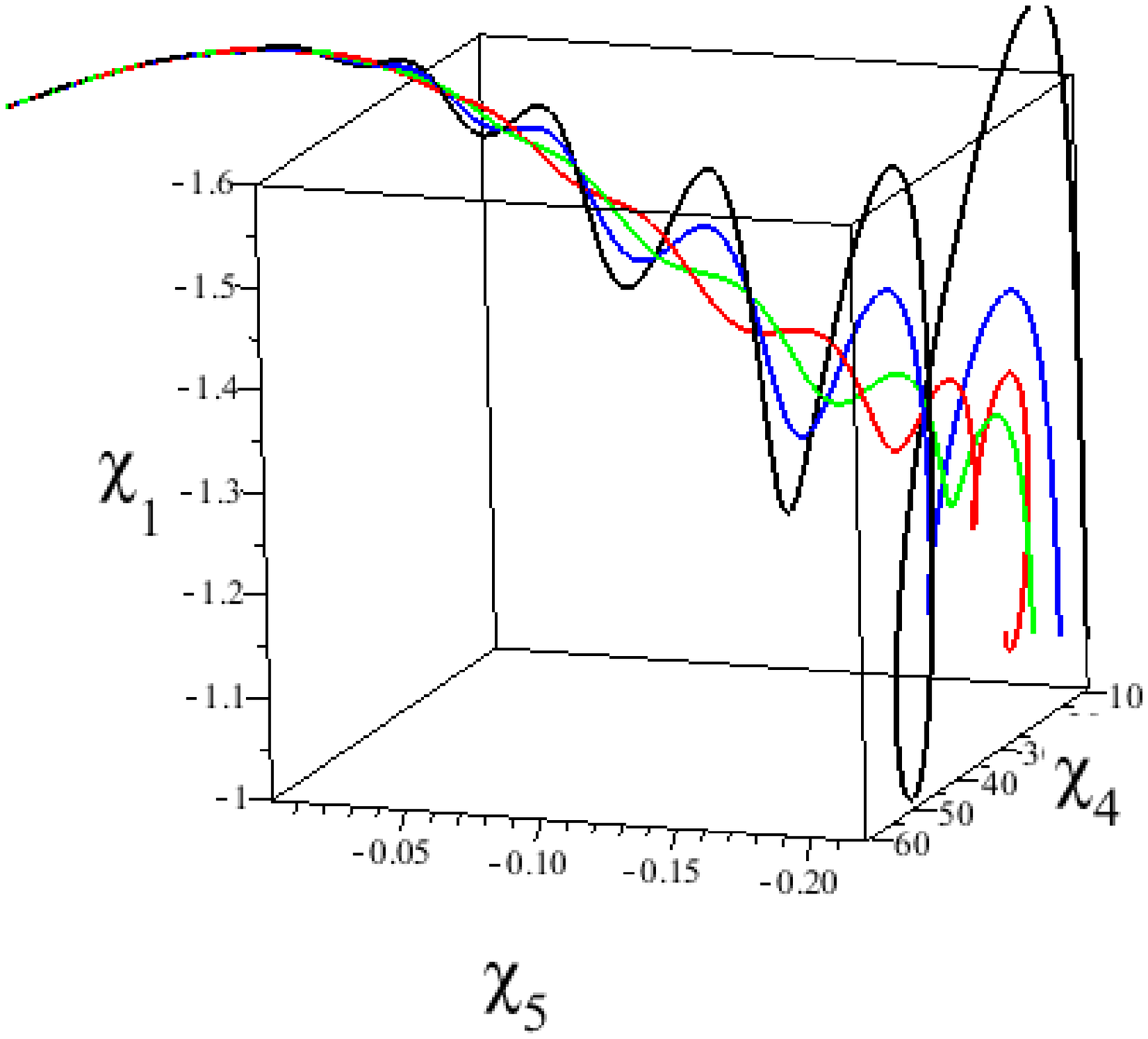} \includegraphics[scale=.5]{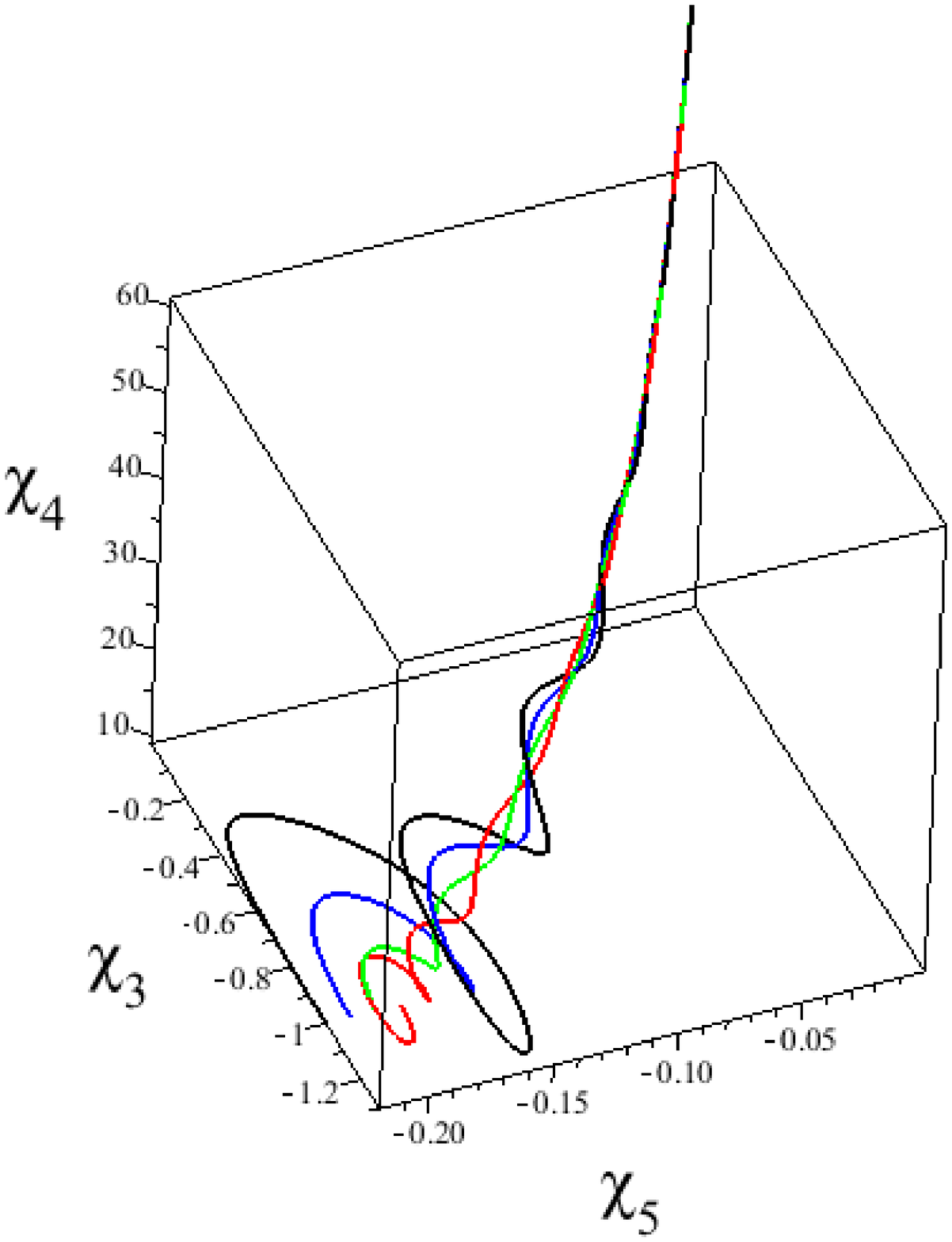} \\
\end{tabular*}\\

\begin{tabular*}{2.5 cm}{cc}
 \includegraphics[scale=.4]{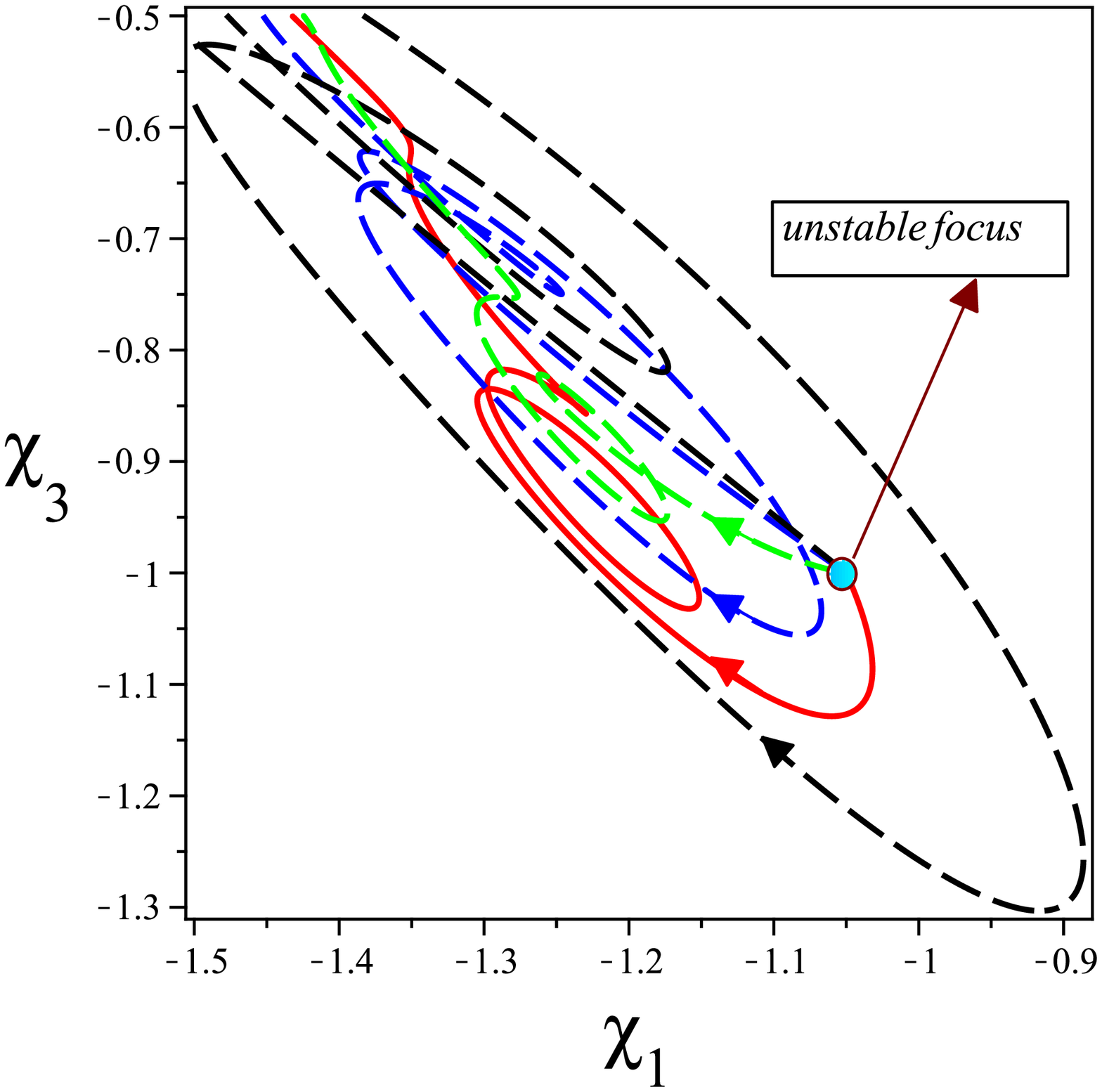} \\
Fig. 9: Attractor property and oscillating behavior of the dynamical system for $f(R)$ theory\\
when $\frac{f_{1R_{0}R_{0}}}{f_{1R_{0}}}\rightarrow0$.\\
\label{Figure:9n}
\end{tabular*}\\

Attractor property and oscillating behavior of the dynamical system in Fig. 9. shows that  the trajectories spirals out from an unstable focus point and moves towards a steady state point.\\

\section{Searching for Dark Energy Dipole using observational data}
There are various ways to investigate possible anisotropy from the data. Generally speaking, there are three important ways:\\
\subsection{Modification of the Luminosity Distance Redshift Relation in a Specific Anisotropic Cosmological Model}
In this method, an expression is derived for the luminosity distance as a function of redshift in a specific anisotropic cosmological model. Many anisotropic cosmological models with modified luminosity distances have been proposed to match observations. Table \ref{table:2} shows modified luminosity distance for some of these models as an incomplete list.
\begin{table*}
\centering
\caption{The incomplete list of modified luminosity distance for some  Anisotropic Cosmological Models}
\begin{minipage}{200mm}
\begin{tabular}{|c |c | c| c|} 
\hline \hline
 &$model$  &  modified luminosity distance    &  $Ref$\ \\ [2ex] 
\hline 
1  &scalar perturbation &($d_{L}=(1+z)\frac{c}{H_{0}}\int_{0}^{z}\frac{(1-d\cos\theta)dz}{\sqrt{\Omega_{m0}(1+z)^{3}+1-\Omega_{m0}
-\frac{4d\cos\theta(1+x)^{5}}{3H_{0}^{2}d^{2}_{L0}}}}$) &  \cite{Li} \ \\
 & & &  \cite{Wang} \ \\
\hline 
2  &Anisotropic $d_{L}$ in the Finslerian space-time &($d_{L}=(1+z)\frac{c}{H_{0}}\int_{0}^{z}\frac{(1-d\cos\theta)^{-1}dz}
{\sqrt{\Omega_{m}(\frac{1-d\cos\theta}{1+z})^{-3}+\Omega_{\Lambda}}}$) &  \cite{Chang3} \\
\hline 
3  & effect of peculiar velocities on$d_{L}$  \ &$\frac{\Delta d_{L}}{d_{L}}=\hat{n}.[\vec{v}_{pec}-(\vec{v}_{pec}-\vec{v}_{obs}).\frac{(1+z)^{2}}{H(z)d_{L}}]$&  \cite{Hui1}-- \\
\hline 
  & \ & &   \cite{Bonvin}\\
4  & "wind" scenario to the bulk flow&($d_{L}=(1+z)\int_{0}^{t}\frac{dt'}{a(t')}(1+d \cos\theta)=\bar{d_{L}}(1+d \cos\theta)$)\ &  \cite{Li3}\\
\hline 
5  & luminosity-distance
&$d_{L}(z,\theta)=\frac{1+z}{H0}\int_{A(z)}^{1}\frac{dA}{A^{2}\bar{H}}\frac{(1-e^{2})^{1/6}}{(1-e^{2}cos\theta)^{1/2}}$&  \cite{a122}\\
 &in ellipsoidal
universe& $1+z=\frac{1}{A}\frac{(1-e^{2}sin\theta)^{1/2}}{(1-e^{2})^{1/3}}$&\\
\hline 
6&measured (perturbed) luminosity-distance&$D_{L}=(1+2\hat{n}.\vec{v_{s}})D_{0L}$ ,$v_{s}$=peculiar velocities &  \cite{De-Chang} \\
\hline 
7& Bianchi I Cosmology & $1+ z(\theta,\phi,a,b,c) =  \left[\frac{a(t_{0})}{a}\right]^{2} \sin^{2}\theta\cos^{2}\phi \left[\frac{b(t_{0})}{b} \right]^{2} $ &   \\
& &$\sin^{2}\theta \sin^{2}\phi + \left[\frac{c(t_{0})}{c}\right]^{2} \cos^{2}\theta  $ &  \cite{Stephen} \\
 & &$d_L(\hat n) = \frac{1+z}{H_{0}} \int_0^z \frac{dz'}{\sqrt{ \Omega_{m, 0}
(1+z')^{3} + \Omega_{de, 0}(1+z')^{3[1+w(\hat n)]}}} $ &   \\
 \hline 
\end{tabular}\\
\label{table:2} 
\end{minipage}
\end{table*}\\

The Bianchi I type cosmological model (\cite{a122},\cite{Sch¨ucker}) and the Rinders-Finsler cosmological model (\cite{Chang1}--\cite{Chang3}) are two models which are consistent with the SNe Ia data. A scalar perturbation of the $\Lambda CDM $ model may also break the spherical symmetry of the Universe such that a preferred axis arises.
\cite{Li3}, \cite{Li}, \cite{Wang} have presented a scalar perturbation for the $\Lambda CDM $ model. Using a scalar perturbation for the FRW metric \ref{manu17},   modification of the Luminosity Distance Redshift Relation in a specific anisotropic cosmological model will be obtain as \cite{Flanagan}

\begin{eqnarray}\label{finalapp}
&&{d}_L({z},\bold{n})  \approx(\chi_s-\chi_o)(1+{z})
\Biglb\{1+\bold{v}_s\cdot\bold{n}
-\frac{ (\bold{v}\cdot\bold{n})^{\chi_s}_{\chi_o}  }{(\chi_s-\chi_o)\mathcal{H}_s}
  \\\nonumber
&&-\frac{1}{2}\int_{\chi_o}^{\chi_s} \nabla^2(\Phi+\Psi)
\frac{(\chi-\chi_o)(\chi_s-\chi)}{\chi_s-\chi_o}d\chi
\Bigrb\}\,.
\end{eqnarray}
Here the luminosity distance is expressed in terms of the observed redshift ${z}$, and the direction to the source, where $\bf{n}$ denotes
 a unit spatial vector from the observer to the source. The notation $\approx$ denotes an approximate equality accurate up to first-order in the potentials $\Phi$,  $\Psi$, and the peculiar velocities (in the conformal spacetime) $\bold{v}$.  The subscripts $s$ and $o$ refer to the
source and the observer, respectively. The affine parameter ${\chi}$ is given by
\[
{\chi}_s=\int_0^{z} \frac{1}{H({z})}d{z}+{\chi}_o\,,
\]

This relation is appropriate for investigation of the peculiar velocities. It should be mentioned that the aim of our study is to investigate Dark Energy Dipole in the model.  

\subsection{Hemisphere Comparison (HC) method}
The HC method divides the data points into two subsets according to their position in the sky and fit the subsets to an isotropic cosmological model (e.g.,$\Lambda CDM$ model). Several groups such as \cite{Schwarz}--\cite{Kalus} have applied the hemisphere comparison method to study the anisotropy of $\Lambda CDM$, $\omega CDM$ and the dark energy model with $CPL$ parametrization. More recently, \cite{Antoniou} have applied the hemisphere comparison method to the standard $ \Lambda CDM$ model and found that the hemisphere of maximum accelerating expansion is in the direction of $(l, b) =(309^{-3}_{+23}, 18^{-10}_{+11})$.
\cite{Schwarz} took use of the hemisphere comparison method to fit the $\Lambda CDM$ model to the supernovas data on several pairs of opposite hemispheres, and a statistically significant preferred axis was found. Some of studies which have used this method are listed in Table \ref{table:3}.\\
\subsection{Dipole-Fitting (DF) method}
Using the DF method, we can directly fit the data to a dipole (or dipole plus monopole) model. If the Universe is really intrinsically anisotropic and
there exists a preferred direction, it should directly affect the expansion rate of the Universe, leading to the anisotropic luminosity distance and anisotropic distance modulus. In fact, this method corresponds to the fluctuation  of the distance modulus. Anisotropic Dipole-fitting method has been used for searching the anisotropy of fine structure constant using quasars data on cosmological scale. \cite{Antoniou} firstly applied this method to anisotropic study using SNe Ia dataset. \cite{Yang} have applied this method to investigate dipolar asymmetry of the Universe. \cite{Chang4} have made a comprehensive comparison between the HC method and the DF method using the Union2 dataset.\\
Several studies payed attention to  the fluctuation of the distance modulus in order to find the preferred axis of the Universe using the DF method (See Table \ref{table:3}). It is worth to mention that the anisotropic property of the Universe directly affect the luminosity distance and leading to the anisotropic luminosity distance. Therefore, besides the DF method for the distance modulus, we have used this method for Luminosity distance. We have explained three types of Dipole-fitting method which are based on deviation of distance modulus and Luminosity distance from their best values in  isotropic model to investigate the anisotropic expansion of the Universe.\\
\begin{table*}
\centering
 \begin{minipage}{200mm}
  \caption{Incomplete list of previous studies using HC and DF method}
\begin{tabular}{|c |c | c| c|} 
\hline \hline 
&$ Method$  &  $Equation$ &  $Ref$\ \\ [2ex] 
\hline 
1  &$Hemisphere Comparison (HC) method$&$\frac{\Delta\Omega_{0m}}{\bar{\Omega_{0m}}}=
2(\frac{\Omega_{0m,u}-\Omega_{0m,d}}{\Omega_{0m,u}+\Omega_{0m,d}}$)\ &  \cite{Yang},\cite{Cai}\\
\hline 
2  &$Hemisphere Comparison (HC) method$&$\frac{\Delta q_{0}}{\bar{q_{0}}}=
2(\frac{q_{0,u}-q_{0,d}}{q_{0,u}+q_{0,d}}$)&  \cite{Cai0} \\
\hline 
3  & $\alpha(Dipole+Monopole) Fit$&$\frac{\Delta\alpha}{\alpha}=A \cos\theta +B$ &  \cite{Antoniou} \\
\hline 
4  & $Dipole+Monopole Fitting for Distance Modulus(DMFDM) $ &($\frac{\Delta\mu}{\mu}=d_{1} cos\theta +m_{1}$) &   \cite{Yang},\ \\
 &   & &,\cite{Wang}\ ,\\
 &   & &\cite{Antoniou}\ \\
\hline 
5  &$Generalized Dipole Fitting  for Luminosity Distance(GDFLD)$&$\frac{d_{L}(z)-d^{0}_{L}(z)}{d^{0}_{L}(z)}=g(z)(\hat{z}.\hat{n}$)=$g(z) cos \theta$) &  \cite{Cai0} \\
\hline 
\end{tabular}\\
\label{table:3}
\end{minipage}
\end{table*} \\\\

\textbf{A. Dipole+Monopole Fitting for Distance Modulus(DMFDM)}\\

\begin{figure}
\centering
\includegraphics[scale=.5]{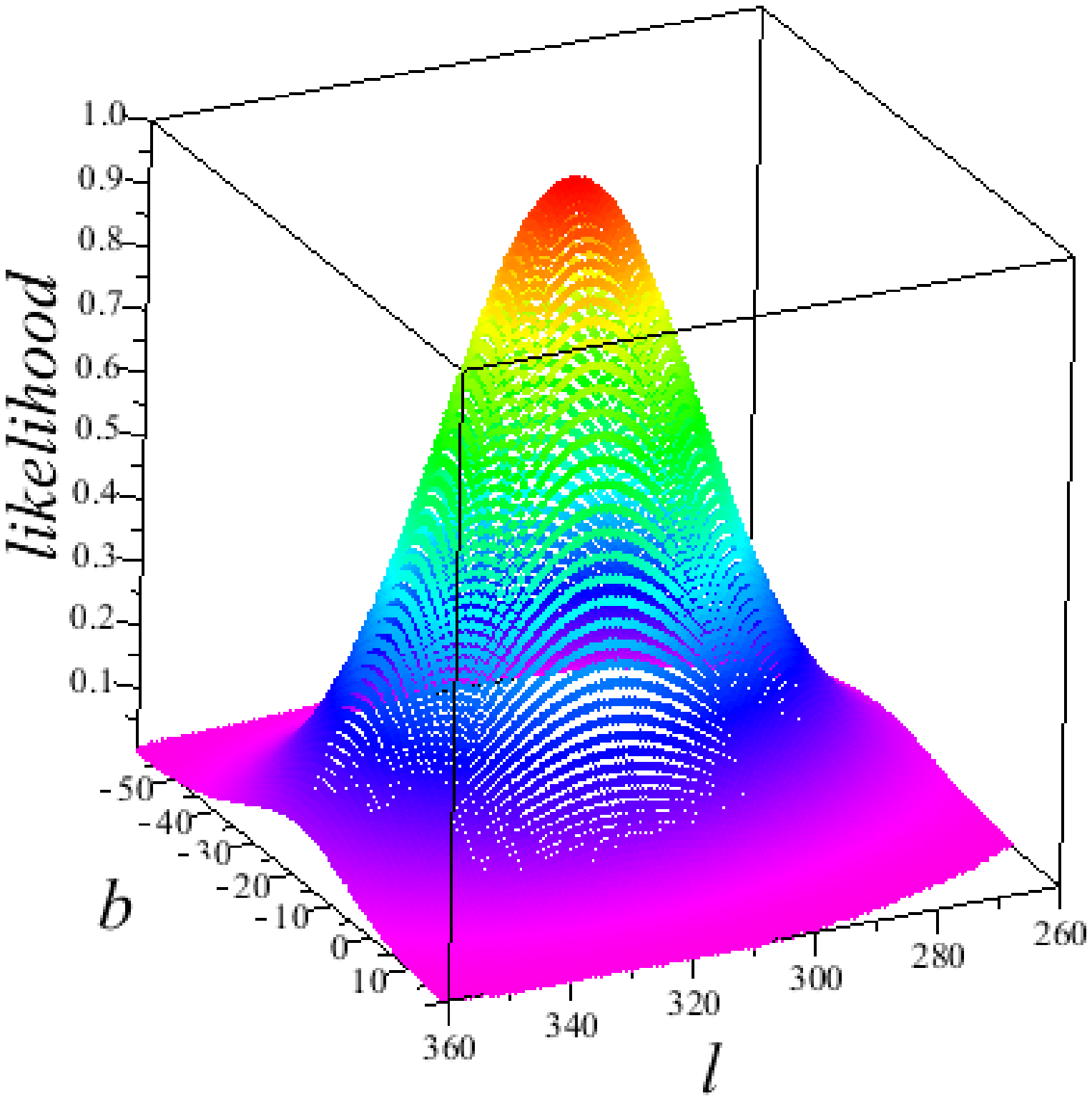}\hspace{0.1 cm}\\
Fig. 10: Two dimensional likelihood for parameters ($l,b$) in f(R,T) model using $DMFDM$ method (used $\chi^2$ analysis with $10^{5} $ datapoints)
\label{Figure:4}
\end{figure}

\begin{figure}
\centering
\includegraphics[scale=.4]{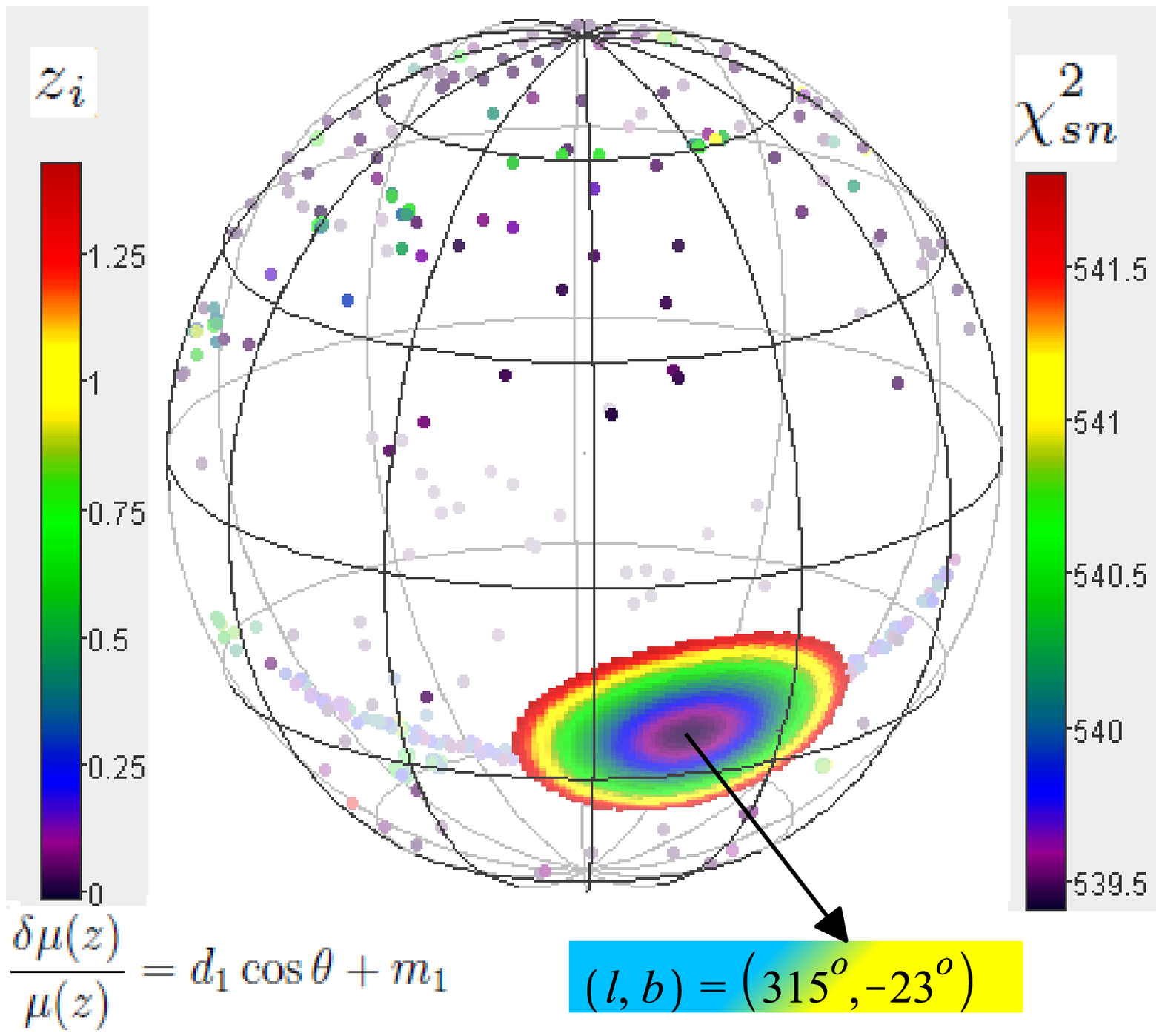}\hspace{0.1 cm}\\
Fig. 11: Union2 dataset and ($1-\sigma$) confidence level for Dark Energy Dipole direction ($l,b$) in f(R,T) model using $DMFDM$ method (used $\chi^2$ analysis with $10^{5}$ datapoints)
\label{Figure:5}
\end{figure}

The main steps of the DMFDM are shown as follows:\\
(I) \emph{Convert the equatorial coordinates of SNe Ia to galactic coordinates}\\
(II)\emph{Calculate the angle of each supernova with respect to the dipole axis, which is determined by}
\begin{eqnarray}
\cos\theta_{i}=\widehat{\textbf{\emph{z}}_{i}}.\widehat{n}
\end{eqnarray}

\emph{where} $\widehat{\textbf{z}}_{i}$ \emph{is the unit direction vector of the supernova, which can be expressed by using the Galactic coordinate system.}
\begin{eqnarray}
\widehat{\emph{z}_{i}}= cos(l_{i})sin (b_{i})\hat{i}+sin (l_{i}) sin( b_{i})\hat{j}+cos (b_{i})\hat{k}
\end{eqnarray}
  \emph{and} $\widehat{\textbf{\emph{n}}}$ \emph{is the direction of dark energy dipole, which is the maximal expanding direction,}
\begin{eqnarray}
\widehat{\textbf{\emph{n}}}= cos(l)sin (b)\hat{i}+sin (l) sin( b)\hat{j}+cos (b)\hat{k}
\end{eqnarray}
\emph{where} $(l,b)$ \emph{is the Galactic coordinate  direction of dipole axis}\\
(III) \emph{Define the angular distribution model with dipole and
monopole}\\
\begin{eqnarray}\label{delmu}
\left(\frac{\Delta\mu}{\bar{\mu}}\right)_{i}=d_{1}\cos\theta_{i}+m_{1}
\end{eqnarray}

 \emph{where $m_{1}$ and $d_{1}$ denote the monopole
and dipole magnitude, respectively} \emph{, $\bar{\mu}$ is the distance modulus predicted by the isotropic $f(R,T)$ model and ${\mu}$ is the true luminosity distance of the supernova  }\\
 (IV) \emph{Fit the SNIa data by minimizing the $\chi^{2}_{sn}$ value of the distance modulus.The $\chi^{2}_{sn}$ for SNIa is obtained by comparing theoretical distance modulus with observed $\mu^{obs}$ of supernovae}.\emph{we suppose the experiment error between each measurement is completely independent, so the covariance matrix can be simplified as the diagonal component, and the $\chi^{2}_{sn}$ can
be written as}
\begin{eqnarray}
\chi^{2}_{sn}= \sum_{i=1}^{557}\frac{[\mu^{obs}(z_{i})-\mu^{th}(\overrightarrow{z_{i}})]^{2}}{\sigma^{2}(z_{i})}.
\end{eqnarray}
\emph{where $\mu^{th}(z_{i})$ is the theoretical distance modulus which  it will be obtain  from  equation  (\ref{delmu}) as}
    \begin{align}
\mu^{th}(z_{i})=\bar{\mu}(z_{i})(1+m+d\cos\theta_{i})
\end{align}
\emph{and $\bar{\mu}(z_{i})=5log_{10}[\bar{d_{L}}(z)]+42.38-5log_{10}h$ also for a FRW cosmological model, one has}
\begin{eqnarray}\label{conser}
\bar{d_L}(z)= (1+z) \int_{0}^{z} \frac{H_{0}}{H(z')} dz'.
\end{eqnarray}
\emph{to match the equation (\ref{conser}) with set of equations (\ref{minima 1} to \ref{minima 5}), we can express the equation (\ref{conser}) by two new differential equations as}
\begin{align}
\frac{d(\bar{d_L}(z))}{dN}=-\left(\bar{d_L}(z)+\frac{e^{2N}}{H(z)}\right)\label{d1}
\end{align}
\begin{align}
\frac{d(H(z))}{dN}=H(z)\left(\frac{\dot{H}}{H^{2}}\right)=H(z)\left(\vartheta-2\right)\label{h1}
\end{align}
\emph{where $\mu_{0}=42.384-5 \log_{h}$, $ H_{0}=100h$ $km.s^{-1}.Mpc^{-1}$ and $\mu^{obs}(z_{i})$ is the measured distance modulus from the Union2 data.}\\

\begin{figure*}
\includegraphics[scale=.3]{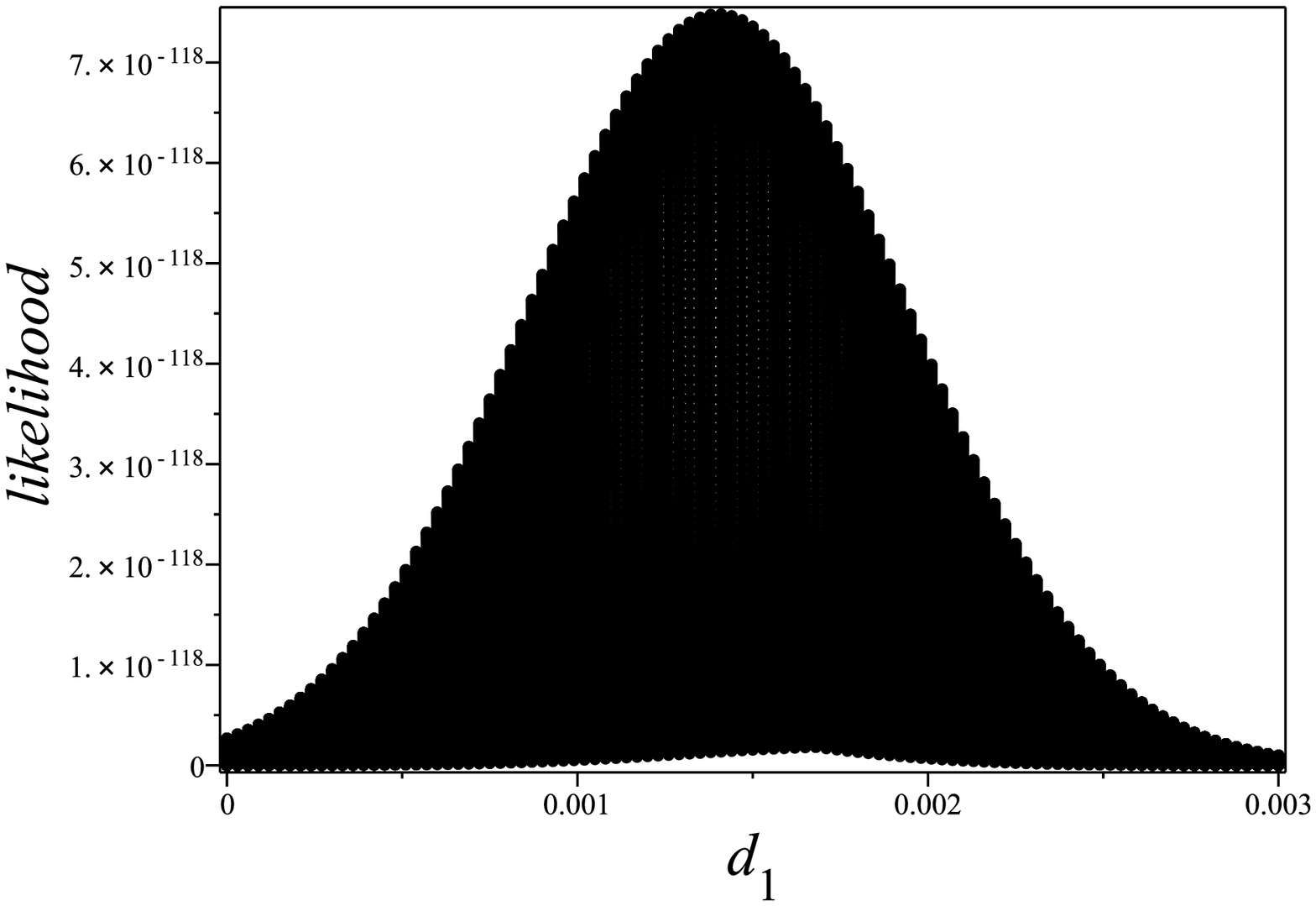}\hspace{0.1 cm}\includegraphics[scale=.3]{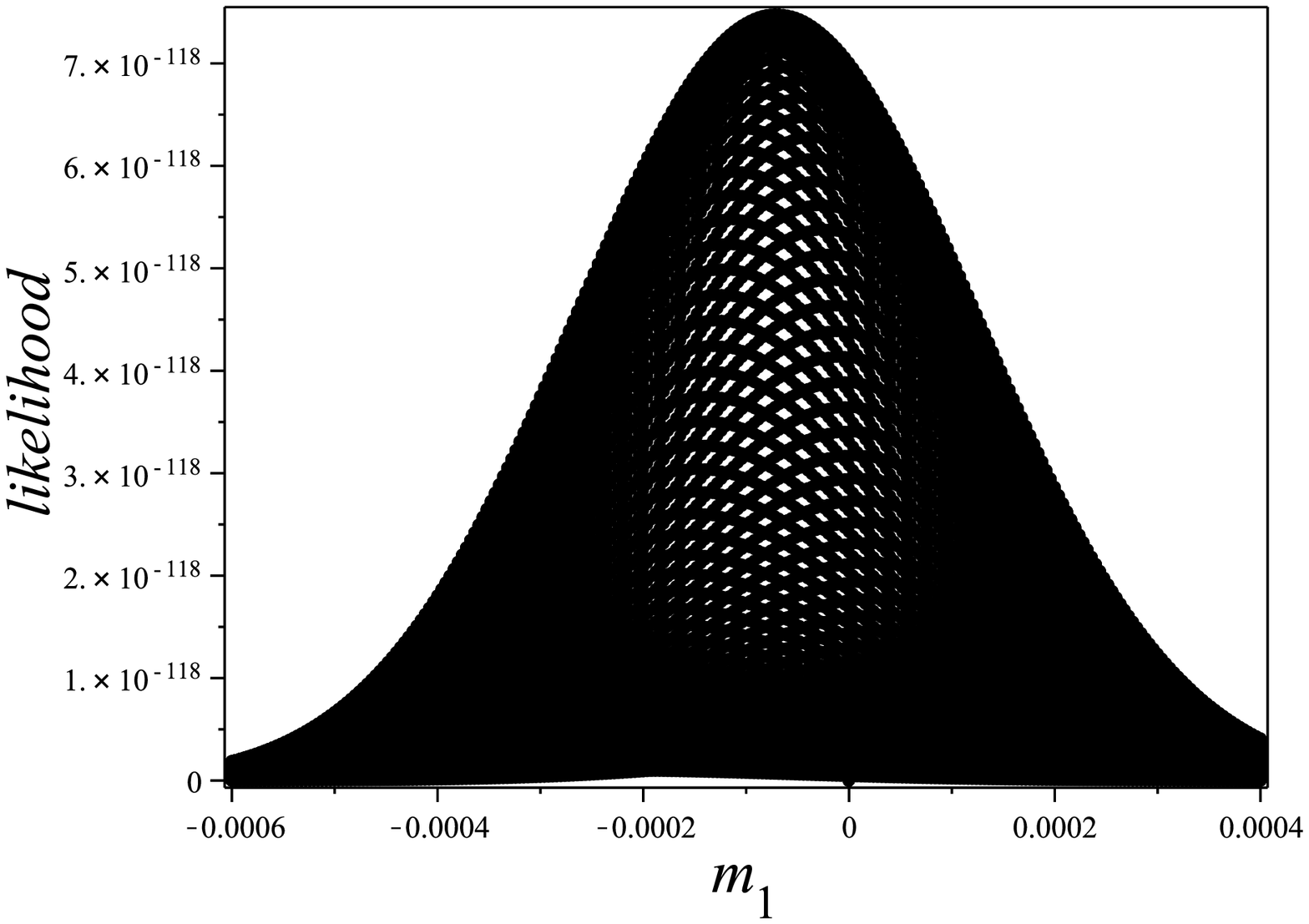}\hspace{0.1 cm}\\
Fig. 12: One dimensional likelihood for parameters($d_{1},m_{1}$) in f(R,T) model using $DMFDM$ method (used $\chi^2$ analysis with $10^{5} $ datapoints)
\label{Figure:6}
\end{figure*}

\begin{figure*}
\includegraphics[scale=.35]{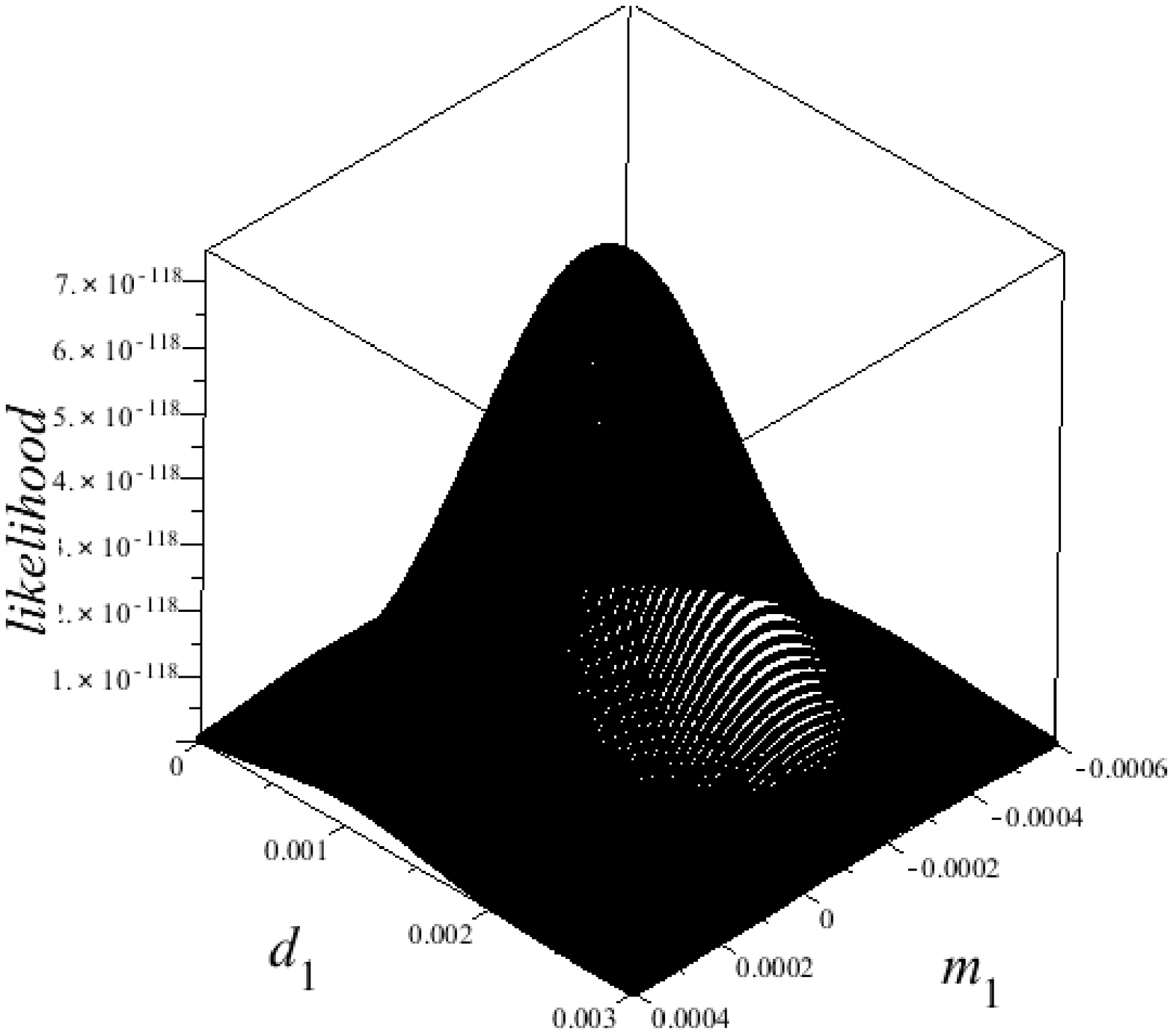}\hspace{0.1 cm}\\
Fig. 13: Two dimensional likelihood for parameters($d_{1},m_{1}$) in f(R,T) model using $DMFDM$ method (used $\chi^2$ analysis with $10^{5} $ datapoints)
\label{Figure:7}
\end{figure*}

In this step, we employ the Union2 dataset to constrain the anisotropic dark energy model. The directions of the SNIa that we have used here are given in \cite{matin} work, and are described in the equatorial coordinates (right ascension and declination). In order to make comparisons with other results, we convert these coordinates to the galactic coordinates $(l,b)$ (\cite{matin1}).\\
The parameters need to be constrained are $(d_{1}, m_{1}, l, b)$. Using the least $\chi^{2}_{sn}$ method, we can find the best-fit parameters$ (d_{1},m_{1}, l, b)$. The best-fit dipole direction is found to be towards
\begin{align}
(l,b)=(315^{0}\pm25^{0}  ,-23^{0}\pm15^{0})
\end{align}
Fig. 10. shows the two dimensional likelihood for parameters ($l,b$) in f(R,T) model using $DMFDM$ method. The distribution of Union2 SnIa Datapoints in galactic coordinates along with the dark energy dipole direction $(l,b)$ are shown in Fig.11. The magnitude of the dipole and the monopole have obtained as
\begin{align}
d_{1}=(1.4\pm 0.8)\times \times 10^{-3}  ,m_{1}=	(-0.72\pm 2.2)\times10^{-4}
\end{align}
We can see that the magnitude of the monopole is one order of magnitude smaller than that of the dipole. This is consistent to the result of \cite{Mariano}, who obtained
\begin{align}
d_{1}=(1.3\pm 0.6)\times \times 10^{-3}  ,m_{1}=	(2\pm 2.2)\times10^{-4}
\end{align}
the result of  \cite{Chang4}, who fitted  the data with a dipole only and obtained
\begin{align}
d_{1}=(1.0\pm 0.5)\times \times 10^{-3}
\end{align}
the result of \cite{Wang} for $0.015<z<8$ with
\begin{align}
d_{1}=(1.4\pm 0.6)\times  10^{-3}  ,m_{1}=	(2.7\pm 2.2)\times10^{-4}
\end{align}
the result of \cite{Yang} with
\begin{align}
d_{1}=(1.2\pm 0.5)\times  10^{-3}  ,m_{1}=	(1.9\pm 2.1)\times10^{-4}
\end{align}

We have obtained the likelihood function of each parameter by performing the $\chi^2$ analysis using $10^5$ data point. The results are shown in Fig.12. and Fig.13.\\

\textbf{B. Dipole+Monopole Fitting for Luminosity Distance (DMFLD)}\\
We perform a similar dipole+monopole fit using the Union2 data. Instead of $(\frac{\Delta\mu(z)}{\bar{\mu}(z)})$ which corresponds  to distance modulus deviations from its isotropic $f(R,T)$ value, we use the luminosity distance deviation from its best fit isotropic $f(R,T)$ value
\begin{eqnarray}\label{d22}
\left(\frac{\Delta d_{L}(z) }{\bar{d_{L}}(z)}\right)_{i}=\frac{d_{L}(z)-\bar{d_{L}}(z)}{\bar{d_{L}}(z)}=d_{2} cos\theta_{i} +m_{2}
\end{eqnarray}

\begin{figure}
\centering
\includegraphics[scale=.4]{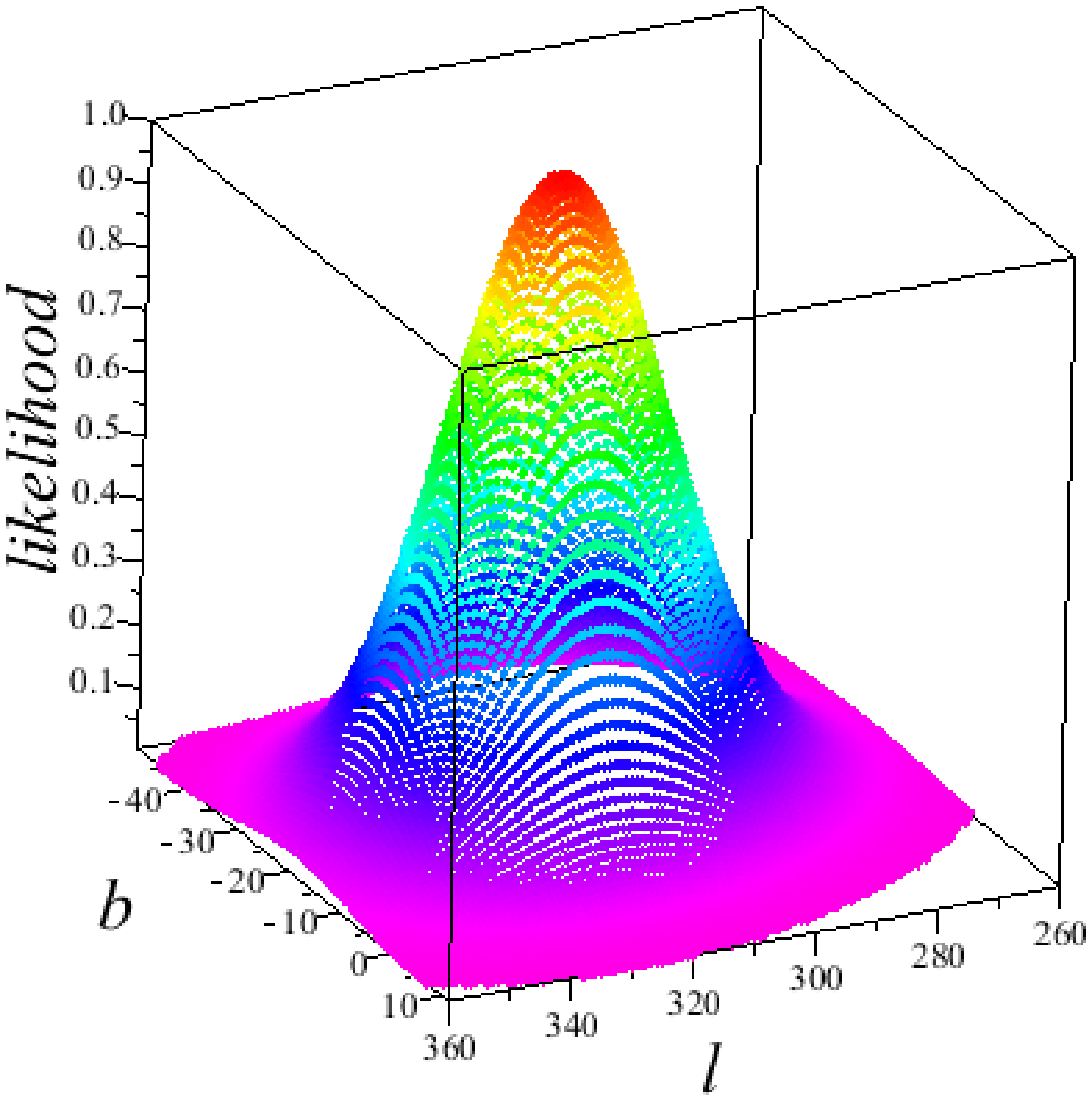}\hspace{0.1 cm}\\
Fig. 14: Two dimensional likelihood for parameters($l,b$) in f(R,T) model using $DMFLD$ method (used $\chi^2$ analysis with $10^{5} $ datapoints)
\label{Figure:8}
\end{figure}
\begin{figure}
\centering
\includegraphics[scale=.4]{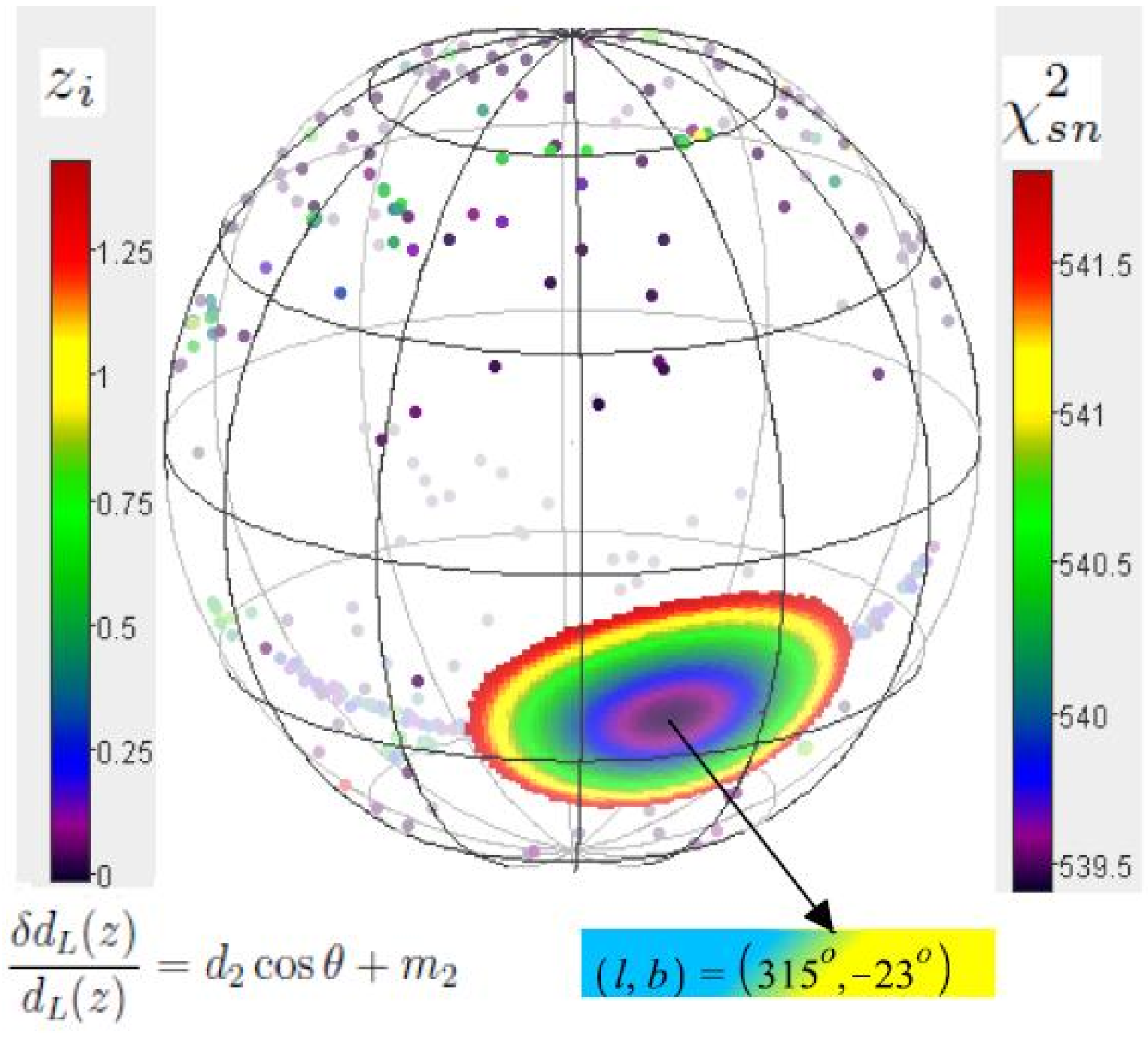}\hspace{0.1 cm}\\
Fig. 15: Union2 datapoints and ($1-\sigma$) confidence level for Dark Energy dipole direction($l,b$) in f(R,T) model using $DMFLD$ method (used $\chi^2$ analysis with $10^{5}$ datapoints)
\label{Figure:9}
\end{figure}
where,$\bar{d_{L}}(z)$ is the luminosity distance of the supernova in isotropic background and $d_{L}(z)$ is the true luminosity distance or anisotropic luminosity distance of the supernova .therfore we can use the following expression
\begin{eqnarray}\label{correspo}
 d^{anis}_{L}(z) \equiv d_{L}(z), \ \ \ d^{iso}_{L}(z)\equiv \bar{d_{L}}(z)
 \end{eqnarray}
 using (\ref{correspo}) we can rewrite the equation (\ref{d22})as
 \begin{eqnarray}\label{d222}
d^{anis}_{L}(z)=d^{iso}_{L}(z)(d_{2} cos\theta_{i} +m_{2}+1)
\end{eqnarray}
 also $\mu^{th}(z_{i})=5log_{10}[d^{anis}_{L}(z)]+42.38-5log_{10}h$
\begin{figure*}
\includegraphics[scale=.3]{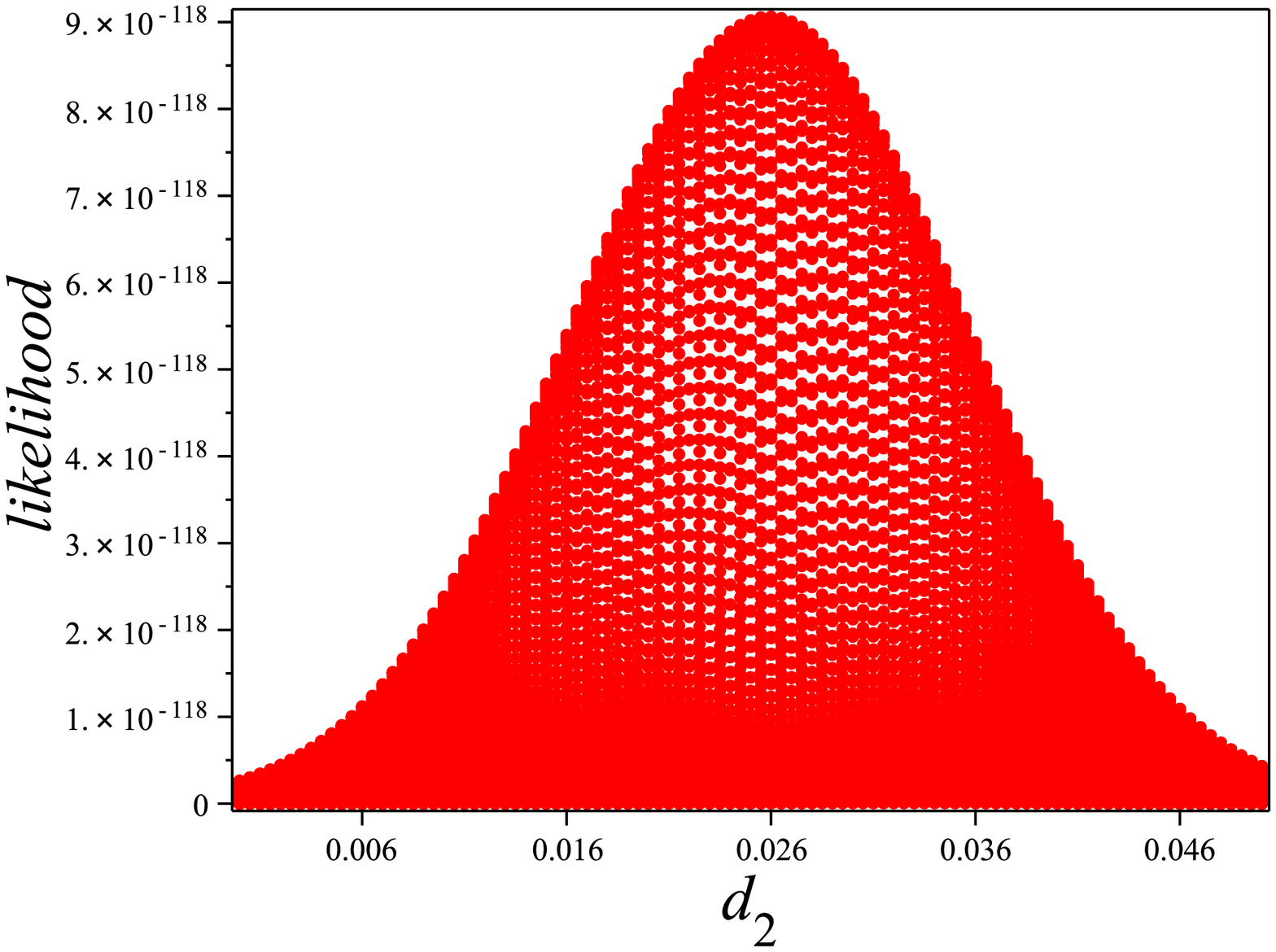}\hspace{0.1 cm}\includegraphics[scale=.3]{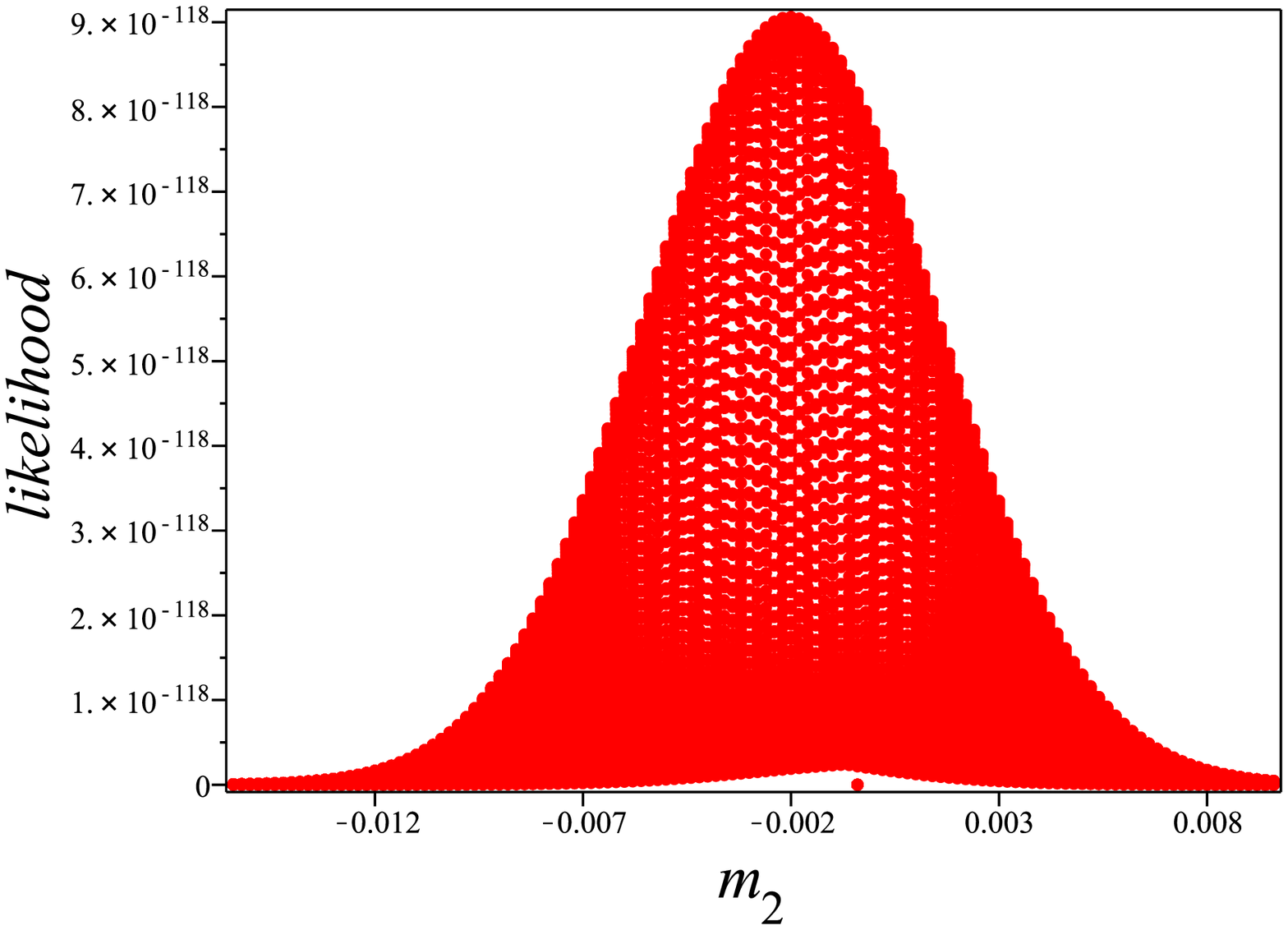}\hspace{0.1 cm}\\
Fig. 16: One dimensional likelihood for parameters($d_{2},m_{2}$) in f(R,T) model using $DMFLD$ method (used $\chi^2$ analysis with $10^{5} $ datapoints)
\label{Figure:10}
\end{figure*}

\begin{figure*}
\includegraphics[scale=.35]{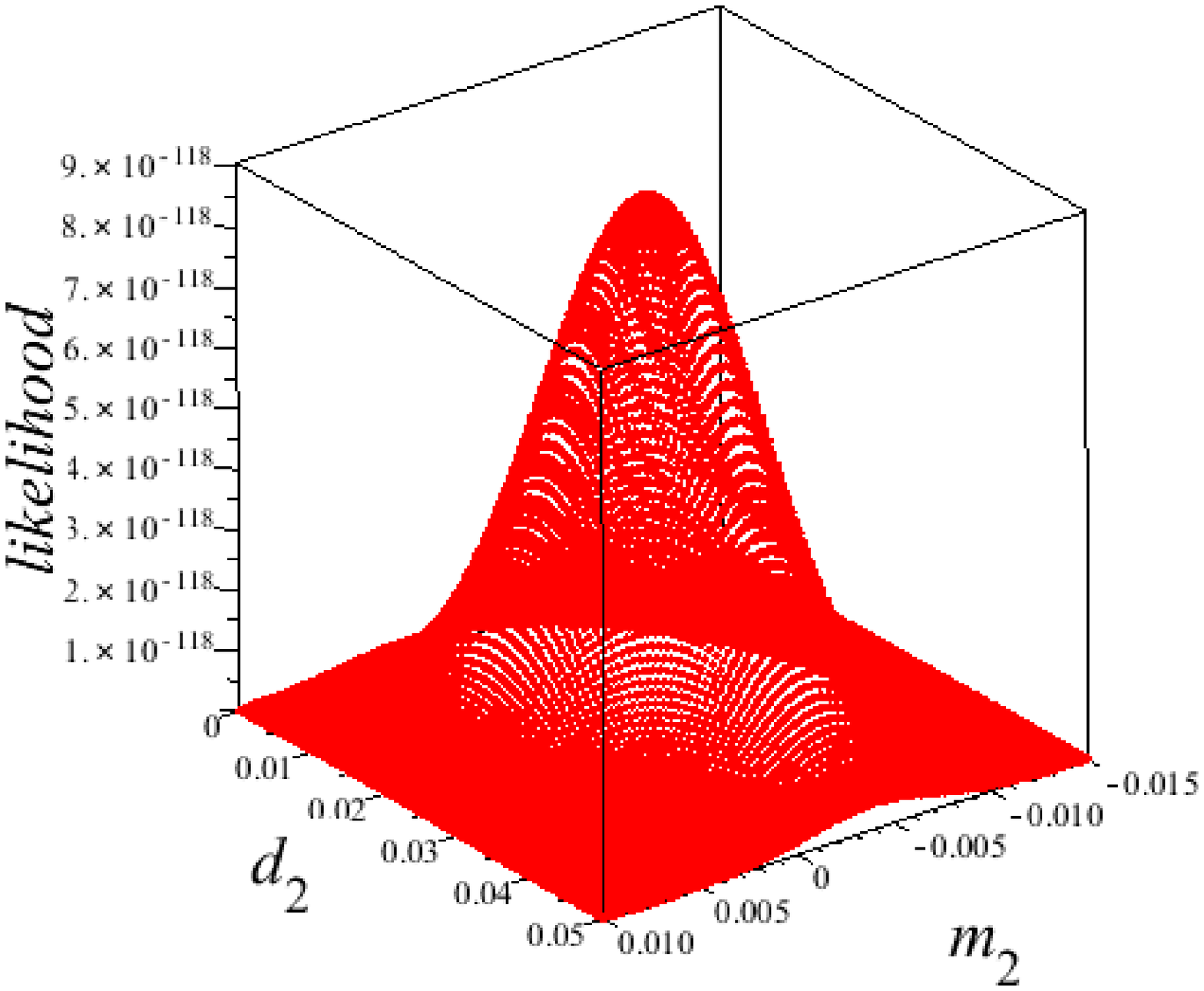}\hspace{0.1 cm}\\
Fig. 17: Two dimensional likelihood for parameters($d_{2},m_{2}$) in f(R,T) model using $DMFLD$ method (used $\chi^2$ analysis with $10^{5} $ datapoints)
\label{Figure:11}
\end{figure*}
we have found the  best fitted dipole direction as
\begin{align}
(l,b)=(315^{0}\pm37^{0}  ,-23^{0}\pm 18^{0})
\end{align}
Fig.14. shows the two dimensional likelihood for parameters ($l,b$) in f(R,T) model using $DMFLD$ method. The distribution of Union2 SnIa Datapoints in galactic coordinates along with the dark energy dipole direction $(l,b)$ are shown in Fig.15. The magnitudes of the dipole and monopole have obtained as
\begin{align}
d_{2}=(0.026\pm 0.014)  ,m_{2}=	(-1.6\pm 5.4)\times10^{-3}
\end{align}
We have obtained the likelihood function of each parameter by performing the $\chi^2$ analysis using $10^5$ data point. The results are shown in Fig.16. and Fig.17.

\textbf{C. Generalized Dipole Fitting for Luminosity Distance (GDFLD)}\\
\begin{figure}
\centering
\includegraphics[scale=.5]{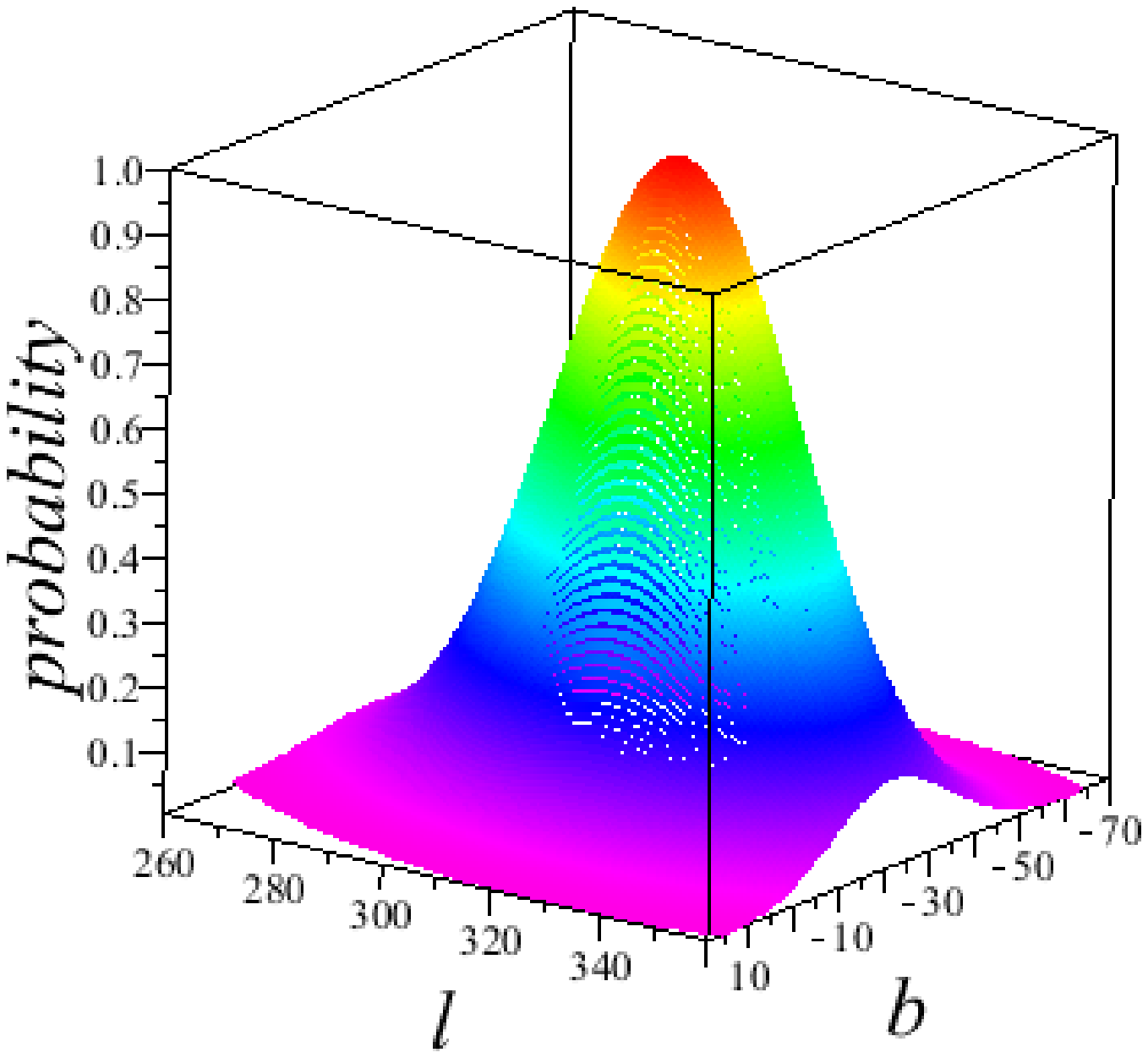}\hspace{0.1 cm}\\
Fig. 18: Two dimensional likelihood for parameters($l,b$) in f(R,T) model using $GDFLD$ method (used $\chi^2$ analysis with $10^{5} $ datapoints)
\label{Figure:12}
\end{figure}
\begin{figure}
\centering
\includegraphics[scale=.4]{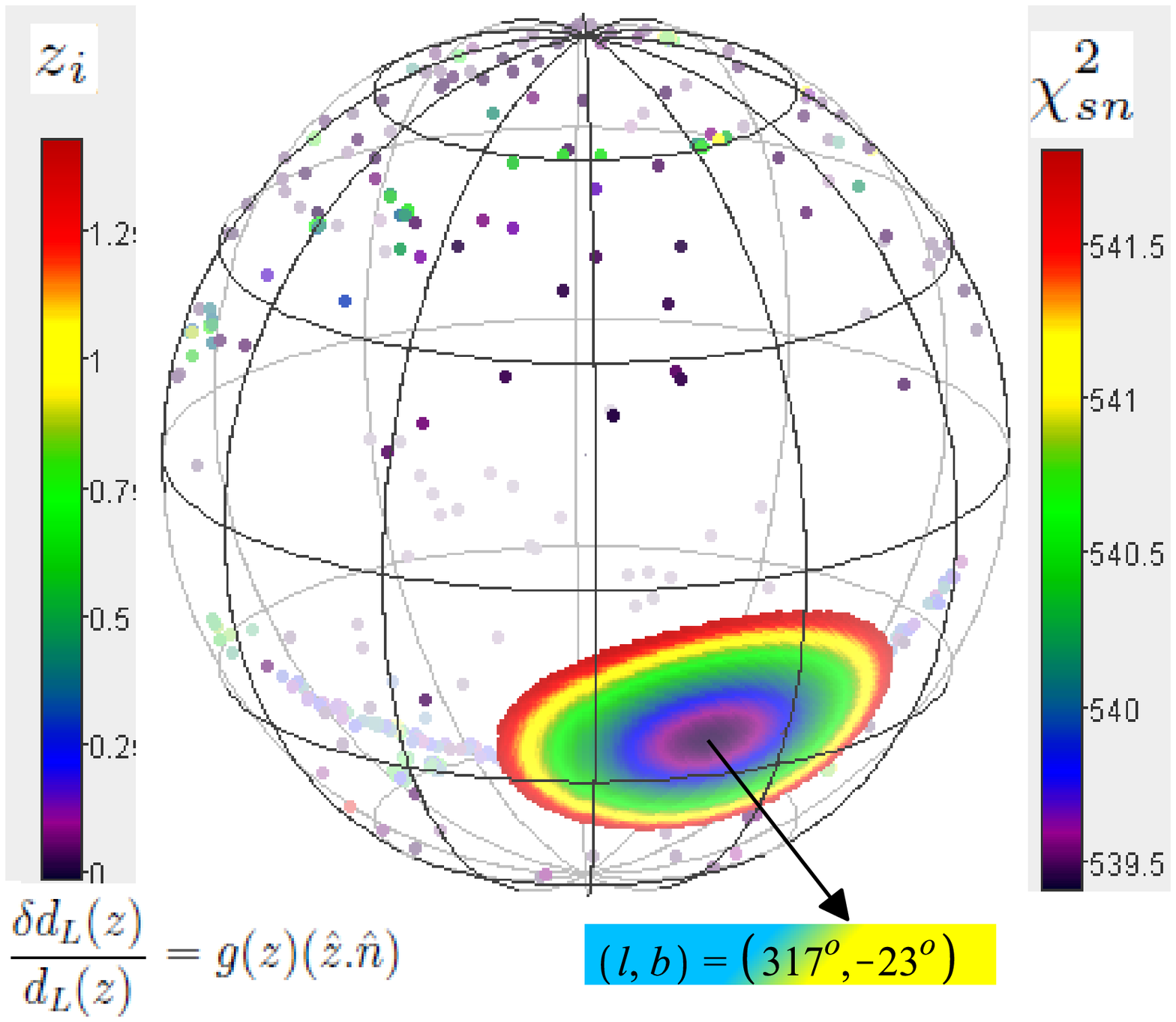}\hspace{0.1 cm}\\
Fig. 19: Union2 datapoints and ($1-\sigma$) confidence level for Dark Energy dipole direction($l,b$) in f(R,T) model using $GDFLD$ method (used $\chi^2$ analysis with $10^{5} $ datapoints)
\label{Figure:13}
\end{figure}

Some studies have shown that the monopole is not significant (monopole magnitude is $\simeq 10^{-4}$)(\cite{Wang},\cite{Antoniou}). We have also obtained (m $\simeq 10^{-4}$). Therefore, neglecting $m$  and by considering  dipole magnitude as a function of $z$ ,the general case of Luminosity Distance dipole fit will be
\begin{eqnarray}\label{d2}
\frac{d_{L}(z)-d^{0}_{L}(z)}{d^{0}_{L}(z)}=g(z)cos\theta=g(z)(\hat{z}.\hat{n})
\end{eqnarray}
\cite{Cai} first applied this method to $\Lambda CDM $ model by assuming  linear function
\begin{figure*}
\includegraphics[scale=.3]{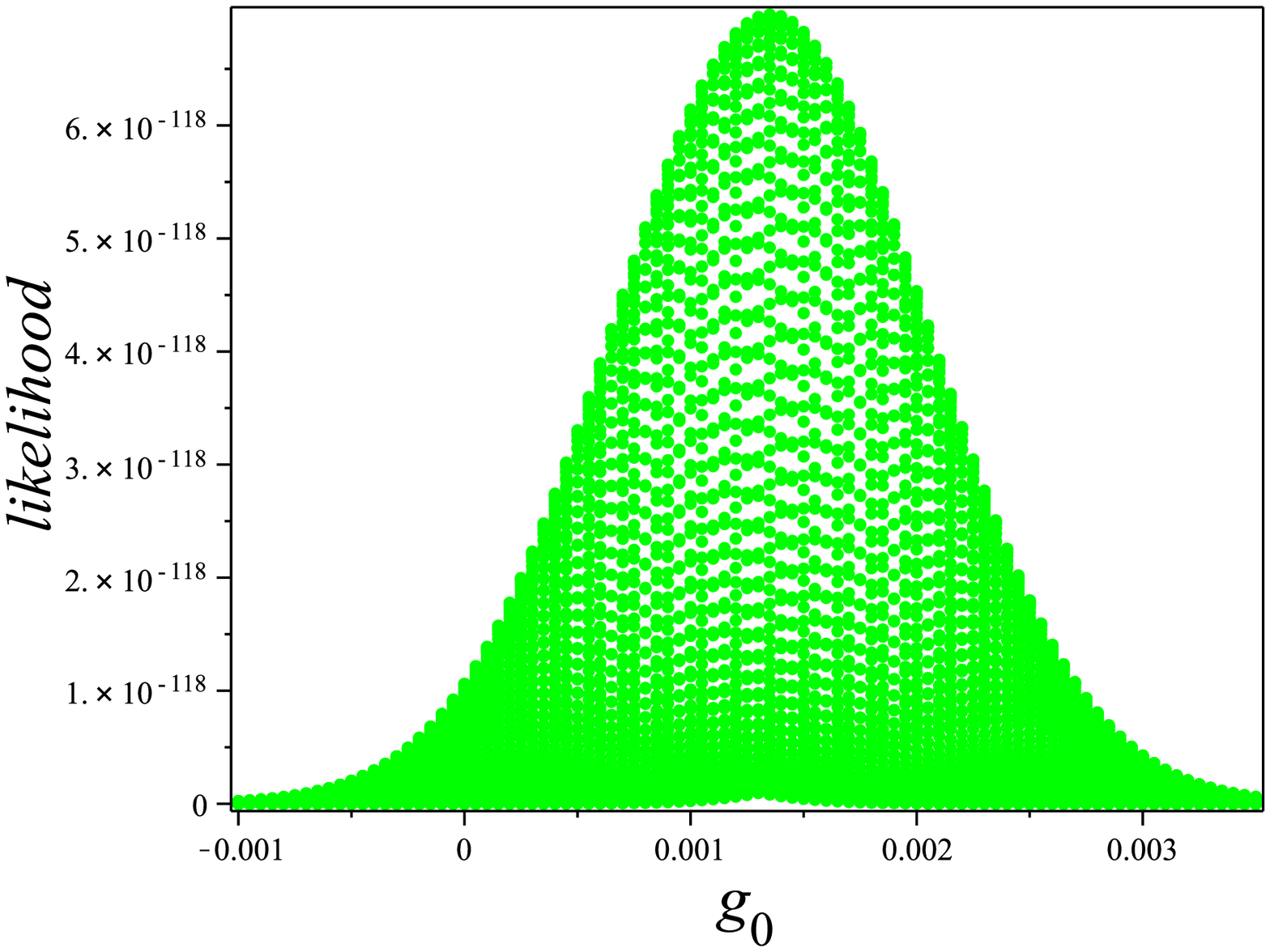}\hspace{0.1 cm}\includegraphics[scale=.3]{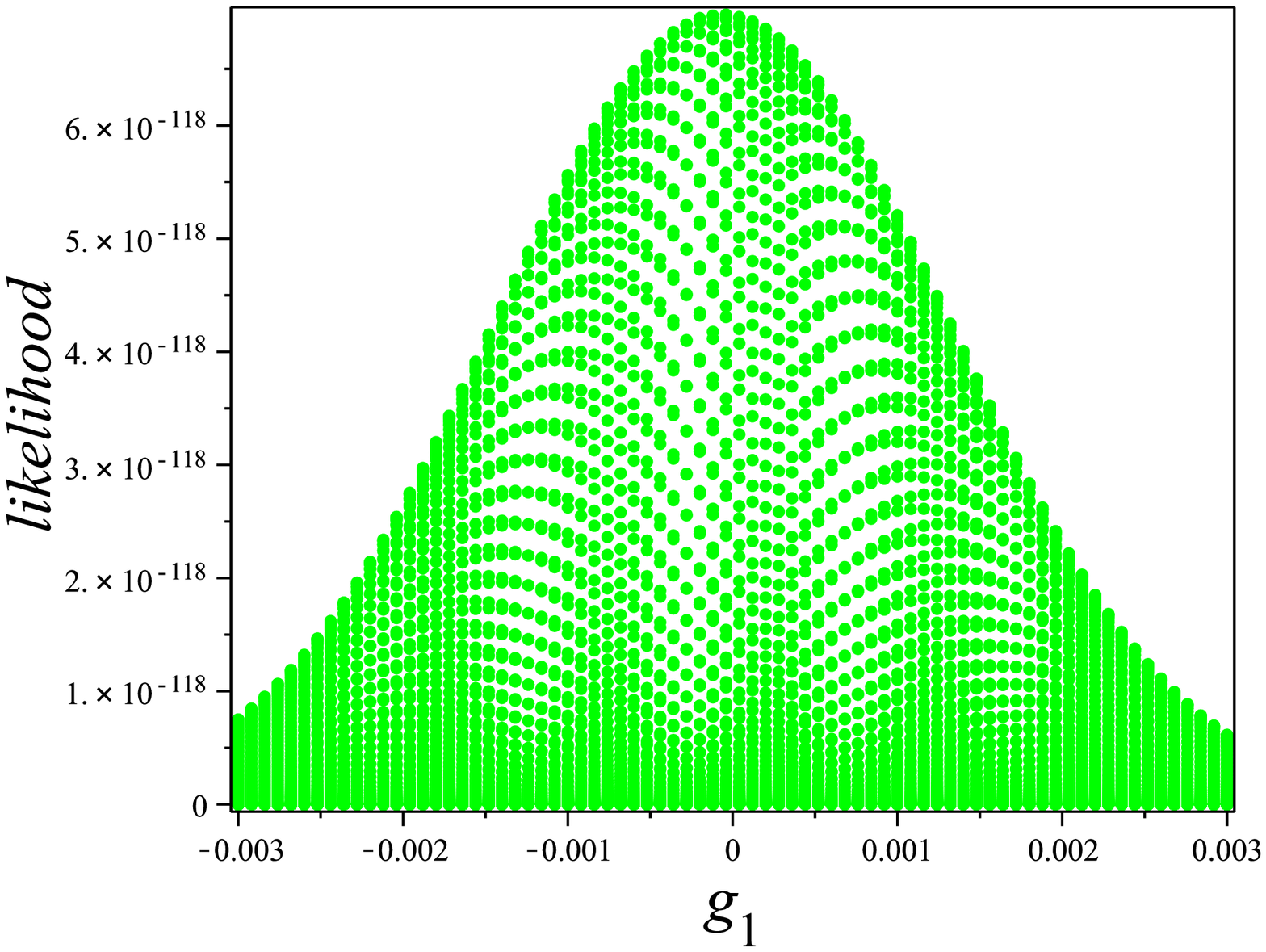}\hspace{0.1 cm}\\
Fig. 20: One dimensional likelihood for parameters($g_{0},g_{1}$) in f(R,T) model using $GDFLD$ method (used $\chi^2$ analysis with $10^{5} $ datapoints)
\label{Figure:14}
\end{figure*}

\begin{figure*}
\includegraphics[scale=.35]{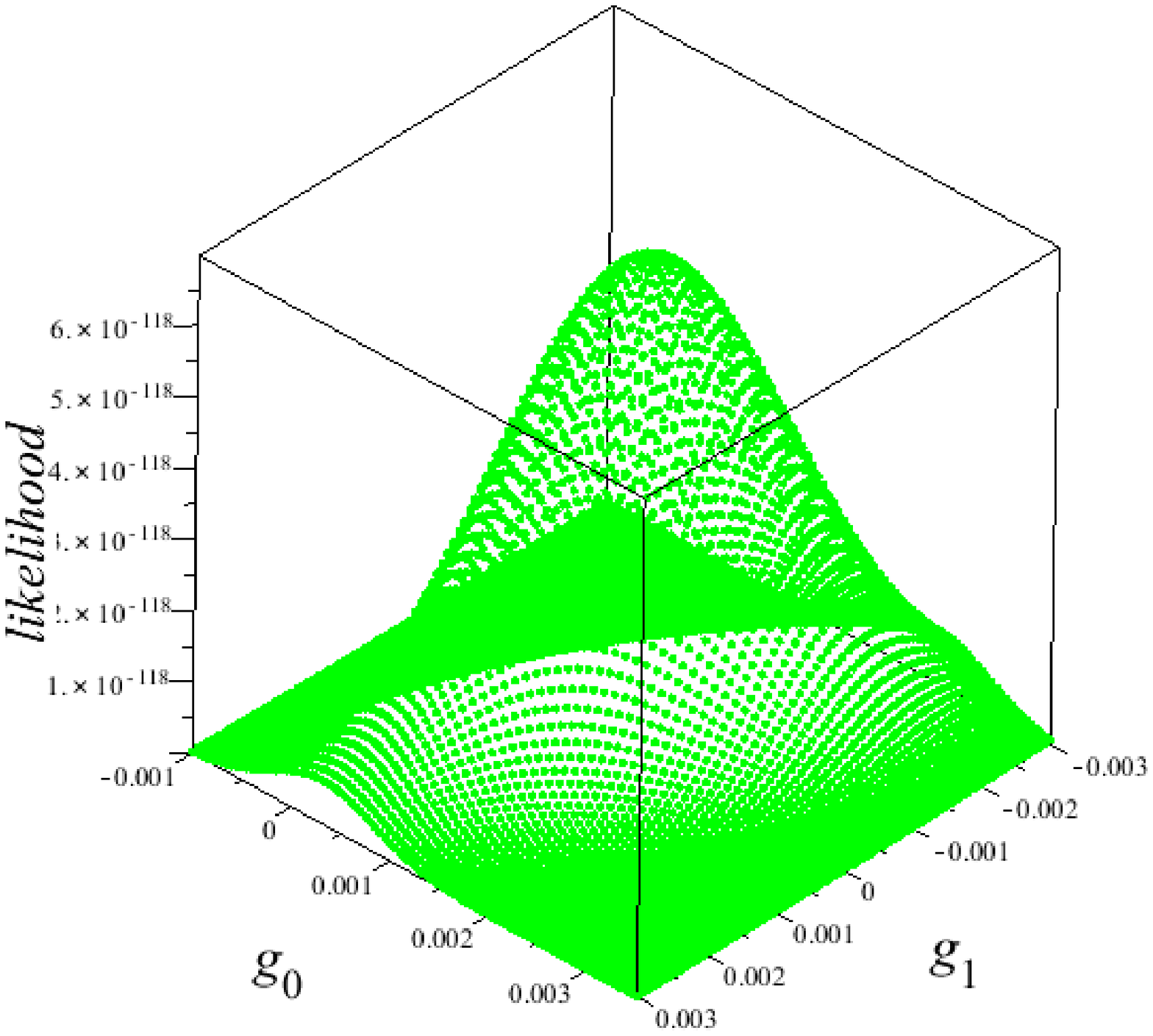}\hspace{0.1 cm}\\
Fig. 21: Two dimensional likelihood for parameters($g_{0},g_{1}$) in f(R,T) model using $GDFLD$ method (used $\chi^2$ analysis with $10^{5} $ datapoints)
\label{Figure:15}
\end{figure*}
of $z$ as
\begin{equation}
g(z) = g_{0} + g_{1}z:
\end{equation}
we have found the best fitted dipole direction as
\begin{align}
(l,b)=(317^{0}\pm32^{0}  ,-23^{0}\pm 18^{0})
\end{align}
Fig.18. shows the two dimensional likelihood for parameters ($l,b$) in f(R,T) model using $GDFLD$ method. The distribution of Union2 SnIa Datapoints in galactic coordinates along with the dark energy dipole direction $(l,b)$ are shown in Fig.19. The magnitudes of the $g_{0}$ and $g_{1}$ have obtained as

\begin{align}
g_{0}=(1.35\pm 1)\times10^{-3}  ,g_{1}=	(-0.4\pm 2)\times10^{-4}
\end{align}
We have obtained the likelihood function of each parameter by performing the $\chi^2$ analysis using $ 10^5$ data point. The results are shown in Fig.20. and Fig.21.

\section{Comparison Of Three DF Models}
In the previous section, we have described three types of dipole-fitting (DF) method which has been used for statistical analysis in order to find the preferred cosmological axis of the Universe in $f(R,T)$ model. These three types are as follows:\\
(I) Dipole + Monopole Fitting for Distance Modulus (DMFDM),\\
(II) Dipole + Monopole Fitting for Luminosity Distance (DMFLD),\\
(III) General Dipole Fitting for Luminosity Distance (GDFLD).\\

\begin{figure*}
\centering
\includegraphics[scale=.4]{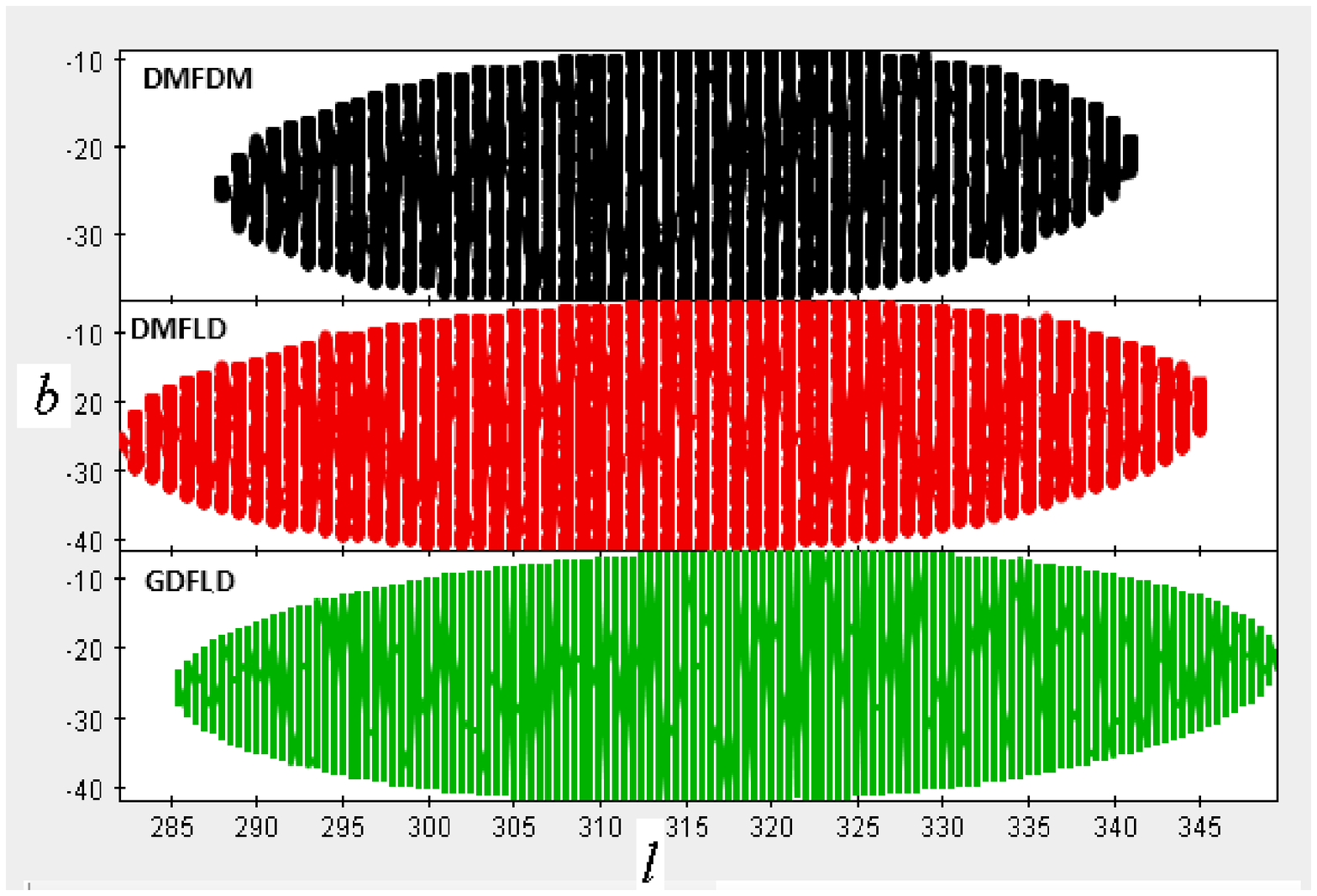}\hspace{0.1 cm}\includegraphics[scale=.4]{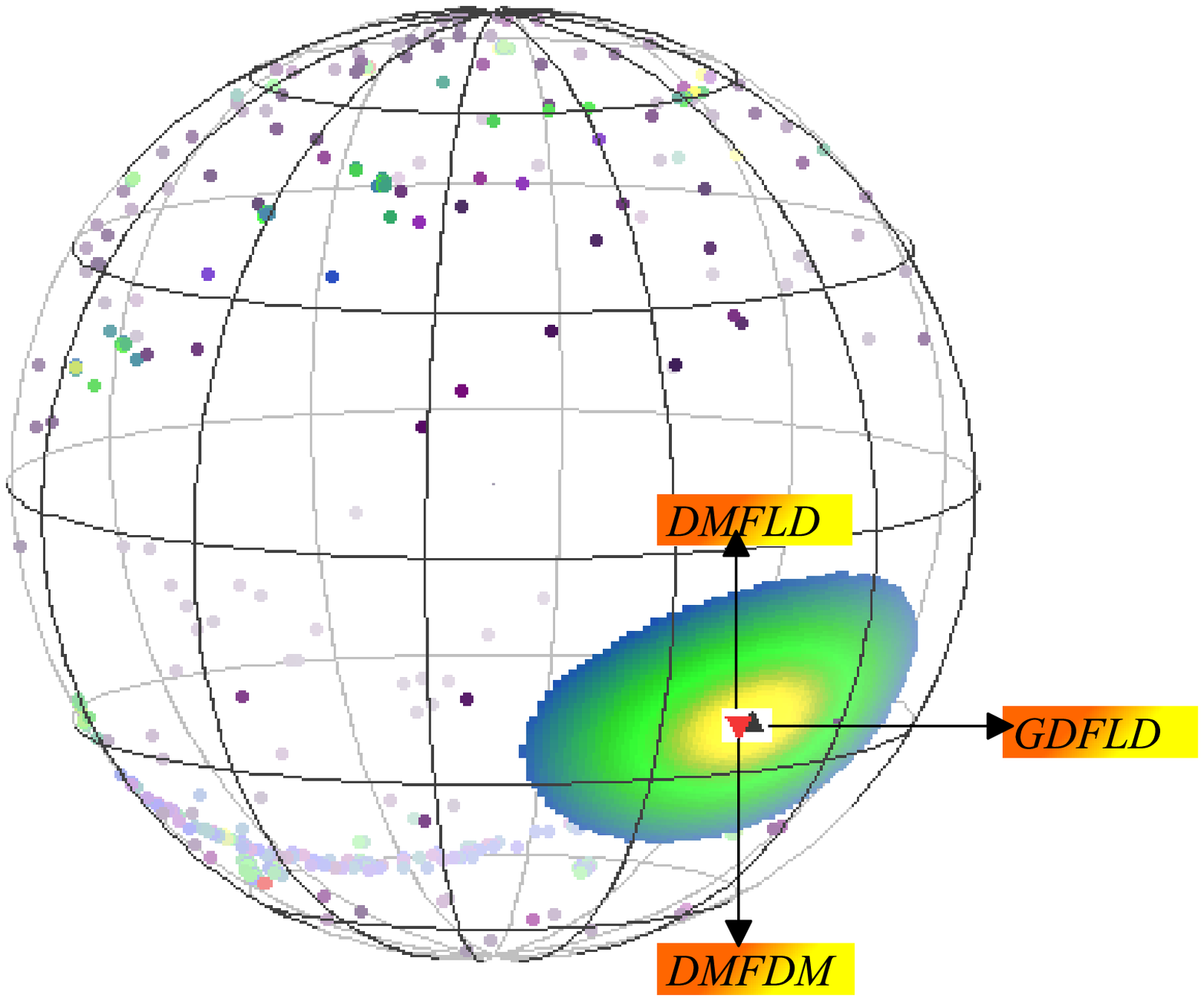}\hspace{0.1 cm}\\
Fig. 22: Comparison of ($1-\sigma$) confidence level of parameteres ($l,b$) for $DMFDM$, $DMFLD$ and $GDFLD$ models
\label{Figure:16}
\end{figure*}

Several groups have applied $DMFDM$ method to study the anisotropy of $\Lambda CDM$, $\omega CDM$ and the dark energy model with $CPL$ parametrization. \cite{Cai} have applied $ GDFLD $ method to study the anisotropy of $\Lambda CDM$, $\omega CDM$ and the dark energy model with $CPL$ parametrization. We have applied all of these DF  methods to study privilege axis of the universe in $f(R,T)$ model.\\
At first, it seems that these methods have a same origin (because of the direct relation between $\mu$ and $d_{L}$). Also, the best fitted  direction of preferred axis of these methods are very close to each other. In fact, $DMFDM$ ($(l,b)=(315^{0}\pm25^{0},-23^{0}\pm15^{0})$) and $DMFLD$ ($(l,b)=(315^{0}\pm37^{0}, -23^{0}\pm 18^{0})$) methods have resulted exactly the same value for the privilege axis of the universe in $f(R,T)$ model. However, their $1-\sigma$ confidence level are different. As left panel of Fig.16. shows, the ($1-\sigma$) confidence region of $DMFDM$ is smaller than $DMFLD$ (right panel of Fig.22.). Moreover, they give different values of dipole magnitude which is interesting to note. The dipole magnitude obtained using  $DMFDM$ method ($d_{1}=(1.4\pm 0.8)\times \times 10^{-3}$) is close to previous studies of \cite{Chang4}, \cite{Wang}, \cite{Yang} as it has been mentioned in $DMFDM$ method section. However, the dipole magnitude obtained using $DMFLD$ method ($d_{2}=(0.026\pm 0.014)$) is different from the value obtained using $DMFDM$ method and also previous studies. Interestingly, the magnitude of anisotropy ($d_{2}=(0.026\pm 0.014)$) obtained using $DMFLD$ method is approximately equal to that of CMB dipole. The recent released Planck data show that the dipole magnitude of CMB temperature fluctuations is about $A=0.07 ± 0.01$ (\cite{Chang3}).\\
There are two reasons to study the dark energy dipole of the universe using the formula based on deviation on Luminosity distance ($DMFLD$ method) instead of  distance modulus ($ DMFDM $ method). The first is that if dark energy has anisotropic repulsive force, it will directly affect the expansion rate of the Universe, leading to the anisotropic luminosity distance; therefore, in formulating the dipole-fitting method it is more appropriate that the $d_{L}$ be revealed directly in the equation. The later reason is that most of the formulaes for modification of $d_{L}$ presented in Table \ref{table:2} with very small values of dipole magnitude $(d\ll1)$  can be simplified as
\begin{equation}
d_{L}=d^{0}_{L}(1\pm d cos\theta),\ \  \frac{d_{L}-d^{0}_{L}}{d^{0}_{L}}=\pm d cos\theta
\end{equation}
which is the same as the DF equation of  $DMFLD$ and $GDFLD$ methods.\\

\section{redshift tomography analysis for three types of DF method in f(R,T) model}
In order to explore the possible redshift dependence of the anisotropy, we implement a redshift tomography analysis, for the following redshift slices: 0-0.2, 0-0.4, 0-0.6, 0-0.8, 0-1.0, 0-1.2, 0-1.4. the results of redshift tomography analysis for f(R,T) model using three types of the DF method are summarized in Table \ref{table:4}, \ref{table:5} and \ref{table:6}.\\

\begin{table}
\caption{Constraints of the directions and amplitude of maximum anisotropy using $DMFDM$ method for different redshift bins of the
SNIa data. The error-bars quoted is 1$\sigma$ error.} 
\centering 
\begin{tabular}{c c c c c c c  } 
\hline\hline 
range &\ \ $l$&\ \ $b$ &$ m_{1}$ & $d_{1}$ \\ [3ex] 
\hline 
0 - 0.2\ \ &$121^{-27}_{+26}$  & $15^{-17}_{+16}$ \ & $-0.00035^{-0.00042}_{+0.00045}$\ \  & $0.00185^{-0.00110}_{+0.00115}$ \\ 
\hline 
0 - 0.4\ \  &$125^{-24}_{+22}$\ &$19^{-14}_{+13}$\ & $-0.00032^{-0.00033}_{+0.00035}$\ \  & $0.00180^{-.00100}_{+0.00100}$\\ 
\hline 
0 - 0.6\ \  &$128^{-28}_{+28}$\ &$16^{-17}_{+18}$\ & $-0.00021^{-0.00030}_{+0.00034}$\ \  & $0.00135^{-0.00090}_{+0.00090}$\\ 
\hline 
0 - 0.8\ \  &$130^{-25}_{+25}$\ &$21^{-14}_{+15}$\ & $-0.00014^{-0.00029}_{+0.00030}$\ \  &$0.00150^{-0.00090}_{+0.00090}$\\ 
\hline 
0 - 1.0\ \   & $134^{-28}_{+24}$\ &$20^{-16}_{+16}$\ & $-0.00011^{-0.00027}_{+0.00030}$\ \ &$0.00135^{-0.00085}_{+0.00085}$\\ 
\hline 
0 - 1.2\ \  &$132^{-26}_{+26}$\ &$22^{-15}_{+14}$\ & $-0.00010^{-0.00028}_{+0.00027}$\ \  &$0.00140^{-0.00085}_{+0.00080}$\\ 
\hline 
0 - 1.4\ \ &$135^{-25}_{+25}$\ &$23^{-15}_{+14}$\ & $-0.00007^{-0.00028}_{+0.00028}$\ \ &$0.00140^{-0.00080}_{+0.00080}$\\ 
 [1ex] 
\hline\hline 
\end{tabular}
\label{table:4} 
\end{table}

\begin{figure*}
\centering
\includegraphics[scale=.35]{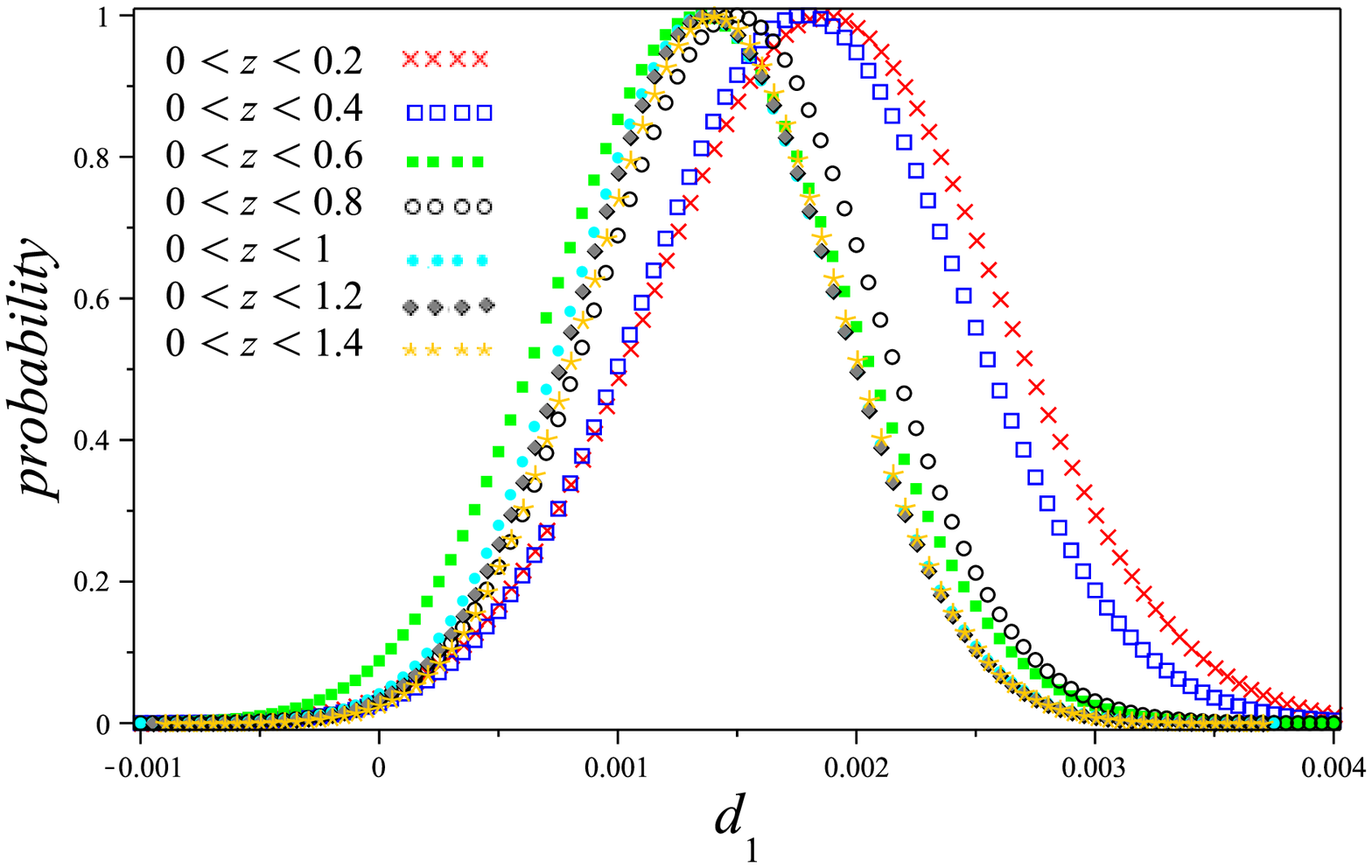}\hspace{0.1 cm}\includegraphics[scale=.35]{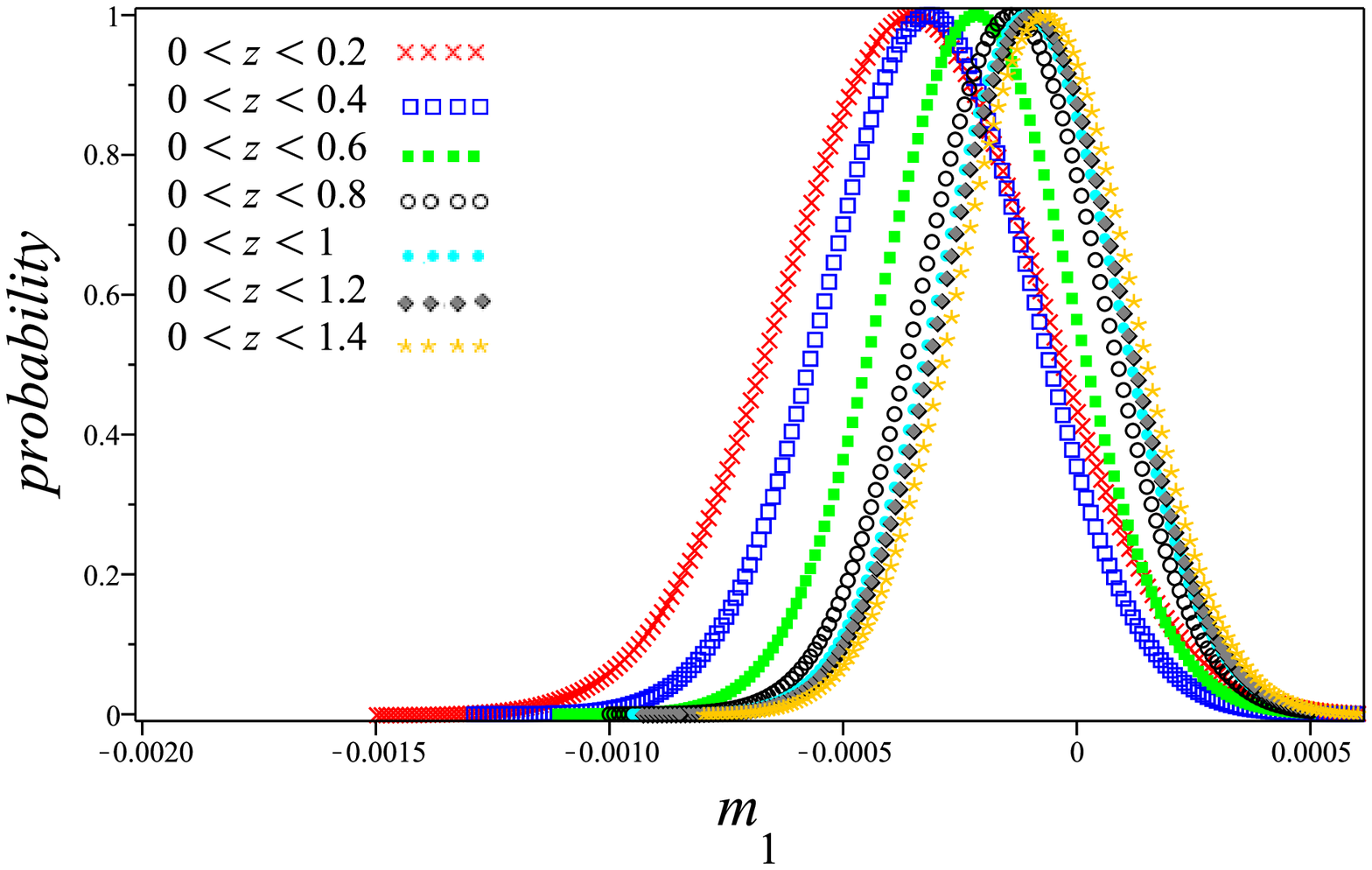}\hspace{0.1 cm}\\
\includegraphics[scale=.35]{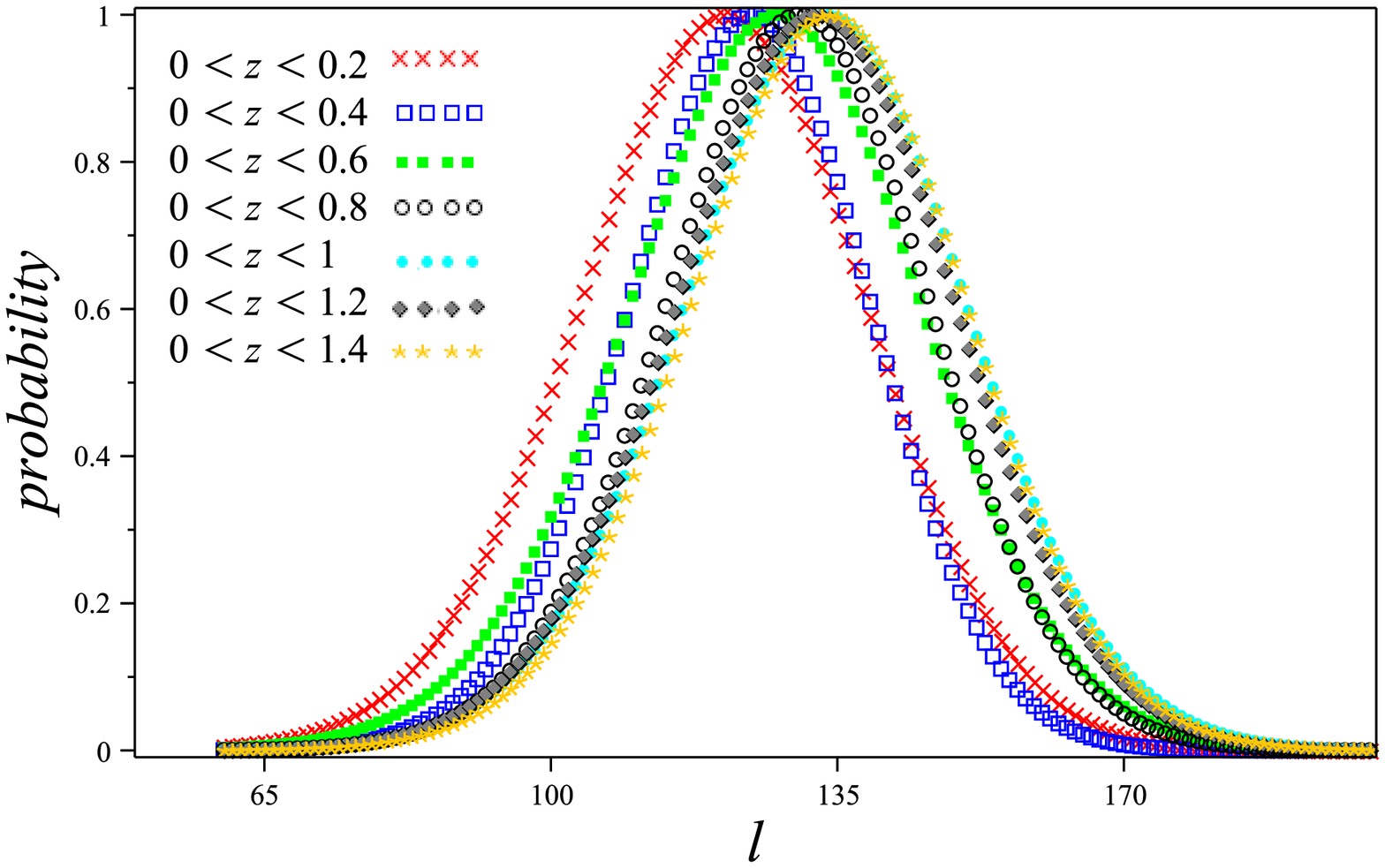}\hspace{0.1 cm}\includegraphics[scale=.35]{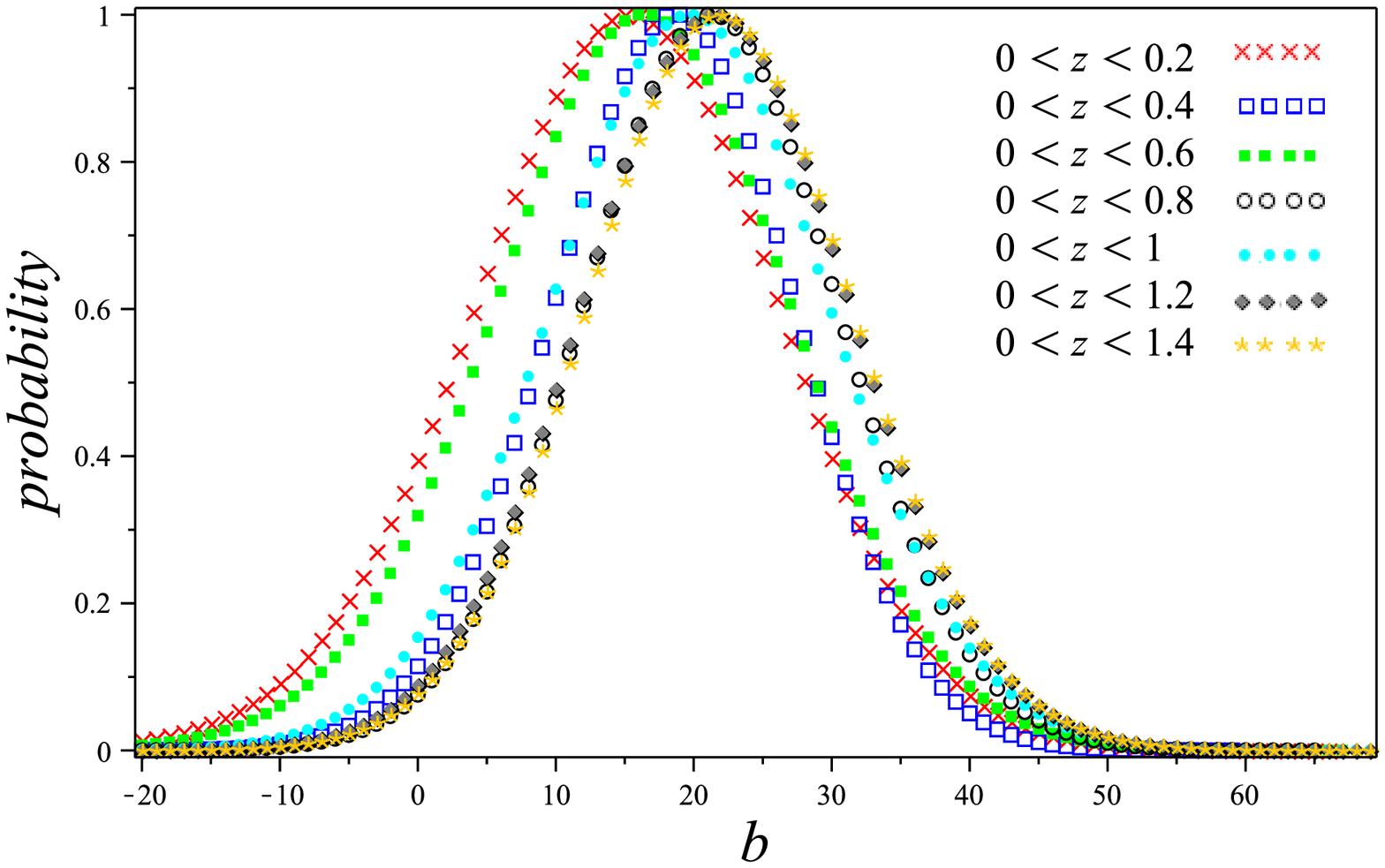}\hspace{0.1 cm}\\
Fig. 23: The redshift tomography analysis for $DMFDM$ method in f(R,T) model.
\label{Figure:17}
\end{figure*}

\begin{figure*}
\centering
\includegraphics[scale=.35]{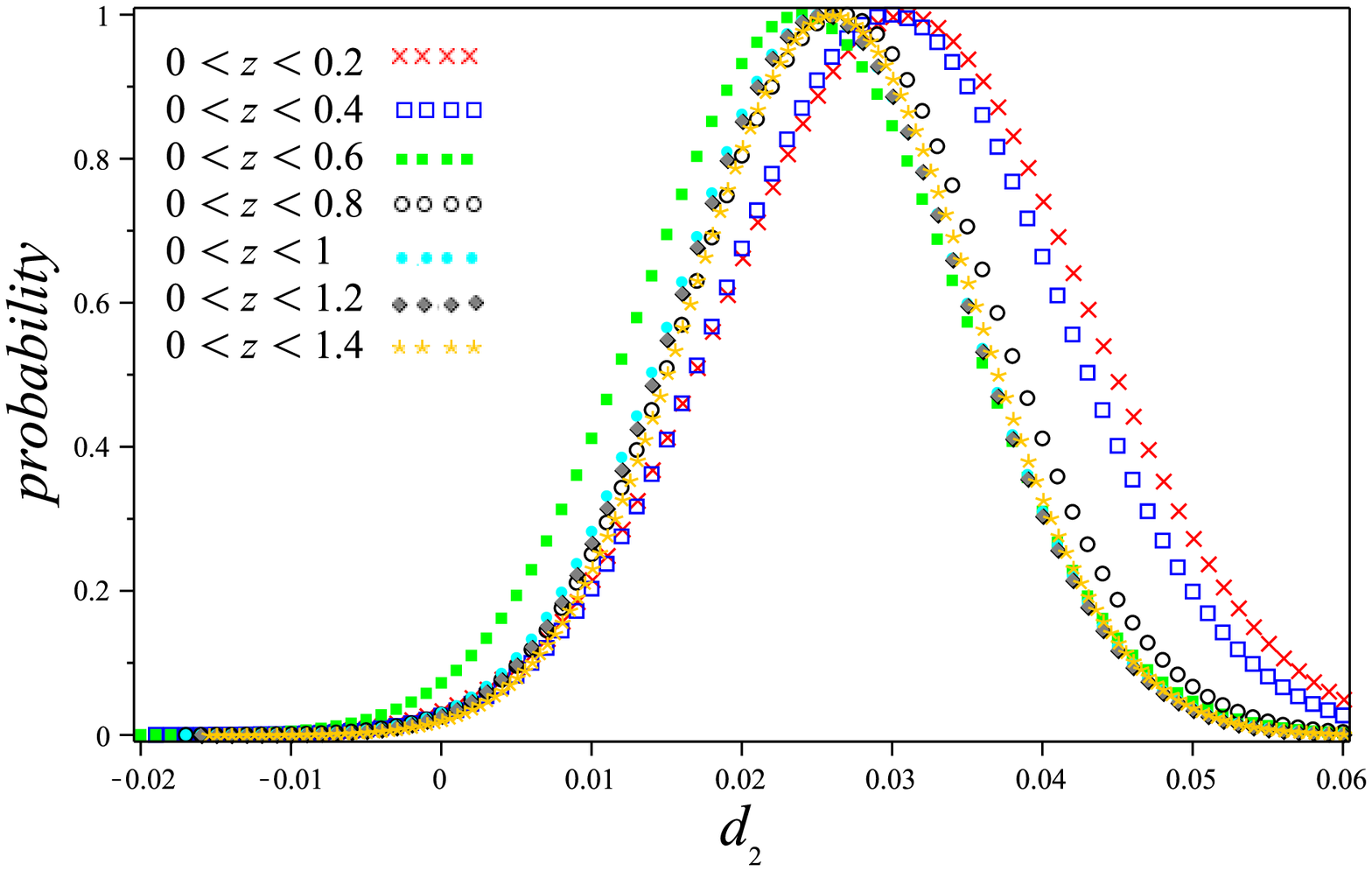}\hspace{0.1 cm}\includegraphics[scale=.35]{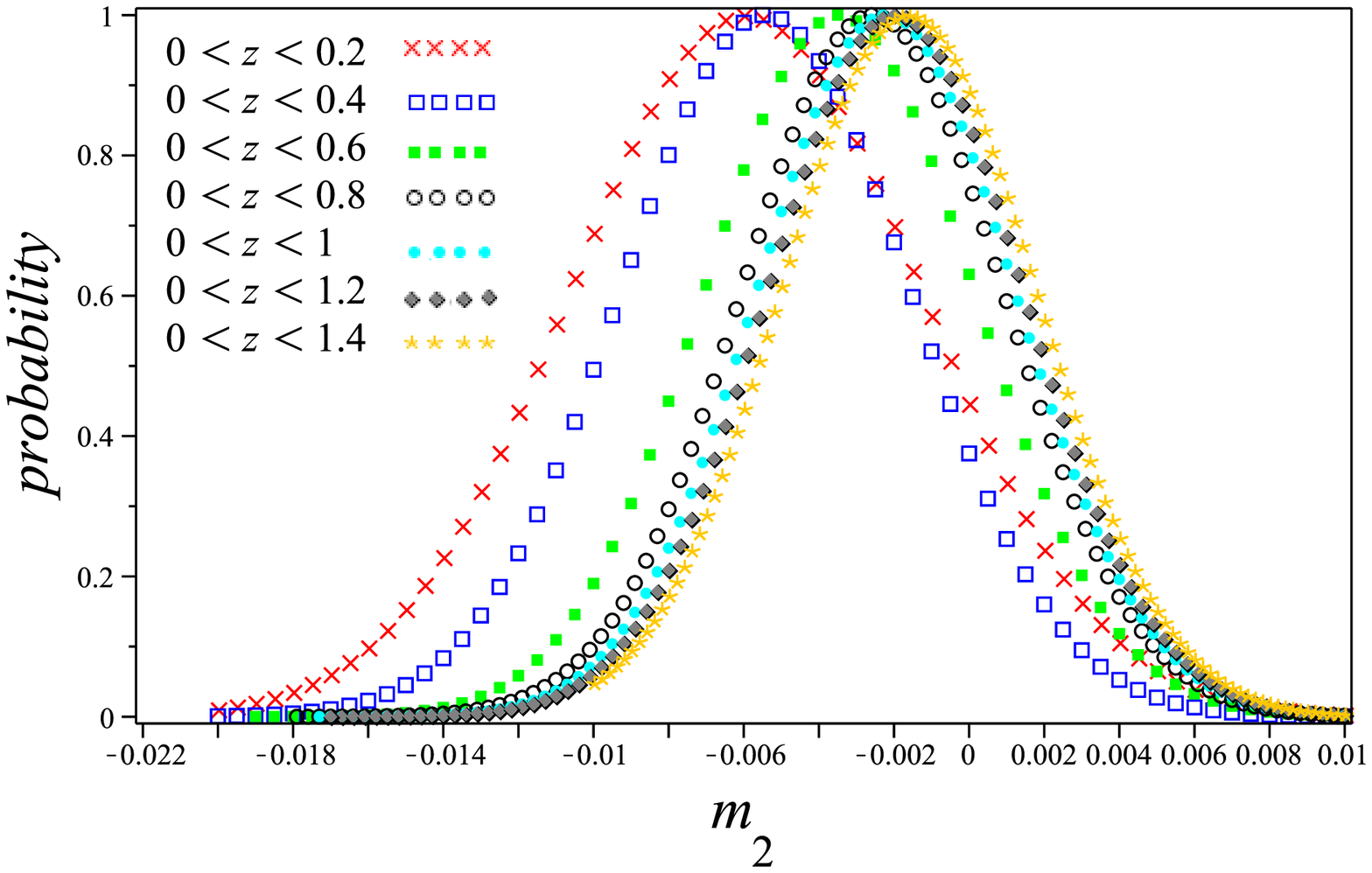}\hspace{0.1 cm}\\
\includegraphics[scale=.35]{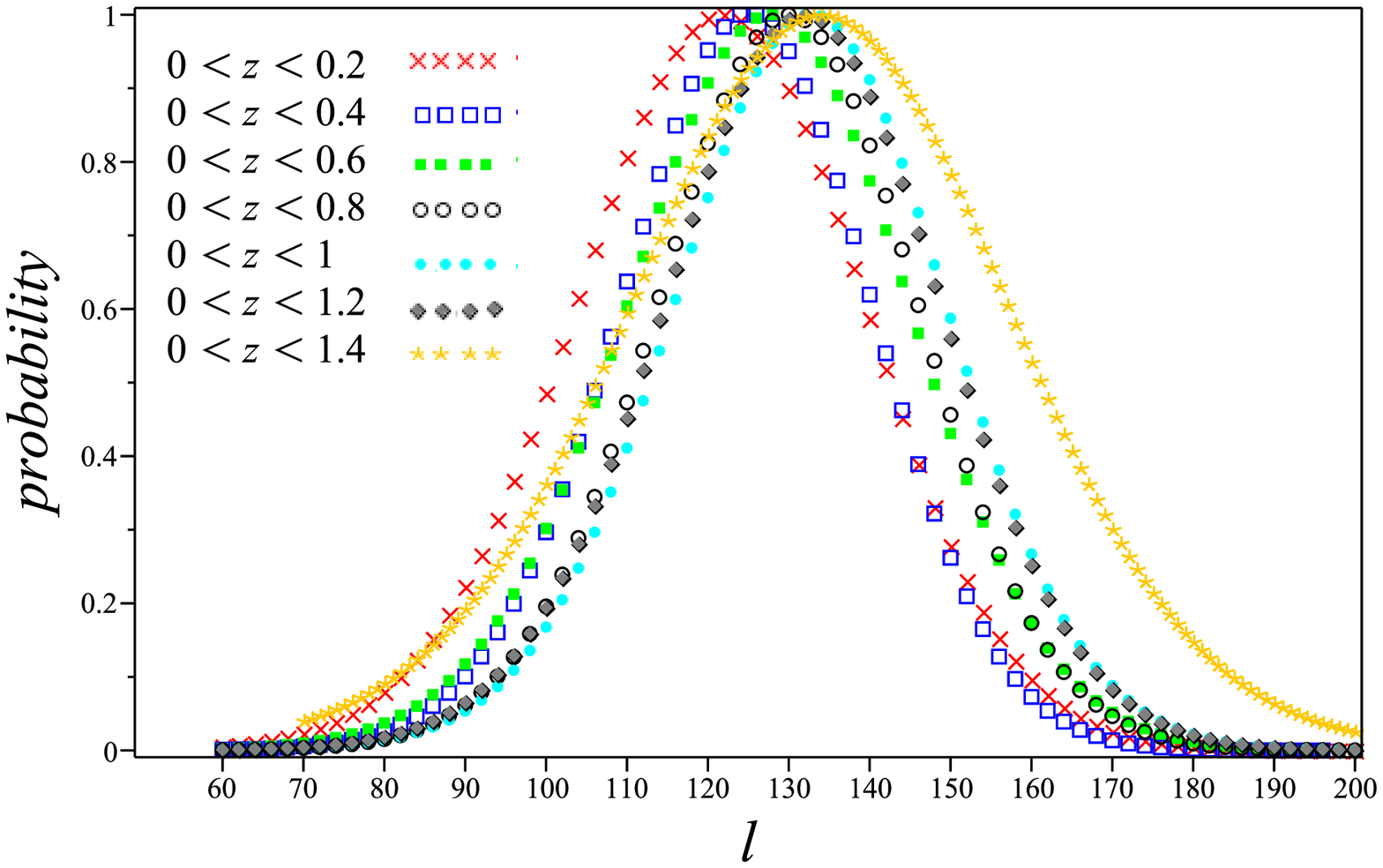}\hspace{0.1 cm}\includegraphics[scale=.35]{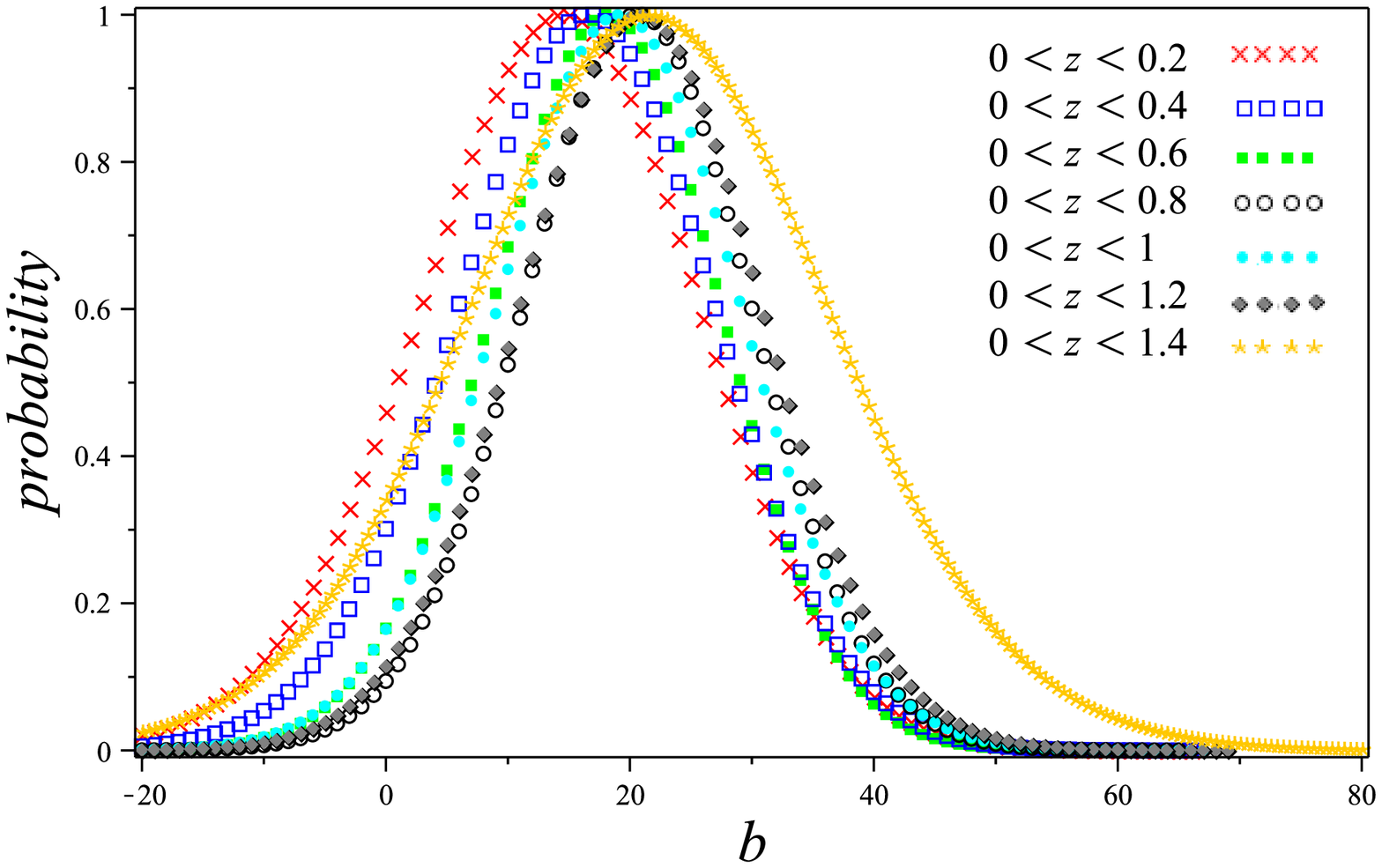}\hspace{0.1 cm}\\
Fig. 24: The redshift tomography analysis for $DMFLD$ method in f(R,T) model.
\label{Figure:18}
\end{figure*}

\begin{table}
\caption{Constraints of the directions and amplitude of maximum anisotropy using $DMFLD$ method for different redshift bins of the
SNIa data. The error-bars quoted is 1$\sigma$ error.} 
\centering 
\begin{tabular}{c c c c c c c  } 
\hline\hline 
range &\ \ $l$&\ \ $b$ &$ m_{2}$ & $d_{2}$ \\ [3ex] 
\hline 
0 - 0.2\ \ &$122^{-28}_{+26}$  & $14^{-17}_{+16}$ \ & $-0.0060^{-0.0075}_{+0.0075}$\ \  & $0.031^{-0.017}_{+0.018}$ \\ 
\hline 
0 - 0.4\ \  &$124^{-24}_{+22}$\ &$18^{-14}_{+13}$\ & $-0.0055^{-0.0055}_{+0.0060}$\ \  & $0.030^{-0.017}_{+0.017}$\\ 
\hline 
0 - 0.6\ \  &$128^{-28}_{+26}$\ &$17^{-16}_{+15}$\ & $-0.0035^{-0.0050}_{+0.0055}$\ \  & $0.024^{-0.015}_{+0.016}$\\ 
\hline 
0 - 0.8\ \  &$130^{-25}_{+24}$\ &$21^{-15}_{+14}$\ & $-0.0026^{-0.0051}_{+0.0052}$\ \  &$0.027^{-0.013}_{+0.013}$\\ 
\hline 
0 - 1.0\ \   & $132^{-24}_{+26}$\ &$19^{-15}_{+15}$\ & $-0.0023^{-0.0051}_{+0.0054}$\ \ &$0.025^{-0.014}_{+0.015}$\\ 
\hline 
0 - 1.2\ \  &$132^{-26}_{+24}$\ &$21^{-22}_{+22}$\ & $-0.0020^{-0.0051}_{+0.0052}$\ \  &$0.025^{-0.013}_{+0.011}$\\ 
\hline 
0 - 1.4\ \ &$135^{-37}_{+35}$\ &$23^{-18}_{+18}$\ & $-0.0016^{-0.0035}_{+0.0038}$\ \ &$0.026^{-0.010}_{+0.010}$\\ 
 [1ex] 
\hline\hline 
\end{tabular}
\label{table:5} 
\end{table}

\begin{figure*}
\centering
\includegraphics[scale=.3]{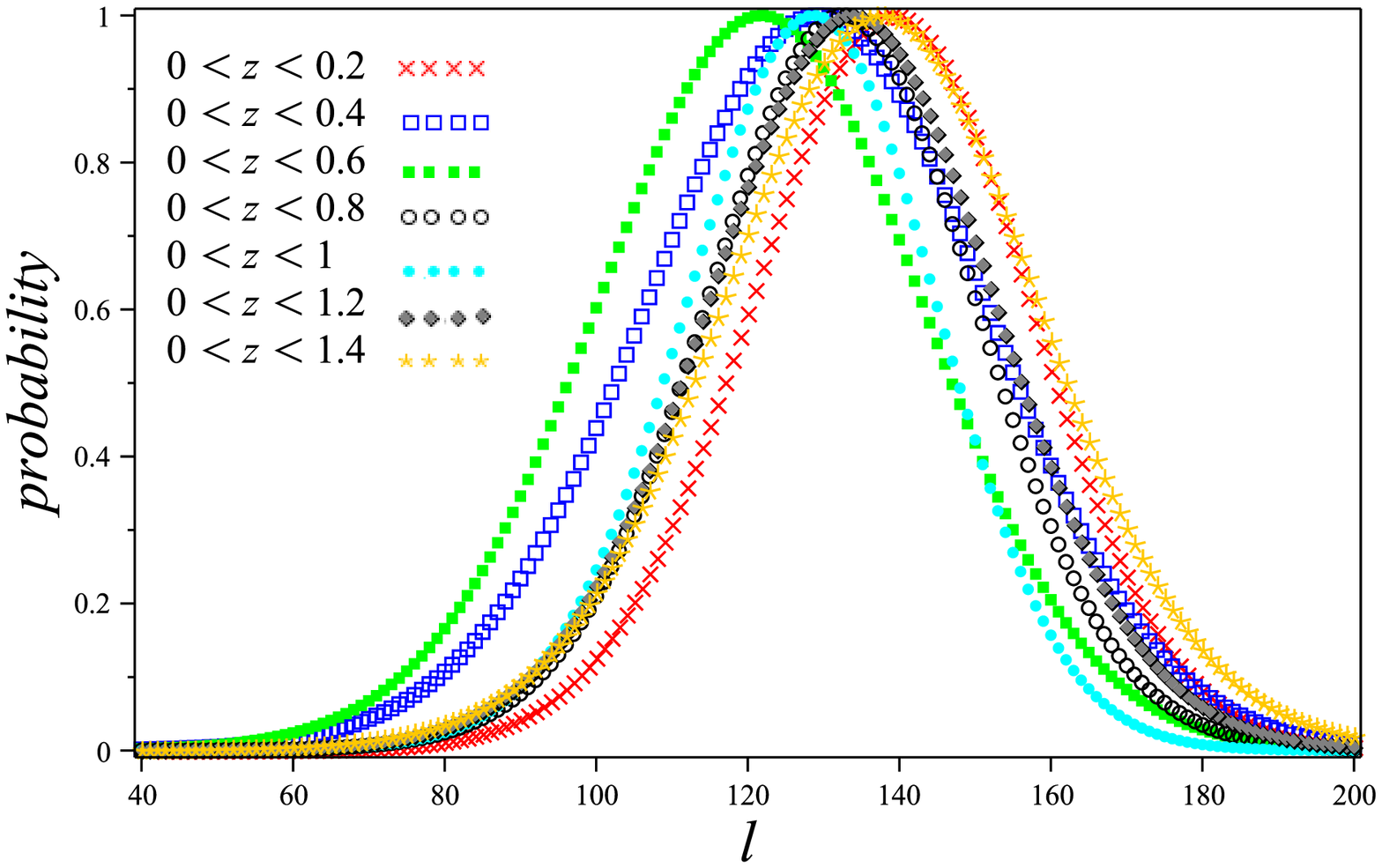}\hspace{0.1 cm}\includegraphics[scale=.3]{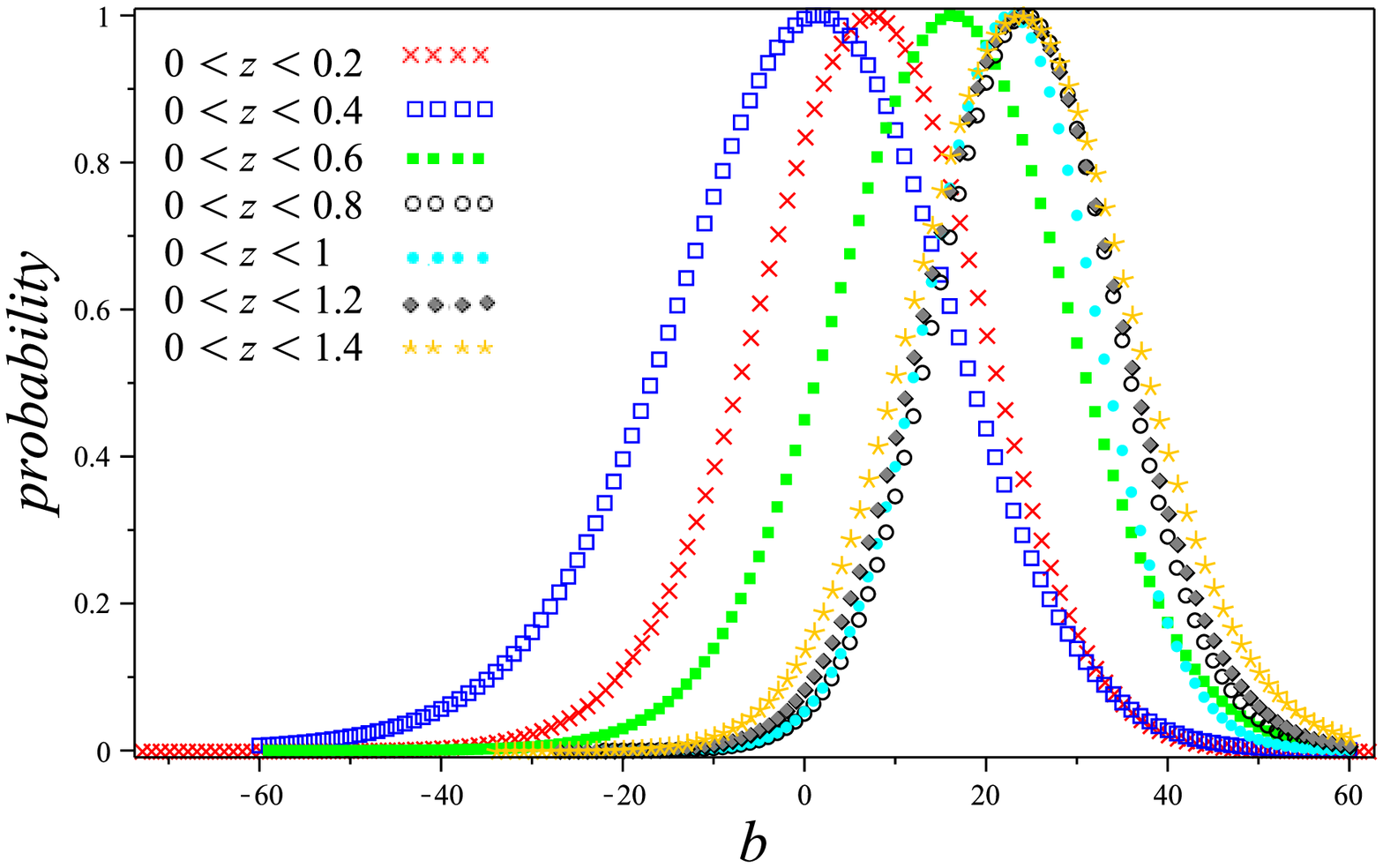}\hspace{0.1 cm}\\
\includegraphics[scale=.3]{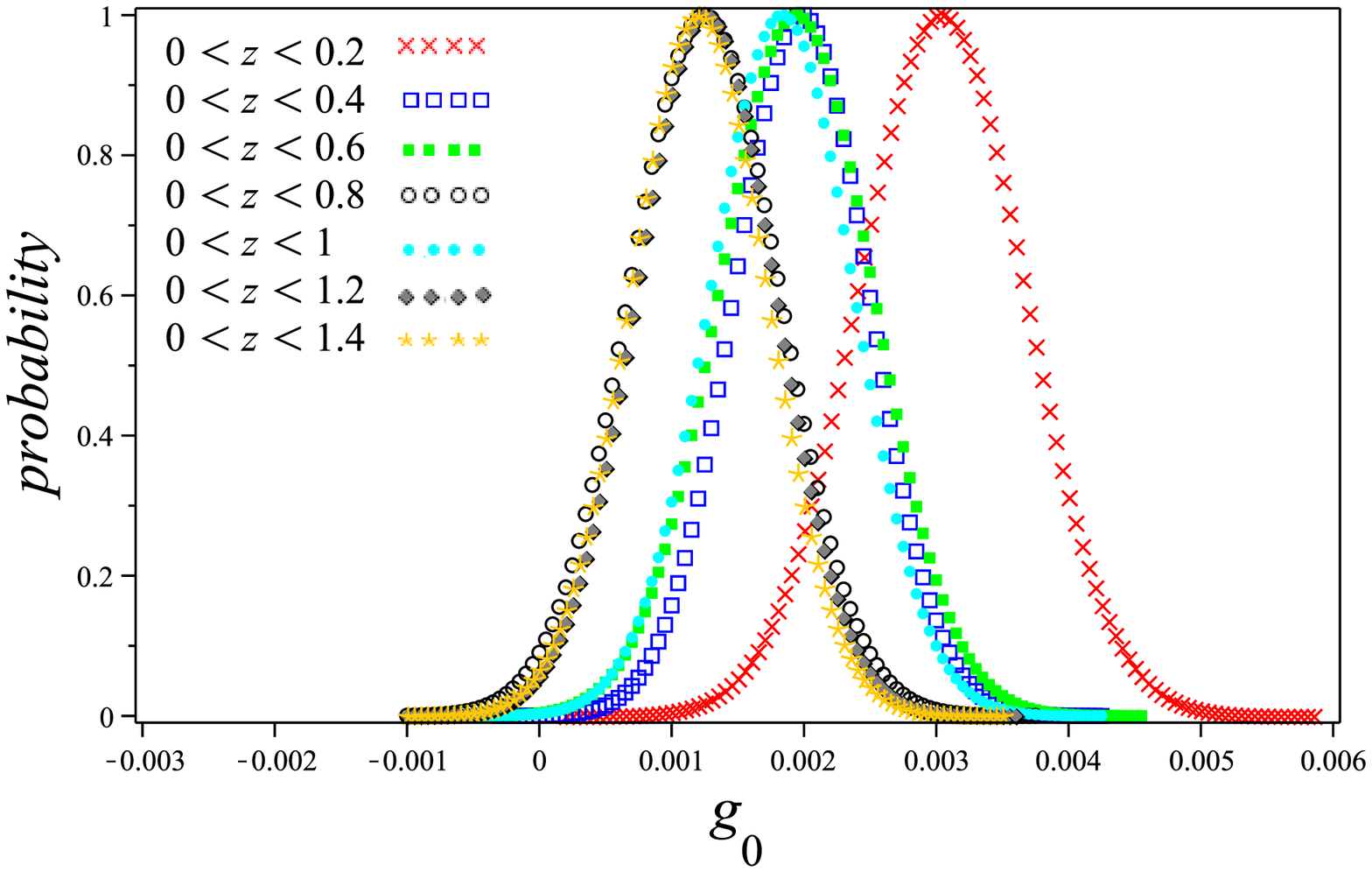}\hspace{0.1 cm}\includegraphics[scale=.3]{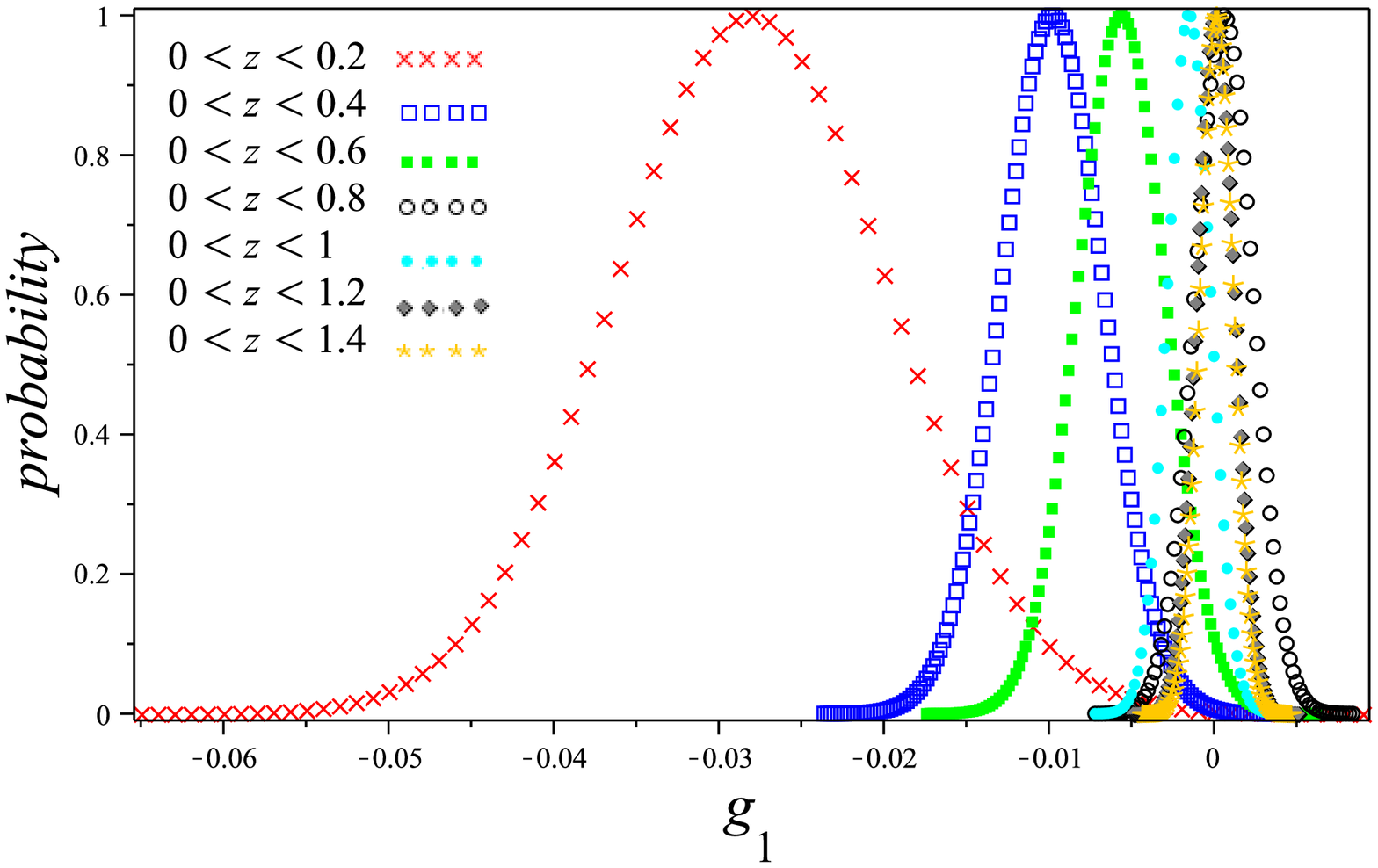}\hspace{0.1 cm}\\
Fig. 25: The redshift tomography analysis for $GDFLD$ method in f(R,T) model.
\label{Figure:19}
\end{figure*}

\begin{table}
\caption{Constraints of the directions and amplitude of maximum anisotropy using $GDFLD$ method for different redshift bins of the
SNIa data. The error-bars quoted is 1$\sigma$ error.} 
\centering 
\begin{tabular}{c c c c c c c  } 
\hline\hline 
range &\ \ $l$&\ \ $b$ &$ g_{0}$ & $g_{1}$ \\ [3ex] 
\hline 
0 - 0.2\ \ &$138.6^{-29}_{+28}$  & $7.3^{-19}_{+18}$ \ & $0.00302^{-0.00095}_{+0.00090}$\ \  & $-0.0278^{-0.0121}_{+0.0120}$ \\ 
\hline 
0 - 0.4\ \  &$129.4^{-34}_{+34}$\ &$1.3^{-21}_{+22}$\ & $0.00201^{-0.00080}_{+0.00075}$\ \  & $-0.0098^{-0.0048}_{+0.0048}$\\ 
\hline 
0 - 0.6\ \  &$121.8^{-33}_{+32}$\ &$16.3^{-20}_{+19}$\ & $0.00194^{-0.00090}_{+0.00090}$\ \  & $-0.0058^{-0.0040}_{+0.0040}$\\ 
\hline 
0 - 0.8\ \  &$133.9^{-30}_{+29}$\ &$24.9^{-16}_{+17}$\ & $0.00125^{-0.00085}_{+0.00090}$\ \  &$0.0006^{-0.0027}_{+0.0027}$\\ 
\hline 
0 - 1.0\ \   & $128.4^{-26}_{+24}$\ &$22.7^{-14}_{+14}$\ & $0.00184^{-0.00085}_{+0.00080}$\ \ &$-0.0015^{-0.0015}_{+0.0015}$\\ 
\hline 
0 - 1.2\ \  &$134.6^{-35}_{+34}$\ &$24.6^{-19}_{+20}$\ & $0.00126^{-0.00080}_{+0.00080}$\ \  &$0.0001^{-0.0018}_{+0.0017}$\\ 
\hline 
0 - 1.4\ \ &$137.7^{-32}_{+32}$\ &$23.7^{-18}_{+18}$\ & $0.00120^{-0.00080}_{+0.00080}$\ \ &$0.0001^{-0.0015}_{+0.0015}$\\ 
 [1ex] 
\hline\hline 
\end{tabular}
\label{table:6} 
\end{table}

In order to elaborate redshift tomography figuratively, we have plotted the likelihood of the parameters $(d_{1}, m_{1}, l_{1}, b_{1})$, $(d_{2}, m_{2}, l_{2}, b_{2})$, $(g_{0}, g_{1}, l_{3}, b_{3})$ for each redshift slice in some figures. The redshift tomography analyses in Fig. 23., Fig. 24. and Fig. 25. show that the preferred axes at different redshifts are all located in a relatively small region of the Galactic Hemisphere. The maximum anisotropic deviation direction is for (DMFDM) method as  $(l, b)=(135^{+25}_{-25}, 23^{+14}_{-15})$, for (DMFLD) as $(l, b)=(135^{+35}_{-37}, 23^{+18}_{-18})$, and  for (GDFLD) method as $(l, b)=(137^{-32}_{+32}, 23^{+18}_{-18})$. Note that these directions are equivalent to $(l, b)=(315^{+25}_{-25}, -23^{+14}_{-15})$ for (DMFDM) method, $(l, b)=(315^{+35}_{-37}, -23^{+18}_{-18})$ for (DMFLD), and $(l, b)=(317^{-32}_{+32}, -23^{+18}_{-18})$ for (GDFLD) method as the maximum axis.

\section{comparison of the f(R,T) model with $CPL$ parametrization, $\omega CDM$ and $\Lambda CDM$ models}
In this section, we compare  $c_{R} R^{\alpha+1}+c_{T}\sqrt{-T}$ Gravity model with $CPL$ parametrization, $\omega CDM$ and  $\Lambda CDM$ models. In the framework of a spatially flat Friedmann universe, the expansion history of the Universe is given by
\begin{align}\label{cp}
H^2(z)=H_{0}^2[\Omega_{m0} (1+z)^3+(1-\Omega_{m0}) f(z)],
\end{align}
\begin{align}
q=\frac{3 w(z) \Omega_{x}(z)+1}{2},
\end{align}

where $H=\frac{\dot{a}}{a}$ is the Hubble parameter, $q$ is the deceleration parameter, $\Omega_{m0}=\frac{\rho_{0}}{\rho_{c}}$ is the current value of the normalized matter density, $\Omega_{x}(z)$ is the normalized dark energy density as a function of redshift which evolves as $\Omega_{x}(z)=\Omega_{x0} f(z) \frac{H_{0}^2}{H^2}$ and
\begin{align}\label{cp2}
f(z)=exp[3 \int_{0}^{z} \frac{1+w{(z^\prime)}}{1+z^\prime} dz^\prime]
\end{align}
Next, we turn to the parametrization of $w(z)$. There are many functional forms of $w(z)$ in the literature. In this work, we consider Chevallier-Polarski-Linder (CPL) parametrization introduced by \cite{Chevalier}, \cite{Chevalier2}, which invokes as barotropic factor the known expression
\begin{eqnarray}
w(z)=w_{0}+w_{1} \frac{z}{1+z}
\end{eqnarray}
In this case, the equation of state becomes $w(z=0) = w_{0}$ at present time and $w(z\rightarrow\infty) = w_{0} + w_{1}$ at earlier time. This simple parametrization is most useful if dark energy is important at late times and insignificant at early times. In addition to its simplicity, this CPL parametrization exhibits interesting properties. However, it cannot describe rapid variations in the equation of state. Using this functional form and Equation (\ref{cp}), Equation (\ref{cp2}) can be written analytically as\\
\begin{eqnarray}
H(z)^{2}&=&H_{0}^{2}[\Omega_{m0}(1+z)^{3}+ (1-\Omega_{m0})(1+z)^{3(1+\omega_{0} + \omega_{1})} \\ \nonumber
&& exp(\frac{-3\omega_{0}z}{1+z})]
\end{eqnarray}
While in the case of $\omega CDM$ model, the equation of state of dark energy is parameterized by a constant
$\omega=\frac{p}{\rho}$; therefore, we have
\begin{eqnarray}
H(z)^{2}=H_{0}^{2}[\Omega_{m0}(1+z)^{3}+(1-\Omega_{m0})(1+z)^{3(1+ \omega)} ]
\end{eqnarray}
\begin{figure*}
\includegraphics[scale=.45]{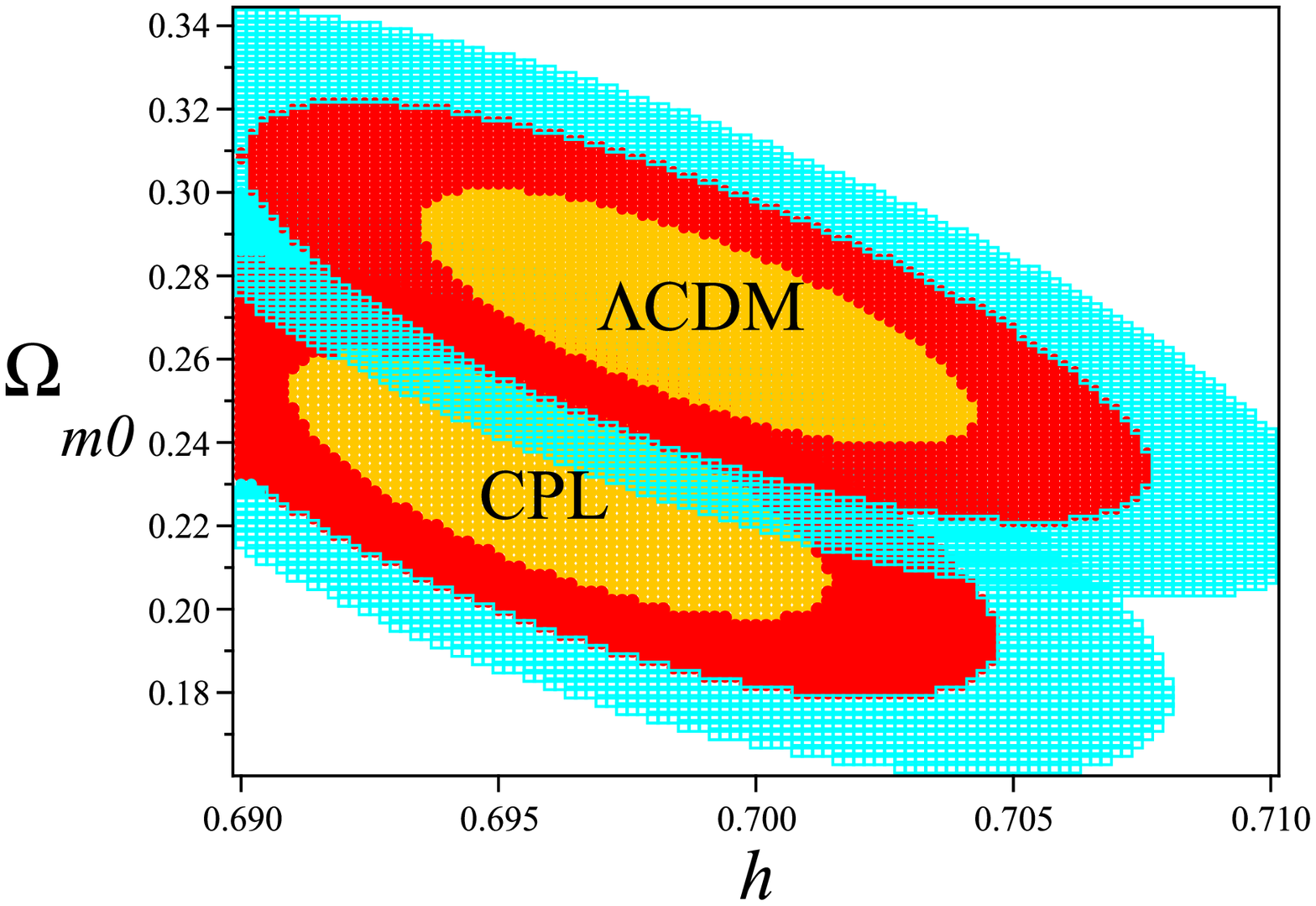}\hspace{0.1 cm}\\
Fig. 26: Confidence levels  for parameters $( \Omega_{m0}, h)$ in $CPL$ and $\Lambda CDM$ models.
\label{Figure:20}
\end{figure*}\\

\begin{table*}
\caption{Best fitted parameters for isotropy background }  
\centering 
\begin{tabular}{|c|c|c|c|c|c|c|c|c|c|c|c|c|c|c||c|} 
\hline\hline 
$Model$   & $\omega$ \ &  $\Omega_{m}$ \ & $\Omega_{\Lambda}$ \ & $\Omega_{CPL}$ \ & $\omega_{0}$& $\omega_{1}$
&   $\alpha$&$\chi^{2}_{min} $& $h$ \\
\hline 
$c_{R} R^{\alpha+1}+c_{T}\sqrt{-T}$  &$ -$ &$ -$ & $-$ & $-$& $-$&$-$&$1^{+0.01}_{-0.01}$& 543.0747981&$0.7^{+0.0147}_{-0.0147}$ \\ 
\hline 
$\omega CDM$  & $-1.05$ & 0.29 & $-$ & $-$& $-$&$-$&$-$& 537.76250&$0.701^{+0.0146}_{-0.0146}$ \\
\hline 
CPL &$ -$ &  $0.23^{+0.03}_{-0.03}$& $-$ & $0.77$ & $-1.23$& $0.18$ & $-$ & $541.0514134$ & $0.6964^{+0.0146}_{-0.0146}$ \\
\hline 
$\Lambda CDM$  & $-$&$ 0.27^{+0.03}_{-0.03}$ & $0.73$& $-$ & $-$& -&$-$& 540.90726 &$0.698^{+0.0148}_{-0.0148}$ \\
\hline 
\end{tabular}\\
\label{table:7} 
\end{table*}

Using Union2 data and by $\chi^{2}$ method, we have best fitted parameters $ \Omega_{m0}, \Omega_{\Lambda}, h$ for $\Lambda CDM$  and parameters $ \Omega_{m0}, \Omega_{CPL}, h, \omega_{0}, \omega_{1}$ for $CPL$ model. Fig. 6 shows the confidence levels for parameters $(\Omega_{m0},h)$ in both $\Lambda CDM$ and $CPL$ models. For $\Lambda CDM$ we have obtained $\Omega_{m0}=0.27$, $\Omega_{\Lambda}=0.73$ and $h=0.698$. For $CPL$ parametrization, we have obtained $\Omega_{m0}=0.23$, $\Omega_{\Lambda}=0.77$, $h=0.6964$, $\omega_{0}=-1.23$ and $\omega_{1}=0.14$ and for $\omega CDM$ parametrization, we have obtained $\Omega_{m0}=0.29$, $\omega=-1.05$ and $h=0.699$. The results summarized in Table \ref{table:7}.\\
We have also considered the $\omega CDM$, $\Lambda CDM$ and the CPL parameterized dark energy models as the isotropic background. We use the isotropic background dark energy parameters in Table \ref{table:7} and fit our anisotropic parameters, respectively. The results are summarized in Table \ref{table:8}. Fig.27., Fig.28. and Fig.29. show the results of constraints on $(l, b)$ which are not much different from the case of the $f(R,T)$ model. This means that the best-fitting value of the maximum deviation direction from the isotropic background is not sensitive to the details of isotropic dark energy models.
The best fitted  trajectories of the effective EoS parameter in isotropic, anisotropic $c_{R} R^{\alpha+1}+c_{T}\sqrt{-T}$ gravity, $ CPL$ and $\omega CDM$ and $\Lambda CDM$ models are shown in Fig. 30. Based on this, the trajectory of anisotropy is not in much difference from the case of the isotropy, also the best fitted trajectory of $CPL$ and $\Lambda CDM$  models are same at late time and different in the future and $\omega CDM$, $\Lambda CDM$ and $c_{R} R^{\alpha+1}+c_{T}\sqrt{-T}$ gravity have  same trajectory at late time and in the future.\\
 \begin{table*}
\caption{the preferred direction of the universe in $\Lambda CDM$ , $CPL$ and $\omega CDM$ models }  
\centering 
\begin{tabular}{|c|c|c|c|} 
\hline\hline 
$Ref$  &  $\Lambda CDM$ \ & $\omega CDM$ \ &  $CPL $ \\
\hline 
 \hline 
this study & $307^{-31}_{+32},-16^{-18}_{+18}$ &$308^{-30}_{+28},-18^{-16}_{+16}$&$309^{-30}_{+27},-17^{-16}_{+17}$ \\
\hline 
\cite{Cai0} & $308^{-23}_{+22},-16^{-14}_{+21}$ &$308^{-28}_{+17},-14^{-30}_{+17}$&$307^{-21}_{+17},-15^{-32}_{+17}$ \\
\hline 
\cite{Yang} & $307^{-16}_{+16},-14^{-10}_{+10}$ & $307^{-16}_{+16},-14^{-10}_{+10}$&$-$ \\
\hline 
\cite{Mariano} & $309^{-18}_{+18},-15^{-12}_{+12}$ & -&- \\
\hline 
\end{tabular}
\label{table:8} 
\end{table*}

\begin{figure*}
\includegraphics[scale=.8]{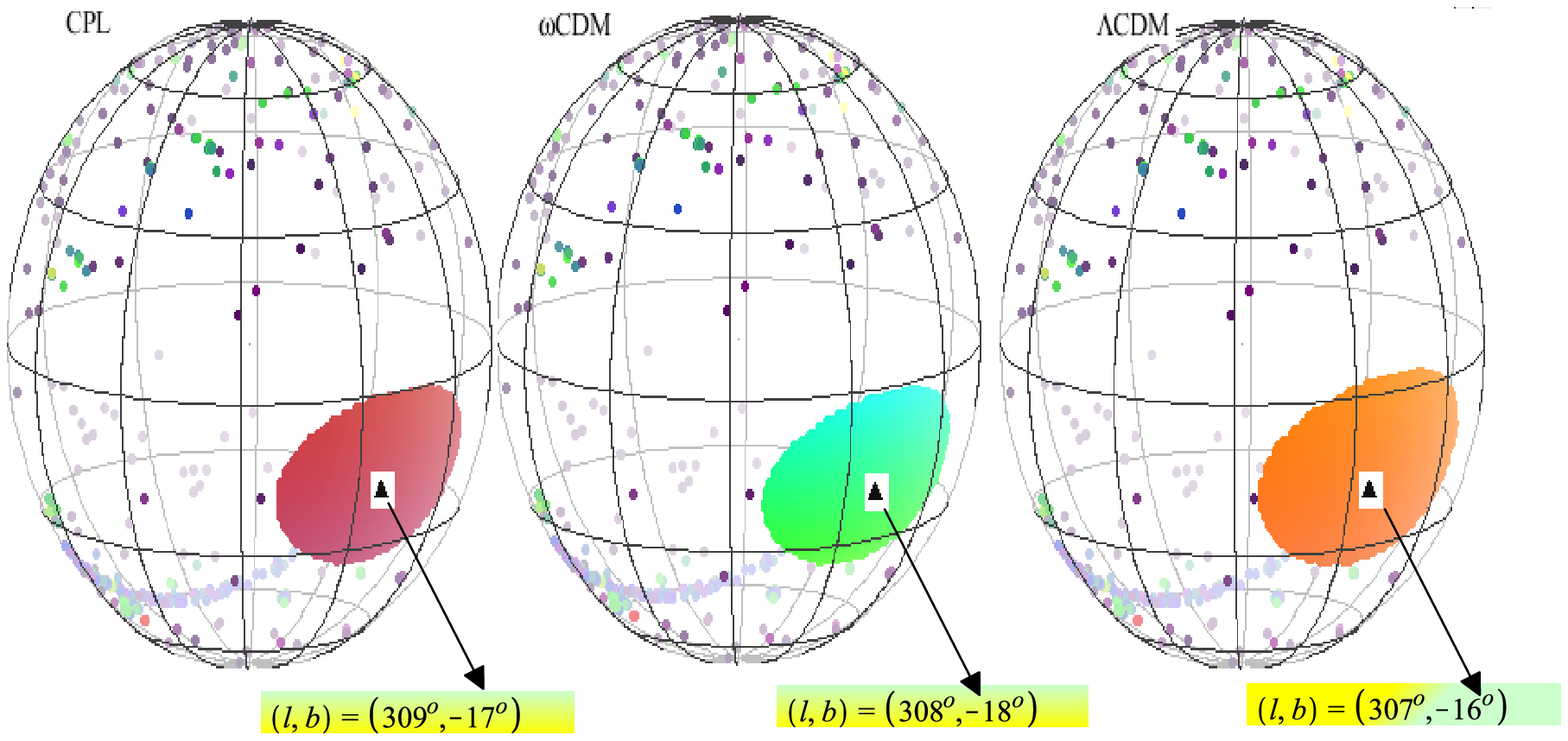}\hspace{0.1 cm}\\
Fig. 27: Union2 datapoints and ($1-\sigma$) confidence level for Dark Energy dipole direction($l,b$) in $CPL$ parametrization , $\omega CDM$ and $\Lambda CDM$ models(used Monte Carlo simulation with $10^{5} $ datapoints)
\label{Figure:21}
\end{figure*}

\begin{figure*}
 \includegraphics[scale=.4]{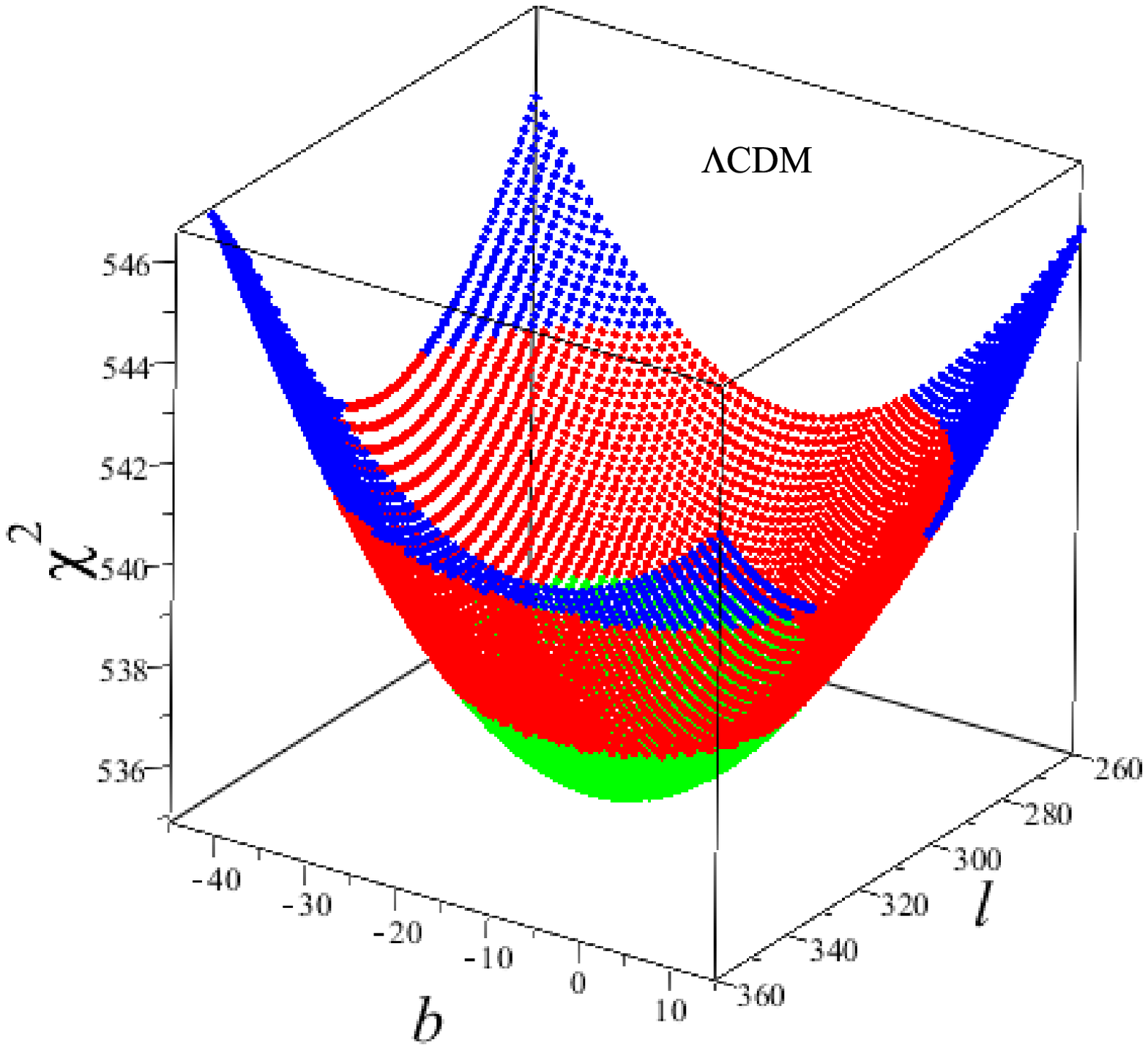}\hspace{0.1 cm}\includegraphics[scale=.4]{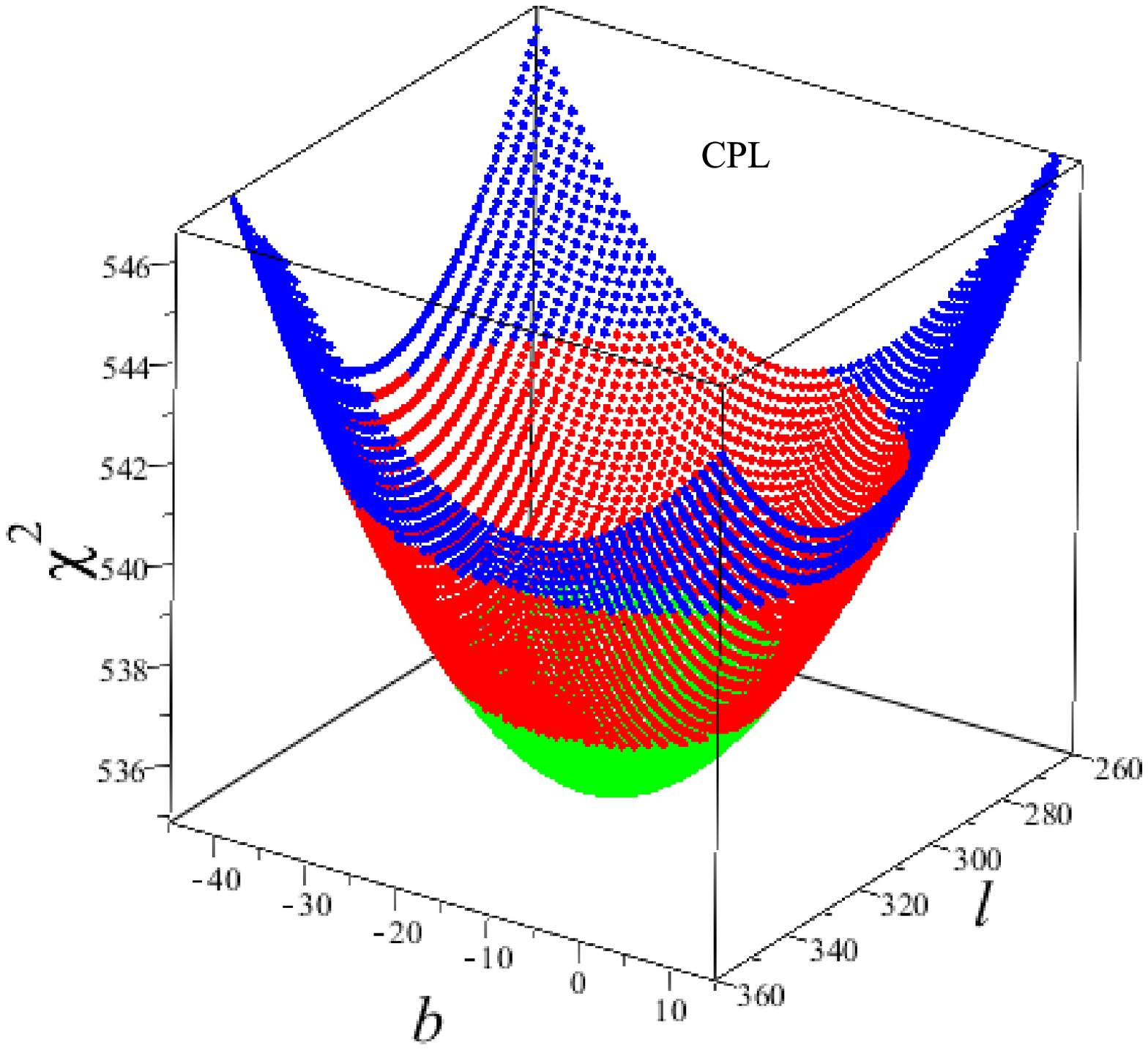}\hspace{0.1 cm}\\
\includegraphics[scale=.4]{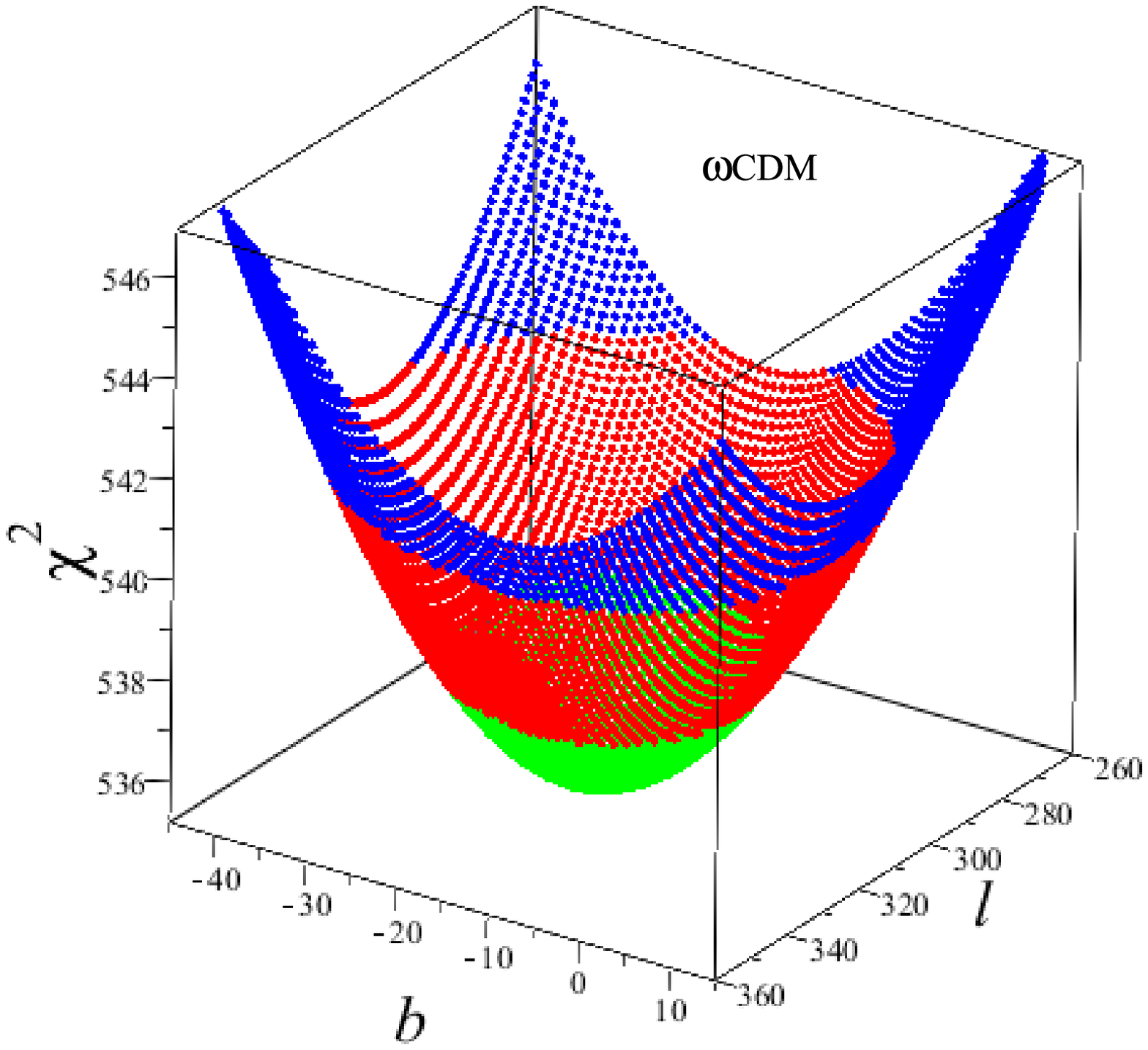}\hspace{0.1 cm}\includegraphics[scale=.4]{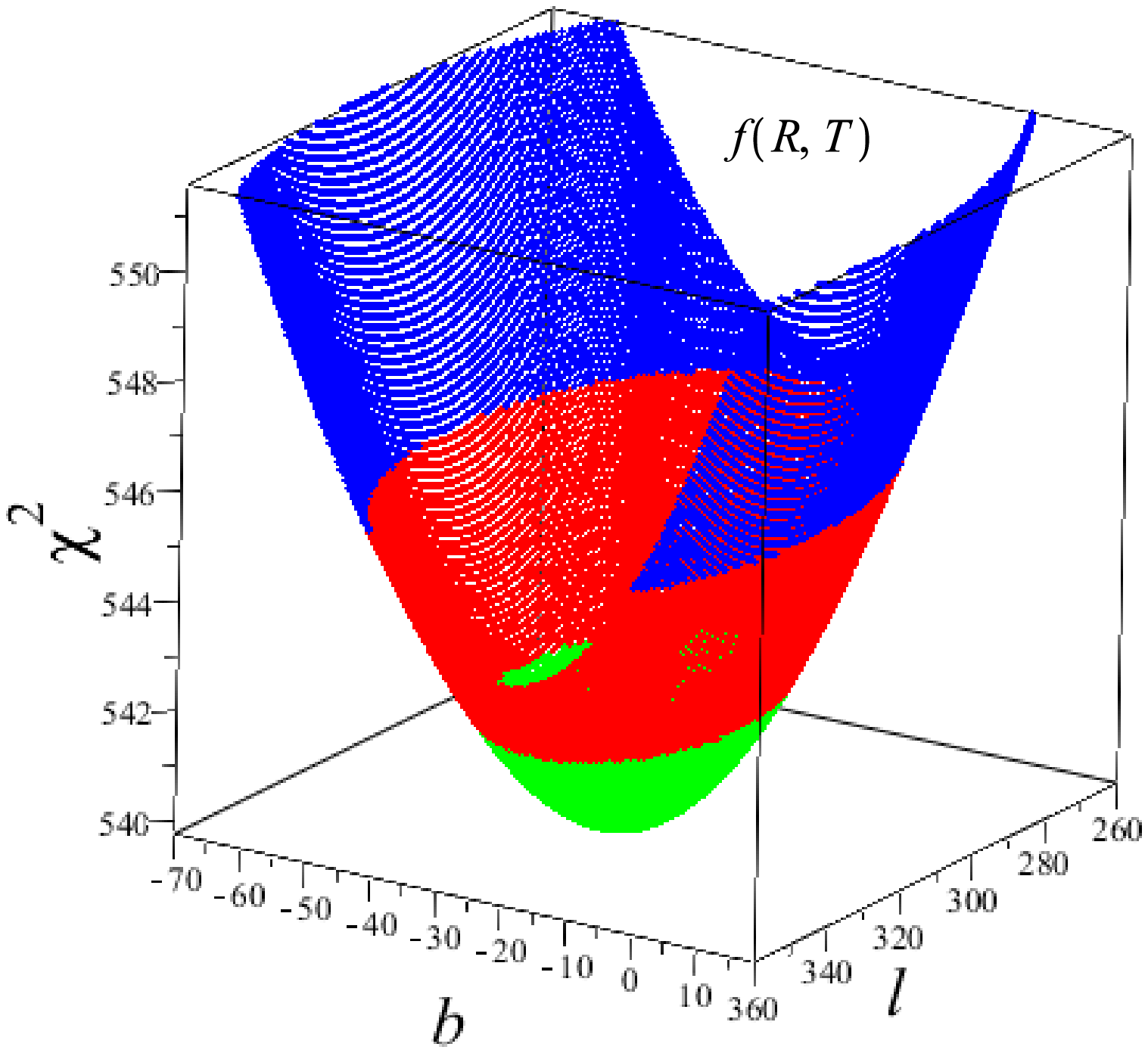}\hspace{0.1 cm}\\
Fig. 28: Two dimensional $\chi^{2}$ of $(l,b)$ for $\Lambda CDM$, $CPL$, $\omega CDM$ and $f(R,T)$ models. \\
\label{Figure:22}
\end{figure*}

\begin{figure*}
\includegraphics[scale=.7]{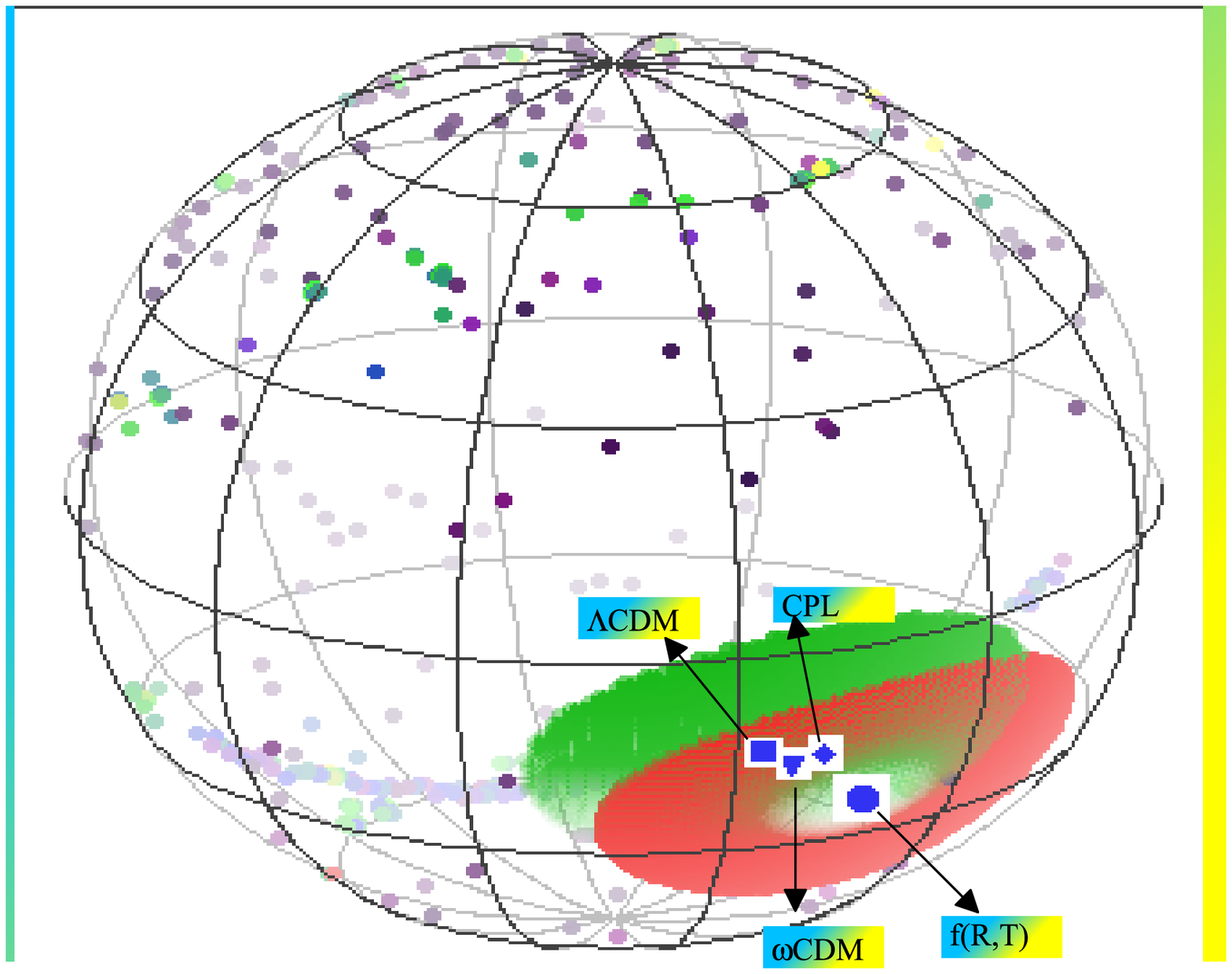}\hspace{0.1 cm}\\
Fig. 29: the $1-\sigma$ errors on the Dark Energy dipole direction, for $\Lambda CDM$, $CPL$, $\omega CDM$\\ and $f(R,T)$ models.\\
\label{Figure:23}
\end{figure*}

\begin{figure*}
\includegraphics[scale=.4]{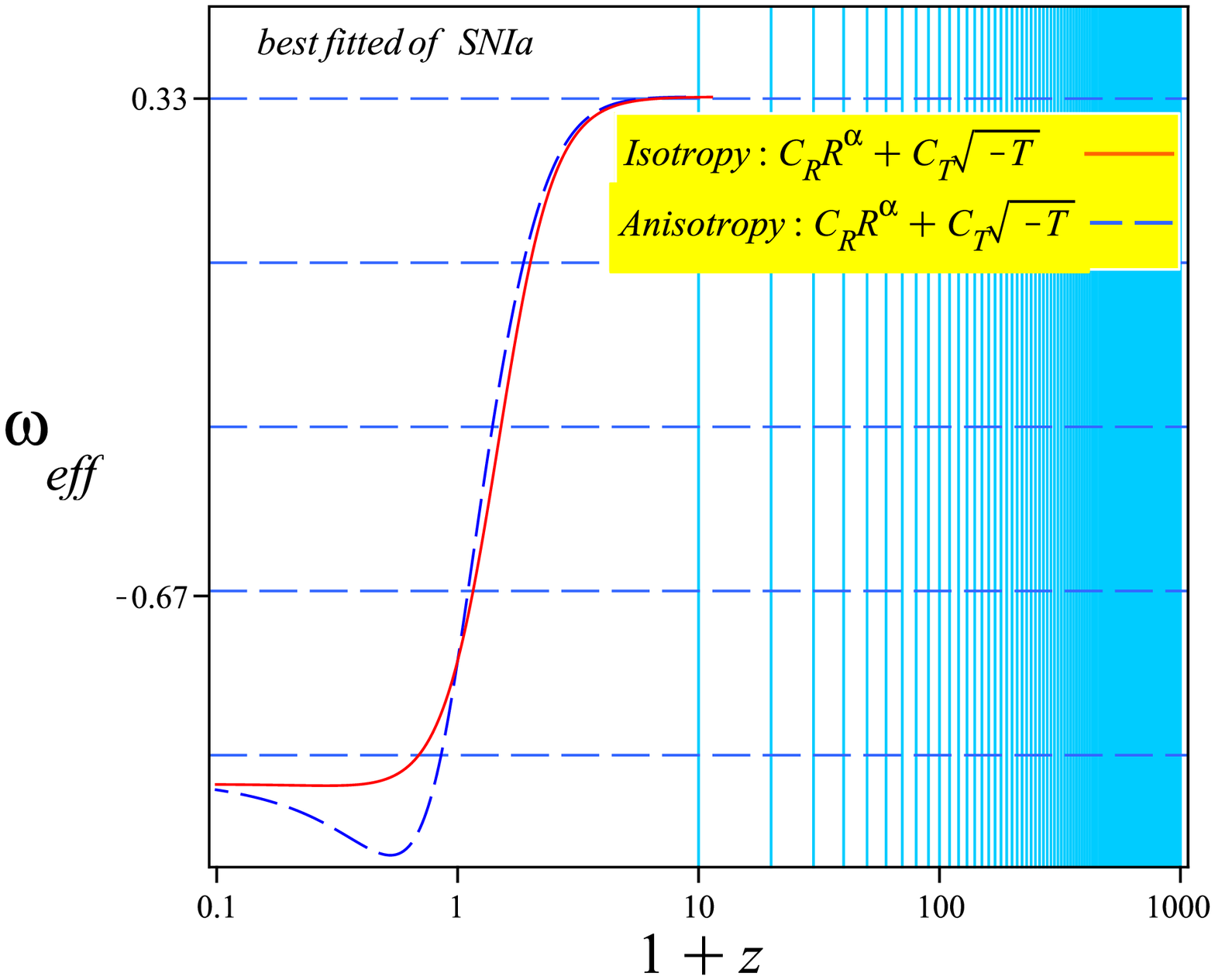}\hspace{0.1 cm}\includegraphics[scale=.4]{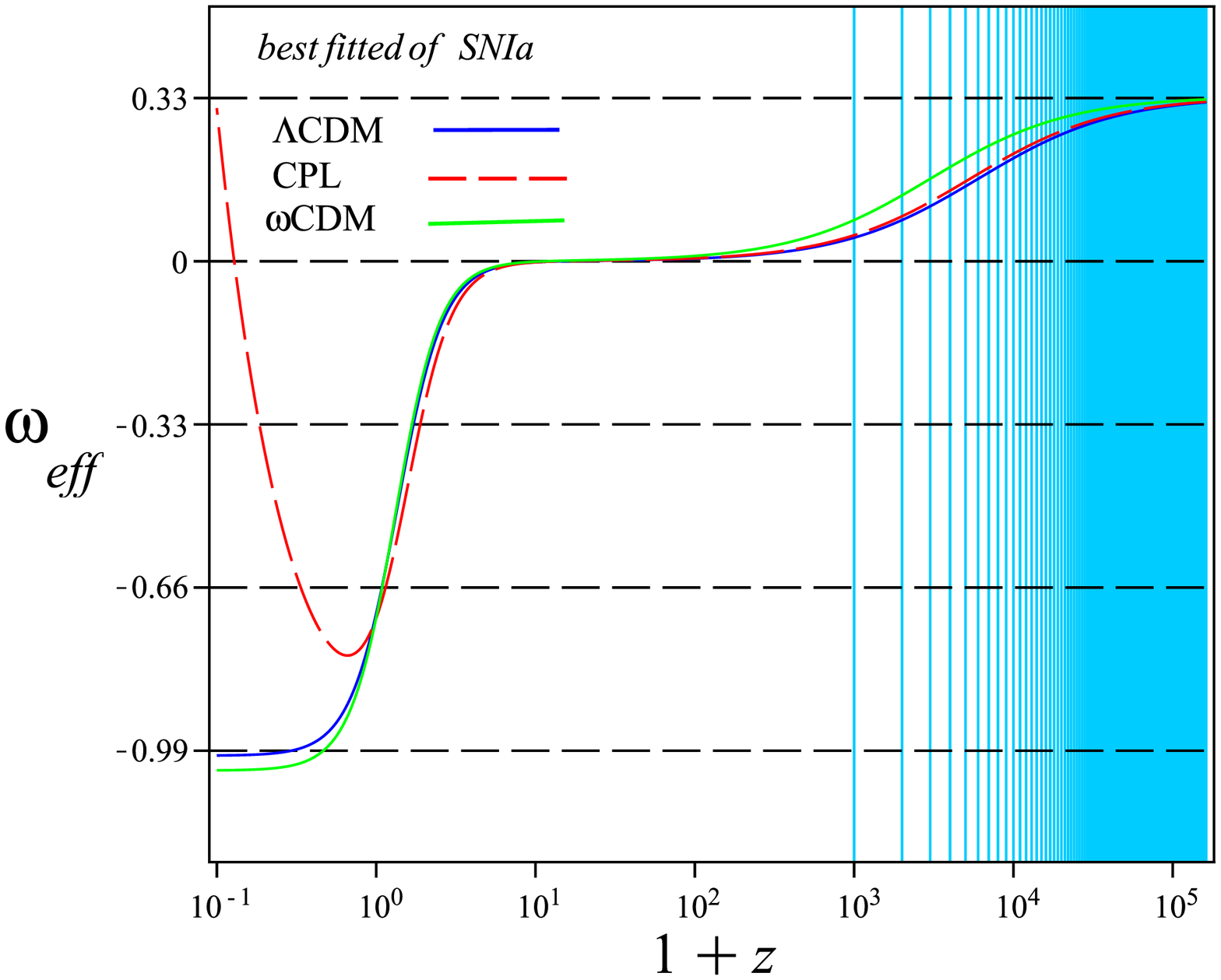}\hspace{0.1 cm}\\
Fig. 30: The best fitted trajectory of Equation of State for\\ (Left) isotropic and anisotropic $f(R,T)$ models.\\ (Right) CPL, $\omega CDM$ and $\Lambda CDM$ models.
\label{Figure:24}
\end{figure*}
\begin{figure*}
\includegraphics[scale=.6]{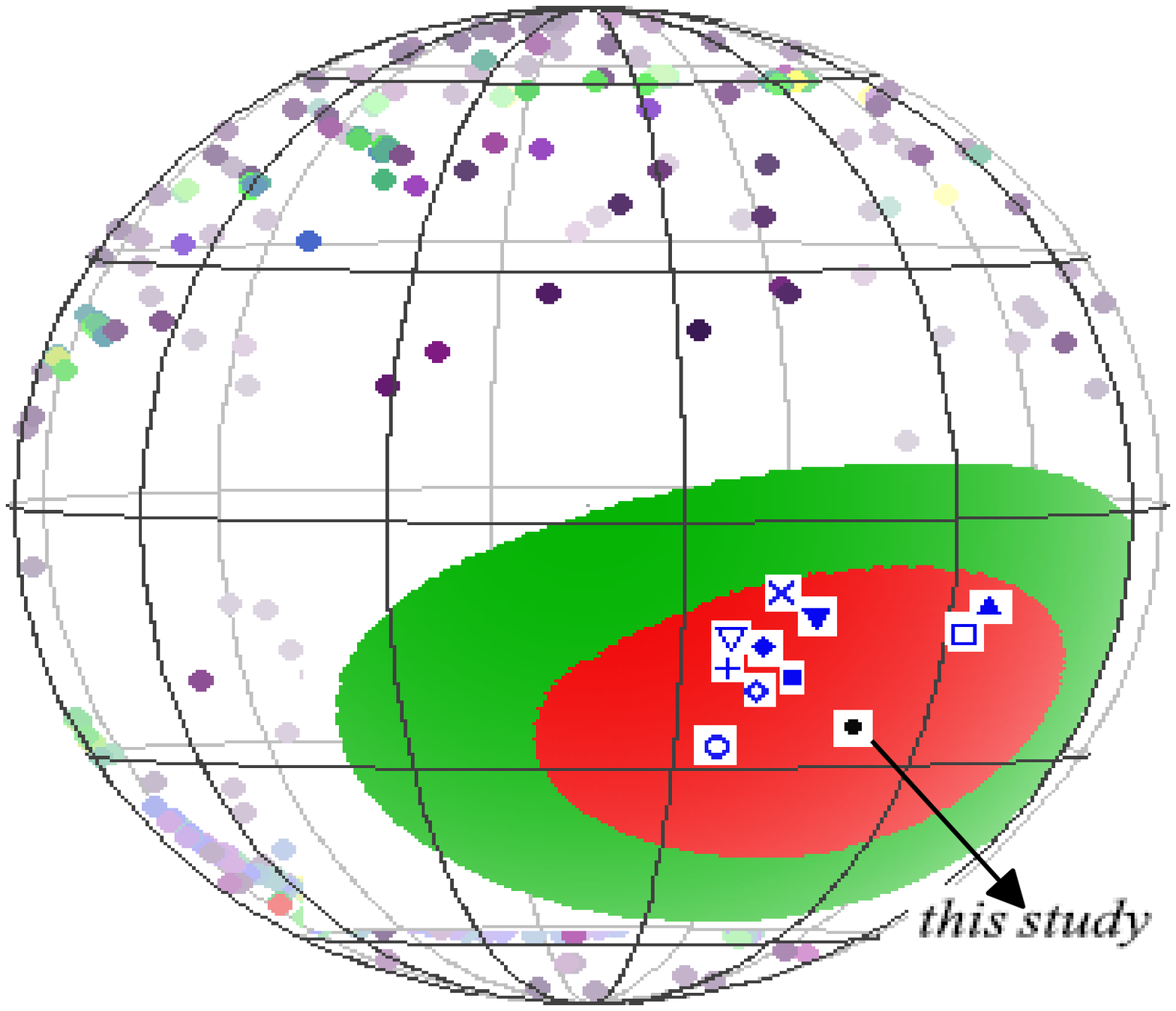}\hspace{0.1 cm}\\
Fig. 31: The direction of preferred axis in galactic coordinate. The point red \\ $\bullet$ denotes our result,
namely,$(l, b)=(317.7^{-32}_{+32},-23.7^{-18}_{+18})$. The results for \\preferred direction in other models are presented for contrast.Point $\times$ denotes\\ the  result of \cite{Wang}, point $\blacklozenge$ denotes the result of \cite{Yang},point $\triangledown$ denotes the result \\of \cite{Cai}, point $\blacktriangledown$ denotes the result of \cite{chang},\\ point $\blacktriangle$ denotes the result of \cite{chang}, point $\diamond$ denotes the result of \cite{Chang1}, point $\square$\\ denotes the result of\cite{Webb1}, point $\circ$ denotes the result of \cite{Chang1} ,  point $\blacksquare$ \\denotes the result of \cite{Mariano}, and point $+$ denotes \\the result of \cite{Cai}. The light green represents \\the 1-$\sigma$ errors on the Dark Energy dipole direction, which includes \\the results for preferred direction in other models.\\
\label{Figure:25}
\end{figure*}

\begin{table*}
\caption{Comparison of the preferred axes of the Universe in f(R,T), CPL, $\omega CDM$ and $\Lambda CDM$ models}  
\centering 
\begin{tabular}{|c|c|c|c|} 
\hline\hline 
$Model$  &  $l$ \ & $b$ \ &  $\chi^{2}_{min} $ \\
\hline 
 \hline 
$c_{R} R^{\alpha+1}+c_{T}\sqrt{-T}$ &$ 317^{\circ}$ & $-23^{\circ}$ & 537.76250 \\
\hline 
CPL &$309^{\circ}$ & $-17^{\circ}$ & 534.87158  \\
\hline 
$\Lambda CDM$ &$307^{\circ}$& $-16^{\circ}$ & 534.88649 \\
\hline 
$\omega CDM$ & $308^{\circ}$ & $-18^{\circ}$ &535.14416 \\
\hline 
\end{tabular}
\label{table:9} 
\end{table*}
\begin{table*}
\caption{Directions of Preferred axes from different cosmological observations} 
\centering 
\begin{tabular}{c c c c c} 
\hline\hline 
Cosmological Obs. &\ \ $l$&\ \ $b$ &\ \ $Reference$&\ \ $Point$  \\ [3ex] 
\hline 
Dark Energy dipole\ \ & $317.7^\circ$ &$-23^\circ$  & This study & red $\bullet$\\ 
\hline 
Dark Energy dipole\ \ & $309.2^\circ$ &$-8.6^\circ$  &\cite{Wang} & $\times$\\ 
\hline 
Dark Energy dipole\ \ & $307.1^\circ$ &$-14.3^\circ$  &\cite{Yang} & $\blacklozenge$\\ 
\hline 
Dark Energy dipole\ \ & $306^\circ$ &$-13^\circ$  &\cite{Cai} & $\triangledown$\\ 
\hline 
Dark Energy dipole \ & $314^\circ$ &$-11^\circ$  & \cite{chang} & $\blacktriangledown$\\ 
\hline 
$\alpha$ dipole \ & $333^\circ$ &$-12^\circ$  & \cite{chang} & $\blacktriangle$\\ 
\hline 
Dark Energy dipole \ & $306^\circ$ &$-18^\circ$  & \cite{Chang1} & $\diamond$\\ 
\hline 
$\alpha$ dipole\ \ & $331^\circ$ &$-14^\circ$  & \cite{Webb1}& $\square$\\ 
\hline 
Dark Energy dipole\ \ & $304^\circ$ &$-27^\circ$  & \cite{Chang1} & $\circ$ \\ 
\hline 
Dark Energy dipole\ \ & $309^\circ$ &$-18^\circ$  & \cite{Mariano}& $\blacksquare$\\ 
\hline 
Dark Energy dipole\ \ & $306^\circ$ &$-13^\circ$  & \cite{Cai}& $+$\\ 
\hline 
\hline\hline 
\end{tabular}\\
\label{table:10} 
\end{table*}

\section{CONCLUSIONS}
We have investigated the cosmological solutions of $f(R,T)$ gravity, in isotropic and anisotropic space-time. In both isotropic and anisotropic cases, our studies are based on the phase–space analysis (the dynamical system approach). In this approach, we convert a set of second order differential equations to a new set of first order ones by defining some dimensionless variables and parameters. There are various reasons for doing this: a first order system is much easier to solve numerically, and also phase planes are useful in visualizing the behavior of dynamical systems, especially in oscillatory systems where the phase paths can "spiral in" towards zero, and "spiral out" towards infinity. Moreover, it gives us useful information  about (in)stability of the system and critical points of the system.\\
At first,  we have obtained the field equations of $f(R, T)$ gravity in isotropy case and have analyzed the stability of the dynamical system for $f(R, T)=f_{1}(R)+f_{2}(T)$. Then, we have studied the evolution of scalar cosmological perturbations in the metric formalism.
The main purpose of scalar perturbations is to find explicit expressions for the parameter  $\Phi$, $\Psi$ and $\delta$ in the framework of nonlinear $f(R,T)$ model. Unfortunately, the system of equations for scalar perturbations is very complicated in the case of nonlinearity. It is hardly possible to solve it directly. Therefore, we have used phase space approach to simplify the nonlinear equations of the  $f(R, T)=f_{1}(R)+f_{2}(T)$ model. We have also reconstructed the parameters $\Phi,\Psi$ and $\delta$ from new variables. In the model, the evolution of matter density perturbations for different cases has been studied and the corresponding results have been shown in Fig. 4, 5, 6, 7, 8 and 9. The attractor property (spiral in and out) of the system leads to an oscillating behavior of the matter perturbations and other variables and parameters. This behavior is predictable for the critical points of the dynamical systems whose eigenvalues are complex.\\

In section 5, we have investigated Dark Energy Dipole in the $f(R, T)$ model using Dipole Fitting method. There is a range of independent cosmological observations which indicate the existence of anisotropy axes. This appears to be one of the most likely directions which may lead to new fundamental physics in the coming years. These cosmological observations along with their preferred directions and the corresponding references are summarized in Table \ref{table:10}.\\
In this paper, we present a detailed analysis of the dark energy dipoles in $f(R,T)=f_{1}(R)+f_{2}(T)$ cosmological model using three types of dipole-fitting (DF) method which are (I)dipole + monopole fitting for distance modulus (DMFDM), (II)dipole + monopole fitting for luminosity distance (DMFLD) and (III) general dipole fitting for luminosity distance (GDFLD).\\
Several groups have applied $DMFDM$ method to study the anisotropy of $\Lambda CDM$, $\omega CDM$ and the dark energy model with $CPL$ parametrization. \cite{Cai} have applied $ GDFLD $ method to study the anisotropy of $\Lambda CDM$, $\omega CDM$ and the dark energy model with $CPL$ parametrization. We have applied all of these DF  methods to study privilege axis of the universe in $f(R,T)$ model.\\
At first, it seems that these methods have a same origin (because of the direct relation between $\mu$ and $d_{L}$). Also, the best fitted  direction of preferred axis of these methods are very close to each other. In fact, $DMFDM$ ($(l,b)=(315^{0}\pm25^{0},-23^{0}\pm15^{0})$) and $DMFLD$ ($(l,b)=(315^{0}\pm37^{0}, -23^{0}\pm 18^{0})$) methods have resulted exactly the same value for the privilege axis of the universe in $f(R,T)$ model. However, their $1-\sigma$ confidence level are different (Fig.22.). The ($1-\sigma$) confidence region of $DMFDM$ is smaller than $DMFLD$. Moreover, they give different values of dipole magnitude which is interesting to note. The dipole magnitude obtained using  $DMFDM$ method ($d_{1}=(1.4\pm 0.8)\times \times 10^{-3}$) is close to previous studies of \cite{Chang4}, \cite{Wang}, \cite{Yang} as it has been mentioned in $DMFDM$ method section. However, the dipole magnitude obtained using $DMFLD$ method ($d_{2}=(0.026\pm 0.014)$) is different from the value obtained using $DMFDM$ method and also previous studies. Interestingly, the magnitude of anisotropy ($d_{2}=(0.026\pm 0.014)$) obtained using $DMFLD$ method is approximately equal to that of CMB dipole. The recent released Planck data show that the dipole magnitude of CMB temperature fluctuations is about $A=0.07 ± 0.01$ (\cite{Chang3}). Also, it is close to the result of (\cite{Chang3}) which have obtained the magnitude of dipolar asymmetry as $|D|=0.044\pm0.018$, using modified luminosity distance in anisotropic cosmological model in the Finsler-Randers spacetime (formula 2 of Table \ref{table:2}).\\

Further results of this paper are  as follows:\\

1. We have found The maximum anisotropic deviation direction for (DMFDM) method as $(l, b)=(315^{+25}_{-25},-23^{+14}_{-15})$, for (DMFLD) as $(l, b)=(315^{+35}_{-37},-23^{+18}_{-18})$, and for (GDFLD) method as $(l, b)=(317^{+32}_{-32},-23^{+18}_{-18})$ which are located very close to each other. Also, the results are consistent with other studies (\cite{Mariano}, \cite{chang},\cite{a122}, \cite{a128}, \cite{Bennett}, \cite{ Cai}). It is interesting that the results of other studies are in $(1-\sigma)$ confidence level of our study (see Fig. 31).\\

2. The dipole directions at high and low redshifts are in agreement. (This is confirmed in the redshift tomography analyses, shown in Fig. 23, 24 and 25.)\\

3.We have also applied $DMFLD$ method to find the preferred direction of the universe in $\Lambda CDM$ , $CPL$ and $\omega CDM$ models using $\chi^{2}$ method. Our results are very close to pervious works which studied anisotropy in these models (see Table\ref{table:8}). It is interesting that the results of constraints on $(l, b)$ in $f(R,T)$ model are not much different from the cases of the $\Lambda CDM$, $\omega CDM$ and $CPL$ models (see Table \ref{table:9} and Fig. 29). This means that the best-fitting value of the maximum deviation direction from the isotropic background is not sensitive to the details of isotropic dark energy models.\\

\end{document}